\documentclass[aps,pra,reprint,twocolumn,superscriptaddress,floatfix,nofootinbib,longbibliography]{revtex4-1}
\usepackage{graphicx,amsmath,amsfonts,amssymb,amsthm,xr}
\usepackage{epsfig,amsmath,amssymb,color,dsfont,upgreek,physics}
\usepackage{mathrsfs}
\usepackage{mathtools}
\usepackage{bbold}
\usepackage{comment}
\usepackage{float}

\usepackage[bookmarks=true,colorlinks,linkcolor=OrangeRed,urlcolor=NavyBlue,citecolor=RoyalBlue]{hyperref}
\usepackage[dvipsnames]{xcolor}
\usepackage{orcidlink}

\definecolor{mygold}{rgb}{0.93,0.69,0.13}

\definecolor{mypurple}{rgb}{0.49,0.18,0.56}

\definecolor{mygreen}{rgb}{0,0.5,0}

\definecolor{mygreen}{rgb}{0,0.5,0}
\definecolor{myred}{rgb}{0.7,0,0}
\definecolor{myblue}{rgb}{0,0,0.5}

\begin{document}
\title{Suppression of $1/f$ noise in quantum simulators of gauge theories}
\author{Bhavik Kumar${}^{\orcidlink{0000-0002-3755-0300}}$}
\affiliation{Department of Physical Sciences, Indian Institute of Science Education and Research (IISER), Mohali, Knowledge City, Sector 81, Punjab 140306, India}
\author{Philipp Hauke${}^{\orcidlink{0000-0002-0414-1754}}$}
\affiliation{INO-CNR BEC Center and Department of Physics, University of Trento, Via Sommarive 14, I-38123 Trento, Italy}
\affiliation{INFN-TIFPA, Trento Institute for Fundamental Physics and Applications, Trento, Italy}
\author{Jad C.~Halimeh${}^{\orcidlink{0000-0002-0659-7990}}$}
\email{jad.halimeh@physik.lmu.de}
\affiliation{Department of Physics and Arnold Sommerfeld Center for Theoretical Physics (ASC), Ludwig-Maximilians-Universit\"at M\"unchen, Theresienstra\ss e 37, D-80333 M\"unchen, Germany}
\affiliation{Munich Center for Quantum Science and Technology (MCQST), Schellingstra\ss e 4, D-80799 M\"unchen, Germany}

\begin{abstract}
In the current drive to quantum-simulate evermore complex gauge-theory phenomena, it is necessary to devise schemes allowing for the control and suppression of unavoidable gauge-breaking errors on different experimental platforms. Although there have been several successful approaches to tackle coherent errors, comparatively little has been done in the way of decoherence. By numerically solving the corresponding Bloch--Redfield equations, we show that the recently developed method of \textit{linear gauge protection} suppresses the growth of gauge violations due to $1/f^\beta$ noise as $1/V^\beta$, where $V$ is the protection strength and $\beta>0$, in Abelian lattice gauge theories, as we show through exemplary results for $\mathrm{U}(1)$ quantum link models and $\mathbb{Z}_2$ lattice gauge theories. We support our numerical findings with analytic derivations through time-dependent perturbation theory. Our findings are of immediate applicability in modern analog quantum simulators and digital NISQ devices. 
\end{abstract}

\date{\today}
\maketitle

\tableofcontents
\section{Introduction}
Quantum simulators are quantum systems implementable in the laboratory onto which quantum many-body models of interest can be mapped and studied \cite{Bloch2008,Hauke2012,Georgescu_review,Altman_review}.
Due to its promise as a probe of phenomena relevant for high-energy and nuclear physics on easily accessible table-top quantum devices, and its potential to calculate time evolution from first principles, the quantum simulation of lattice gauge theories \cite{Rothe_book} has come at the forefront of research in several fields ranging from condensed matter to subatomic physics \cite{Alexeev_review,klco2021standard,Dalmonte_review,Zohar_review,aidelsburger2021cold,Zohar_NewReview,Bauer_review,Catterall2022}. 
Thanks to the advent of high-control and precision synthetic quantum devices, recent years have seen various groundbreaking quantum-simulation experiments of gauge theories \cite{Martinez2016,Muschik2017,Bernien2017,Klco2018,Kokail2019,Schweizer2019,Goerg2019,Mil2020,Klco2020,Yang2020,Zhou2021,Nguyen2021,Wang2022,Mildenberger2022}. 

Of particular interest in this endeavor are gauge theories with both dynamical matter and gauge fields. The characteristic property of gauge theories is their gauge symmetry \cite{Weinberg_book,Gattringer_book,Zee_book}, which imposes local constraints that enforce specific configurations of matter and electric fields, such as Gauss's law from quantum electrodynamics. 
A major issue in quantum simulations is stabilizing gauge symmetry against gauge-breaking terms that will unavoidably arise either due to higher orders in the perturbative mapping or due to experimental imperfections \cite{Halimeh2020a}. These terms allow for processes driving the system dynamics out of the \textit{physical gauge sector} of Gauss's law, in which it should stay in an ideal scenario where such terms are not present. Even when perturbative in strength, gauge-breaking terms can be quite detrimental to gauge-theory quantum simulations, leading to gauge-noninvariant dynamics that cannot be directly related to the target model \cite{Halimeh2020b,Halimeh2020c,Halimeh2020f}.

Various methods have been proposed to suppress coherent gauge-breaking errors \cite{Zohar2011,Zohar2012,Banerjee2012,Zohar2013,Banerjee2013,Hauke2013,Stannigel2014,Kuehn2014,Kuno2015,Yang2016,Kuno2017,Negretti2017,Dutta2017,Barros2019,Halimeh2020a,Kasper2021nonabelian,Lamm2020,Halimeh2020e,Halimeh2021stabilizing,Halimeh2021gauge,Halimeh_BriefReview}, but there has been little work done on suppressing incoherent errors due to decoherence, which can be quite adverse to the stability of gauge-theory implementations \cite{Halimeh2020f,Halimeh2020g}. Indeed, decoherence \cite{Zeh1970,Schlosshauer2005} poses a major roadblock to achieving long evolution times in quantum simulations of quantum many-body models in general, whose key properties of quantum entanglement and superposition are particularly sensitive to interactions with the environment. Prominent examples of the detrimental effects of decoherence on quantum many-body systems include $1/f$ noise in superconducting quantum interference devices (SQUIDs) that undermines superconducting qubits \cite{Yoshihara2006,Kakuyanagi2007,Bialczak2007,Bylander2011,Wang2015,Kumar2016}. Given that superconducting qubits, as well as other platforms, have been of great recent interest in the quantum simulation of gauge theories \cite{Wang2022,Mildenberger2022}, suppressing $1/f$ noise sources using efficient and experimentally feasible schemes becomes of central importance.

In this work, using exact diagonalization calculations and time-dependent perturbation theory, we demonstrate how the principle of \textit{linear gauge protection}, initially devised to control coherent gauge-breaking errors \cite{Halimeh2020e}, can be employed to suppress the growth of the gauge violations due to incoherent errors with spectral form $1/f^\beta$ ($\beta>0$) as $1/V^\beta$, where $V$ is the protection strength. The rest of this paper is organized as follows: We briefly review the concept of linear gauge protection in Sec.~\ref{sec:linpro}, and $1/f^\beta$ noise and the corresponding Bloch--Redfield formalism in Sec.~\ref{sec:BRE}. We present our main numerical results in Sec.~\ref{sec:results}. We finally conclude and provide an outlook in Sec.~\ref{sec:conc}. We include Appendix~\ref{app:BRE} for a derivation of the Bloch--Redfield equation employed for our analysis, Appendix~\ref{app:TDPT} for our derivations in time-dependent perturbation theory, in addition to Appendix~\ref{app:supp} where we provide supplemental numerical results.

\section{Linear gauge protection}\label{sec:linpro}
Let us consider an Abelian gauge theory described by the Hamiltonian $\hat{H}_0$, and whose gauge symmetry is generated by the operator $\hat{G}_j$, where $j$ denotes a site on a lattice of length $L$. The gauge invariance of $\hat{H}_0$ is encoded in the commutation relations $\big[\hat{H}_0,\hat{G}_j\big]=0,\,\forall j$. The set of gauge-invariant states $\{\ket{\psi}\}$ is defined as the simultaneous eigenstates of the generators: $\hat{G}_j\ket{\psi}=g_j\ket{\psi},\,\forall j$. A set of these eigenvalues $\mathbf{g}=(g_1,g_2,\ldots,g_L)$ over the volume of the system defines a unique gauge superselection sector, the projector onto which is $\hat{\mathcal{P}}_\mathbf{g}$. One can further define a \textit{target} or \textit{physical} gauge superselection sector $\mathbf{g}_\text{tar}=(g_1^\text{tar},g_2^\text{tar},\ldots,g_L^\text{tar})$ in which one wishes to restrict the dynamics in an experiment, for example.

In experimental implementations of gauge theories, $\hat{H}_0$ is mapped onto the microscopic degrees of freedom of a quantum simulator. In general, unavoidable gauge symmetry-breaking errors $\lambda\hat{H}_1$ at strength $\lambda$ will arise in this process either due to higher orders in the perturbation theory used to perform the mapping, or in experimental imperfections in equipment. Even when perturbative, these errors can generate gauge violations that grow as $\lambda^2t^2$ over evolution time $t$, which in turn lead to a complete departure from faithful gauge-theory dynamics beyond timescales $t\propto1/\lambda$ \cite{Halimeh2020a}.

In order to suppress these errors in a controlled way, the concept of linear gauge protection was introduced in Ref.~\cite{Halimeh2020e}. It entails adding the protection term
\begin{align}\label{eq:HG}
    V\hat{H}_G=V\sum_jc_j\hat{G}_j,
\end{align}
where $V$ is the protection strength. The sequence $c_j$ can be chosen to be rational and satisfying the condition $\sum_jc_j\big(g-g_j^\text{tar}\big)=0\iff g_j=g_j^\text{tar},\,\forall j$. In this case, the sequence is said to be \textit{compliant}, and, for a volume-independent and sufficiently large $V$, the gauge violation is controlled up to times exponential in $V$ \cite{Halimeh2020e,ARHH2017}. Although $V$ is volume-independent, the sequence $c_j$ would have to grow (not faster than) exponentially with system size in order to satisfy the compliance condition. This renders the compliant sequence somewhat inconvenient for large-scale gauge-theory quantum simulators such as those realized in recent cold-atom setups \cite{Yang2020,Zhou2021}.

However, reality turns out to be more forgiving, and even simple noncompliant sequences such as $c_j=(-1)^j$ can give excellent protection in the target sector against gauge errors up to all accessible evolution times in both finite systems \cite{Halimeh2020e} and the thermodynamic limit \cite{vandamme2021reliability}. This can be explained through the coherent quantum Zeno effect \cite{facchi2002quantum,facchi2004unification,facchi2009quantum,burgarth2019generalized}, which guarantees that upon adding the protection term~\eqref{eq:HG} an effective Zeno Hamiltonian $\hat{H}_Z=\hat{H}_0+\lambda\hat{\mathcal{P}}_{\mathbf{g}_\text{tar}}\hat{H}_1\hat{\mathcal{P}}_{\mathbf{g}_\text{tar}}$ emerges that faithfully reproduces the dynamics of the faulty gauge theory $\hat{H}_0+\lambda\hat{H}_1+V\hat{H}_G$ up to timescales linear in $V$ in a worst-case scenario \cite{Halimeh2020e}.

For certain gauge theories, the full local generator $\hat{G}_j$ may be too challenging to realize in an experiment \cite{Schweizer2019}, in which case the linear gauge protection as given in Eq.~\eqref{eq:HG} becomes impractical. Nevertheless, a powerful workaround exists based on \textit{local pseudogenerators} $\hat{W}_j$, which are identical to the full local generators $\hat{G}_j$ in the target sector, but not necessarily outside of it \cite{Halimeh2021stabilizing}. Formally, they satisfy the relation
\begin{align}
    \hat{W}_j\ket{\phi}=g_j^\text{tar}\ket{\phi}\iff\hat{G}_j\ket{\phi}=g_j^\text{tar}\ket{\phi}.
\end{align}
One can then extend the principle of linear gauge protection to one in terms of the local pseudogenerator, with protection term 
\begin{align}\label{eq:HW}
    V\hat{H}_W=V\sum_jc_j\hat{W}_j,
\end{align}
where the same rules apply for the sequence $c_j$ as in the case of Eq.~\eqref{eq:HG}. Note that even though $\hat{H}_0$ commutes with $\hat{G}_j$, it generally does not commute with $\hat{W}_j$, with the latter associated with a local symmetry richer than that generated by $\hat{G}_j$ \cite{Halimeh2021enhancing}. The resulting Zeno Hamiltonian when protecting with Eq.~\eqref{eq:HW} is $\hat{H}_Z=\hat{\mathcal{P}}_{\mathbf{g}_\text{tar}}\big(\hat{H}_0+\lambda\hat{H}_1\big)\hat{\mathcal{P}}_{\mathbf{g}_\text{tar}}$, under which the dynamics of the faulty gauge theory $\hat{H}_0+\lambda\hat{H}_1+V\hat{H}_W$ can be faithfully reproduced up to times at least linear in $V$ \cite{Halimeh2021stabilizing}.

In terms of purely unitary errors, extensive numerical simulations in exact diagonalization (ED) and infinite matrix product states (iMPS) based on the time-dependent variational principle \cite{Haegeman2011,Haegeman2013,Haegeman2016} have shown that for a compliant or properly chosen noncompliant sequence, linear gauge protection in the full local generator or the local pseudogenerator leads to stabilized gauge-theory dynamics up to all accessible evolution times with the gauge violation settling at a timescale $\propto1/V$ into a plateau of value $\propto\lambda^2/V^2$ \cite{Halimeh2020e,vandamme2021reliability,Halimeh2021stabilizing,vandamme2021suppressing}. Importantly, the linear gauge protection terms~\eqref{eq:HG} and~\eqref{eq:HW} are composed of single and two-body terms at most, and they are local, which renders them experimentally highly feasible.

It is a relevant open question whether linear gauge protection can be employed to protect against incoherent errors due to noise in an experiment. When left unchecked, these errors lead to gauge violations growing $\propto\gamma t$, where $\gamma$ is the strength of the incoherent errors. Even just slowing down the growth of gauge violations due to them would be greatly desirable in near-term quantum simulators.

\section{$1/f$ noise and the Bloch--Redfield master equation}\label{sec:BRE}
We focus here on $1/f$ noise, a decohering process with a noise power spectrum 
\begin{align}\label{eq:spectral}
S(\omega)=\frac{\gamma}{\lvert\omega\rvert^\beta},
\end{align}
where $\gamma$ is the system-environment coupling strength, $\omega$ is the frequency, and $0<\beta<2$. This type of noise is ubiquitous in nature, especially in condensed matter systems in quasi-equilibrium (for $\beta\approx1$) and electronic equipment, but this signal can also be found in biological systems, music, and even in economics \cite{Press1978,kogan1996electronic}. In particular, as mentioned above, it is present in SQUIDs, which can lead to adverse effects on quantum simulation platforms based on superconducting qubits \cite{Yoshihara2006,Kakuyanagi2007,Bialczak2007,Bylander2011,Wang2015,Kumar2016}.

Since we \textit{a priori} know the noise power spectrum of the environment, we employ the Bloch--Redfield formalism \cite{cohen1992atom,breuer2002theory} to derive a master equation from a microscopic perspective. We consider a system $\hat{H}_S$ coupled to a bath (the environment) $\hat{H}_{B}$ with the interaction Hamiltonian $\hat{H}_{SB}=\sqrt{\gamma}\sum_{\alpha}\hat{A}_{\alpha}\otimes \hat{B}_{\alpha}$, where $\hat{A}_{\alpha}$ and $\hat{B}_{\alpha}$ are system and bath operators, respectively, with system-environment coupling strength $\gamma$. 
In general, the system operators $\hat{A}_{\alpha}$ do not preserve Gauss's law. 
Under the assumption of weak system-environment coupling, we obtain a master equation in terms of system operators and correlation functions that characterize the statistical properties of the bath. 

To obtain the master equation in terms of a noise power spectrum that can be numerically implemented, we write the bath correlation function $C_{\alpha \nu}(\tau)=\gamma \operatorname{Tr}_{B}\left[\hat{\tilde{B}}_{\alpha}(t) \hat{\tilde{B}}_{\nu}(t-\tau) \hat{\rho}_{B}\right]$---here, we denote tilde on quantities written in the interaction picture---in terms of the spectral function $S_{\alpha \nu}(\omega)$, after neglecting a small energy shift arising due to the imaginary part in the Fourier transform of $C_{\alpha \nu}(\tau)$:
\begin{align}\label{eq:sfn}
S_{\alpha \nu}(\omega)=2\int_{0}^{\infty}d \tau e^{i\omega\tau}C_{\alpha \nu}(\tau).
\end{align}
Hence, one can show that the final form of the Bloch--Redfield master equation, describing the evolution of the reduced density matrix for the system, after employing the Born, Markov, and the secular approximation as detailed in Appendix \ref{app:BRE} can be written explicitly as,
\begin{align}\label{eq:redfield}
d_t\rho_{a b}(t)=-i \omega_{a b} \rho_{a b}(t)+\sum_{c, d} R_{a b c d} \rho_{c d}(t),
\end{align}
where $R_{a b c d}$ is the Bloch--Redfield relaxation tensor, which can be written in matrix form with $\hat{A}_{\alpha}$ assumed to be Hermitian for ease of numerical implementation,
\begin{align}\nonumber
R_{a b c d}=&-\frac{1}{2} \sum_{\alpha}\bigg[\delta_{b d} \sum_{n} A_{a n}^{\alpha} A_{n c}^{\alpha} S_{\alpha}(\omega_{c n})\\\nonumber
&-A_{a c}^{\alpha} A_{d b}^{\alpha} S_{\alpha}(\omega_{c a})+\delta_{a c} \sum_{n} A_{d n}^{\alpha} A_{n b}^{\alpha} S_{\alpha}(\omega_{d n})\\
&-A_{a c}^{\alpha} A_{d b}^{\alpha} S_{\alpha}(\omega_{d b})\bigg].
\end{align}
The Redfield tensor contains all the information about the dissipative processes that arise due to the coupling of the system with the bath degrees of freedom.

One requirement for the validity of the Bloch--Redfield approach is the smallness of the Bloch--Redfield decay rates that describe the effective incoherent coupling between two eigenlevels $i$ and $f$  against the relevant transition frequencies $\omega_{if}$ \cite{PhysRevA.80.022303}. 
The Bloch--Redfield decay rates, also known as the golden rule rates, are defined as 
$\Gamma_{if}\propto{\sum_{\mathbf{\alpha}}\left|\left\langle i\left|\hat{A}_{\alpha}\right| f\right\rangle\right|^2 S_\alpha\left(\omega_{if}\right)}$. 
We checked for the numerical models we describe throughout our paper that the condition $\Gamma_{if}\ll\omega_{if}$ was always satisfied. 
In particular, 
as the system operators $\hat{A}_{\alpha}$ violate Gauss's law, the relevant incoherent transitions happen on large energy scales of order $V$, where the noise spectrum becomes weak, thus further solidifying our approach for employing this formalism.

As $1/f$ noise and other types of decoherence can drastically undermine performance in an experimental setup, it becomes important to find ways that may ameliorate its effect. Left unchecked, decoherence can lead to a fast buildup in the gauge violation, which renders the quantum simulation of true gauge-theory dynamics unfaithful \cite{Halimeh2020f,Halimeh2020g}.

\section{Results and discussion}\label{sec:results}

We now present our numerical results on the quench dynamics of gauge theories subjected to $1/f$ noise, which we have computed using the exact diagonalization toolkit QuTiP \cite{Johansson2012,Johansson2013}. In all cases, we prepare our system in an initial state $\hat{\rho}_0$ in the target gauge sector $\mathbf{g}_\text{tar}$, and monitor its quench dynamics in the presence of $1/f$ noise with and without linear gauge protection. In particular, we will focus on the dynamics of the gauge violation,
\begin{align}\label{eq:viol}
\varepsilon(t)&=\frac{1}{L}\sum_{j=1}^{L}\Tr\Big\{\hat{\rho}(t)\big(\hat{G}_j-g_j^\text{tar}\big)^2\Big\},
\end{align}
where $\hat{\rho}(t)$ is the time-evolved density operator of the system at time $t$, in addition to calculating the dynamics of relevant local observables. Due to the large evolution times we investigate, we restrict our system size to $L=4$ sites due to computational overhead, and we employ periodic boundary conditions.

\begin{figure}[t!]
	\centering
	\includegraphics[width=0.48\textwidth]{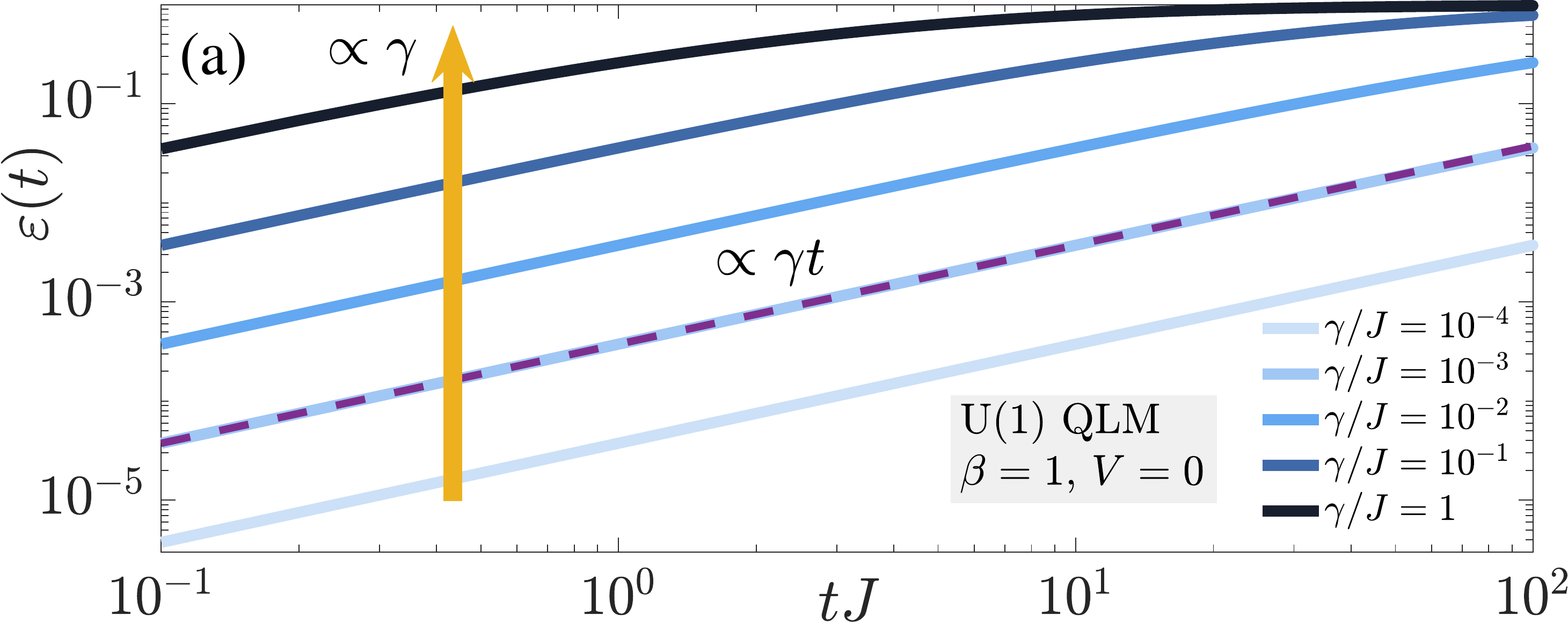}\\
	\vspace{1.1mm}
	\includegraphics[width=0.48\textwidth]{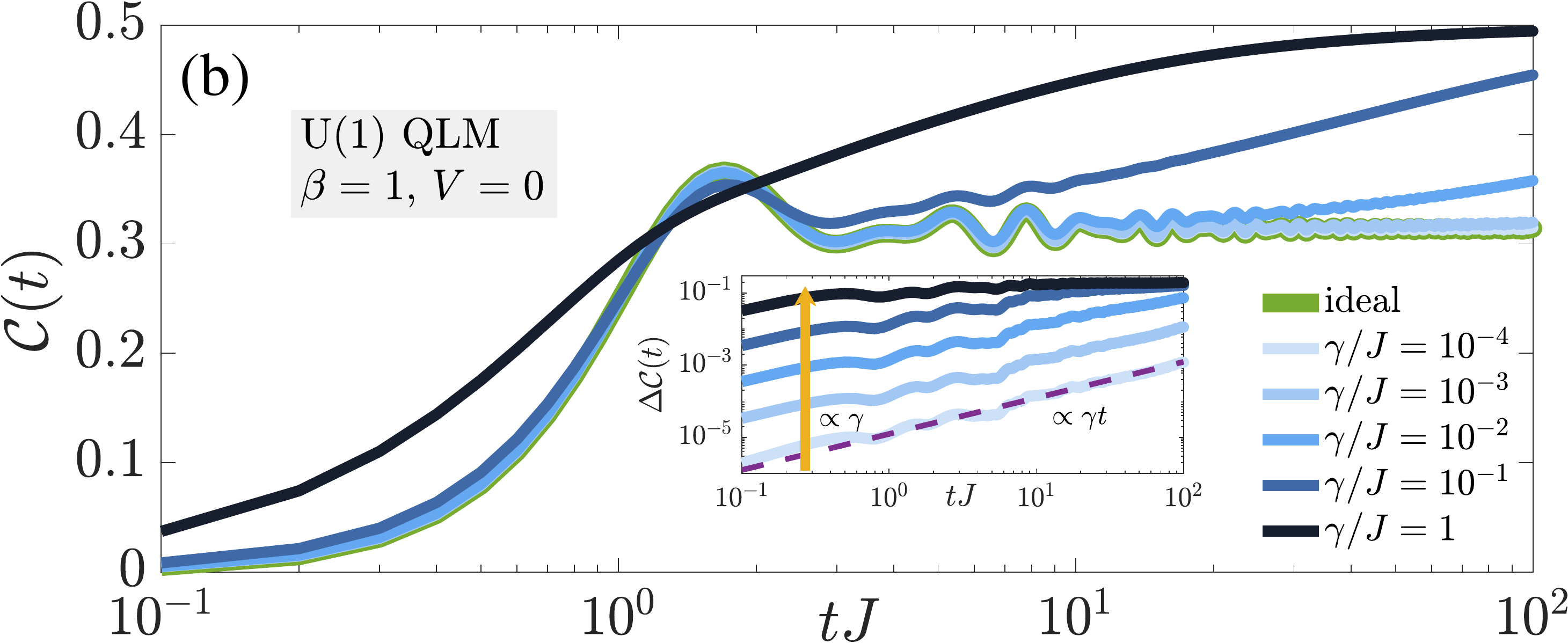}
	\caption{(Color online). (a) Quench dynamics of the gauge violation~\eqref{eq:viol} and (b) the chiral condensate~\eqref{eq:CC} in the presence of incoherent errors generated by the noise spectral function $S(\omega)=\gamma/|\omega|$ for various values of system-environment coupling strength $\gamma$ with quantum jump operators $\hat{A}_j^{m}=\hat{\sigma}^{x}_{j}$ and $\hat{A}_{j,j+1}^{g}=\hat{s}^{x}_{j,j+1}$ for matter and gauge fields, respectively, and without adding any protection terms, i.e., $V=0$. Here, the quench Hamiltonian is the $\mathrm{U}(1)$ quantum link model~\eqref{eq:U1QLM}, and the initial state is the gauge-invariant vacuum, with all matter sites empty while the local electric fields on odd (even) links point down (up). For both quantities, we see that errors evolve $\propto\gamma t$, and already small values of $\gamma$ significantly undermine gauge-theory dynamics.}
	\label{fig:U1_beta1_V0} 
\end{figure}

\subsection{$\mathrm{U}(1)$ quantum link model}
We first consider the $\mathrm{U}(1)$ quantum link model \cite{Chandrasekharan1997,Wiese_review,Hauke2013,Kasper2017}
\begin{align}\label{eq:U1QLM}
\hat{H}_0=\sum_{j=1}^{L}\left[J\left(\hat{\sigma}_j^- \hat{s}^+_{j,j+1}\hat{\sigma}_{j+1}^{-}+\text {H.c.}\right)+\frac{\mu}{2} \hat{\sigma}_j^z\right],
\end{align}
where on site $j$ the matter field is represented by the Pauli operator $\hat{\sigma}^z_j$, with $\mu$ denoting the fermionic mass, the gauge (electric) field on the link between sites $j$ and $j+1$ is denoted by the spin-$1/2$ operator $\hat{s}^+_{j,j+1}$ ($\hat{s}^z_{j,j+1}$), $L$ is the total number of sites with periodic boundary conditions enforced, and the overall energy scale is set by the coupling strength $J=1$. The generator of the $\mathrm{U}(1)$ gauge symmetry of Hamiltonian~\eqref{eq:U1QLM} is given by
\begin{align}\label{eq:GjU1}
\hat{G}_j=(-1)^j\bigg(\hat{s}^z_{j-1,j}+\hat{s}^z_{j,j+1}+\frac{\hat{\sigma}^z_j+1}{2}\bigg).
\end{align}
The model~\eqref{eq:U1QLM} is a quantum link formulation \cite{Chandrasekharan1997} of lattice quantum electrodynamics in $1+1$D, and is experimentally very relevant as it has been the subject of recent large-scale cold-atom quantum simulations \cite{Yang2020,Zhou2021}.

We now prepare the system in a vacuum state, which is one of two doubly degenerate eigenstates of Hamiltonian~\eqref{eq:U1QLM} at $\mu/J\to\infty$. This initial state is in the target sector $g_j^\text{tar}=0,\,\forall j$, i.e., $\Tr\{\hat{\rho}_0\hat{G}_j\}=0,\,\forall j$, where its sites host no matter and the local electric fields are in a staggered formation. We then quench this vacuum state with $\hat{H}_0+V\hat{H}_G$ at $\mu/J=0.5$ in the presence of $1/f$ noise with power spectrum~\eqref{eq:spectral} and jump operators $\hat{A}_j^{m}=\hat{\sigma}^{x}_{j}$ and $\hat{A}_{j,j+1}^{g}=\hat{s}^{x}_{j,j+1}$, which couple the matter and gauge fields to the environment, respectively. Let us first consider the case without protection, i.e., $V=0$, shown in Fig.~\ref{fig:U1_beta1_V0} setting $\beta=1$. The dynamics of the gauge violation~\eqref{eq:viol} is shown for various values of the system-environment coupling strength $\gamma$ in Fig.~\ref{fig:U1_beta1_V0}(a). At early times, the violation grows $\propto\gamma t$, as can be shown in time-dependent perturbation theory, until it begins to settle into a maximal violation plateau at a timescale $\propto1/\gamma$. We observe similar behavior in the chiral condensate, a measure of how strongly the dynamics spontaneously breaks the chiral symmetry associated with fermions in the theory, 
\begin{align}\label{eq:CC}
\mathcal{C}(t)&=\frac{1}{2}+\frac{1}{2L} \sum_{j=1}^{L}\big\{\hat{\rho}(t)\hat{\sigma}^z_j\big\},
\end{align}
shown in Fig.~\ref{fig:U1_beta1_V0}(b). The error with respect to the ideal case, shown in the inset, grows $\propto\gamma t$ before settling into a maximal value at late times for sufficiently large $\gamma$. These results demonstrate the pernicious effect of $1/f$ noise on quantum simulations of gauge theories when left unprotected.

\begin{figure}[t!]
	\centering
	\includegraphics[width=0.48\textwidth]{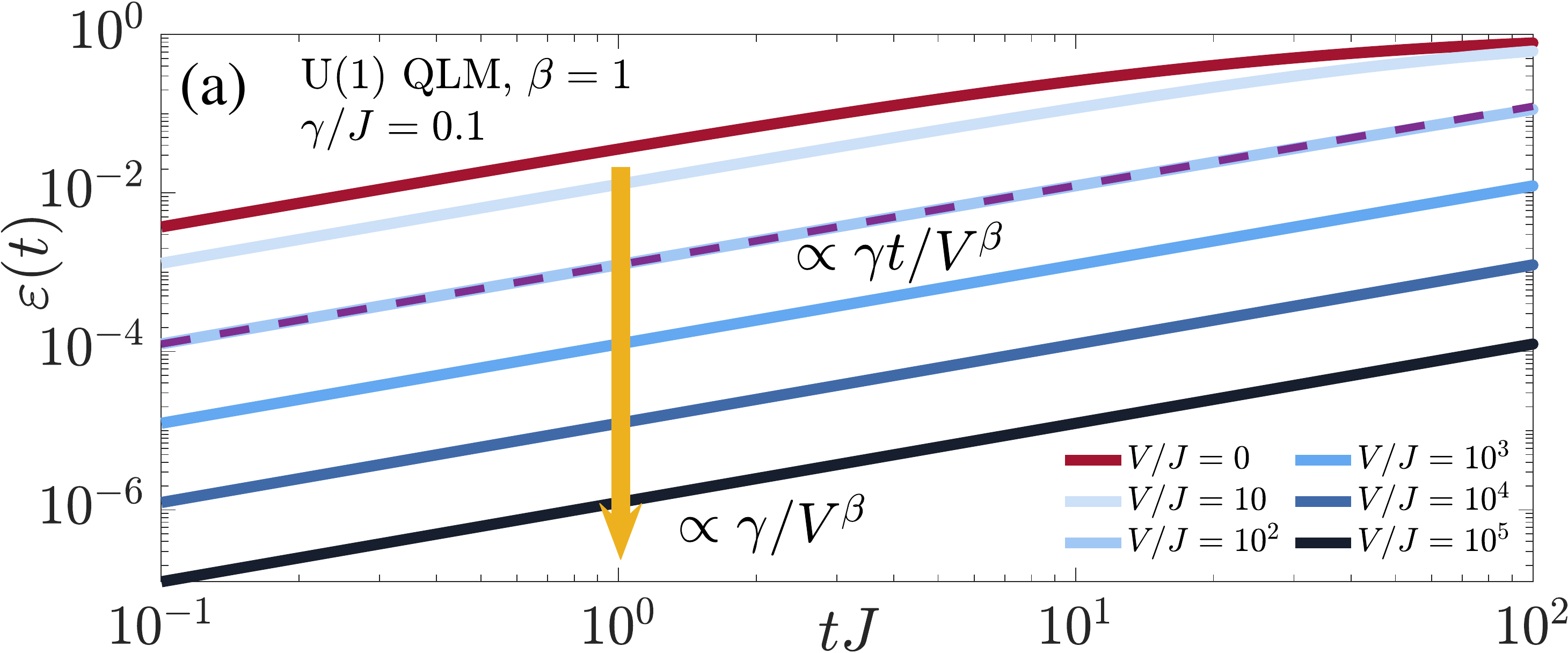}\\
	\vspace{1.1mm}
	\includegraphics[width=0.48\textwidth]{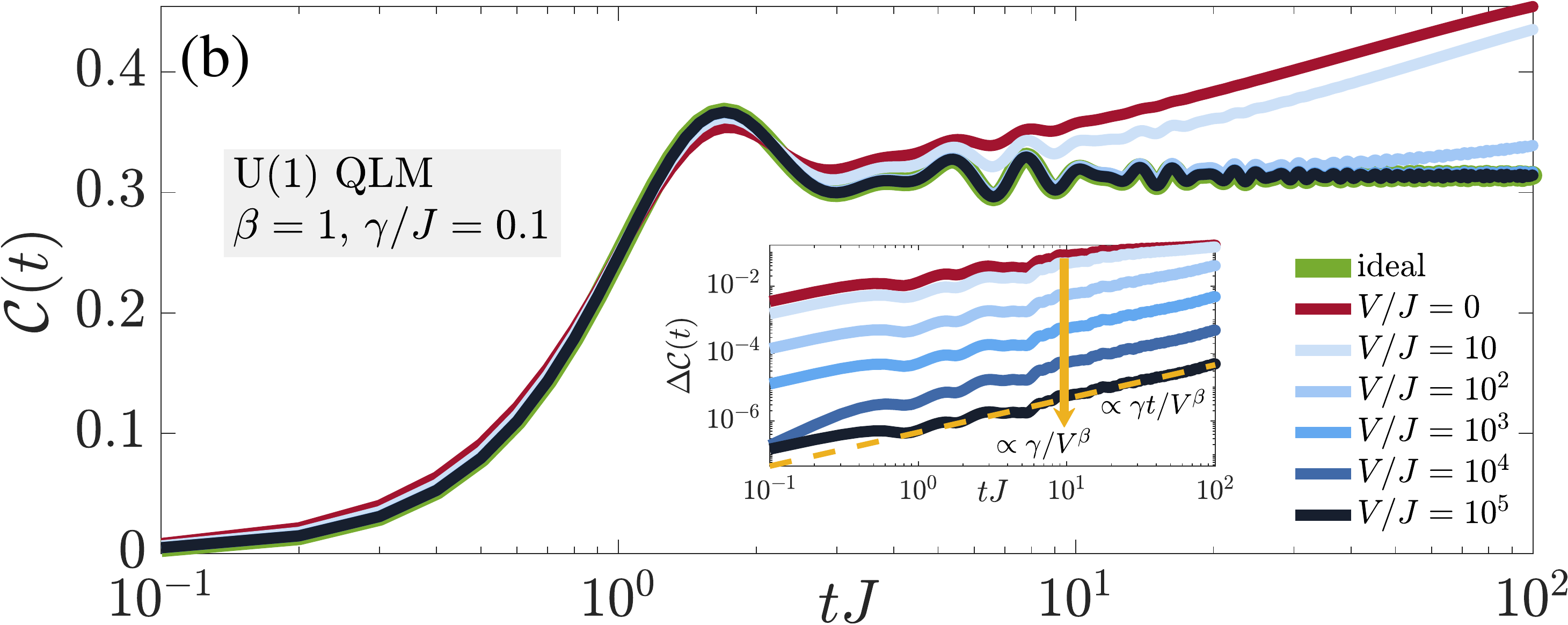}
	\caption{(Color online)(a) Quench dynamics of the gauge violation~\eqref{eq:viol} and (b) the chiral condensate~\eqref{eq:CC} in the presence of incoherent gauge-breaking errors generated by the noise spectral function $S(\omega)=\gamma/|\omega|$ at fixed system-environment coupling strength $\gamma=0.1J$ and with the linear gauge protection term~\eqref{eq:HG} turned on at various values of the protection strength $V$. We employ the compliant sequence $c_j\in\{-115,116,-118,122\}/122$. As we switch on the gauge protection, the growth of the gauge violation is suppressed as $\epsilon(t)\propto{\gamma t/V}$ until it starts to plateau at a timescale $\propto{V/(J\gamma)}$, extending the coherent lifetime of a potential experiment linearly in $V$. Similar conclusions can be drawn for the chiral condensate where in the presence of linear gauge protection, the ideal-theory dynamics is reproduced up to a timescale $\propto{V/(J\gamma)}$ with a deviation $\propto{\gamma/V}$ as shown in the inset.}
	\label{fig:U1_beta1_gamma0.1} 
\end{figure}

We now repeat the same quench protocol as in Fig.~\ref{fig:U1_beta1_V0}, but with fixed $\gamma{=}0.1J$ and the addition of the gauge protection~\eqref{eq:HG} at strength $V$, with $c_j=\{-115,116,-118,122\}/122$ chosen to be a compliant sequence. The corresponding dynamics of the gauge violation is shown in Fig.~\ref{fig:U1_beta1_gamma0.1}(a), where we see a robust suppression in the growth of the gauge violation such that $\varepsilon(t)\propto\gamma t/V$ at short times, in agreement with time-dependent perturbation theory; see Appendix~\ref{app:TDPT}. This suppression is also seen in the dynamics of the chiral condensate, shown in Fig.~\ref{fig:U1_beta1_gamma0.1}(b). Indeed, whereas the unprotected case (red curve) quickly and significantly diverges from the ideal case (green curve), at sufficiently large $V$ the agreement with the ideal case is excellent. The inset shows the deviation from the ideal case for the various considered values of $V$, where we find that the error grows roughly $\propto\gamma t/V$. These results show, therefore, that linear gauge protection extends the timescale of the dynamics during which one can perturbatively connect to a gauge theory from $\propto1/\gamma$ to $\propto V/(J\gamma)$. Even though linear gauge protection does not suppress the gauge violation into a long-lived plateau of constant value as it does in the case of purely coherent errors \cite{Halimeh2020e}, this is nevertheless a positive result that can allow one to significantly enhance the achievable coherent evolution times, and which can thus be of significant benefit to current and near-term gauge-theory quantum simulators.

\begin{figure}[t!]
	\centering
	\includegraphics[width=0.48\textwidth]{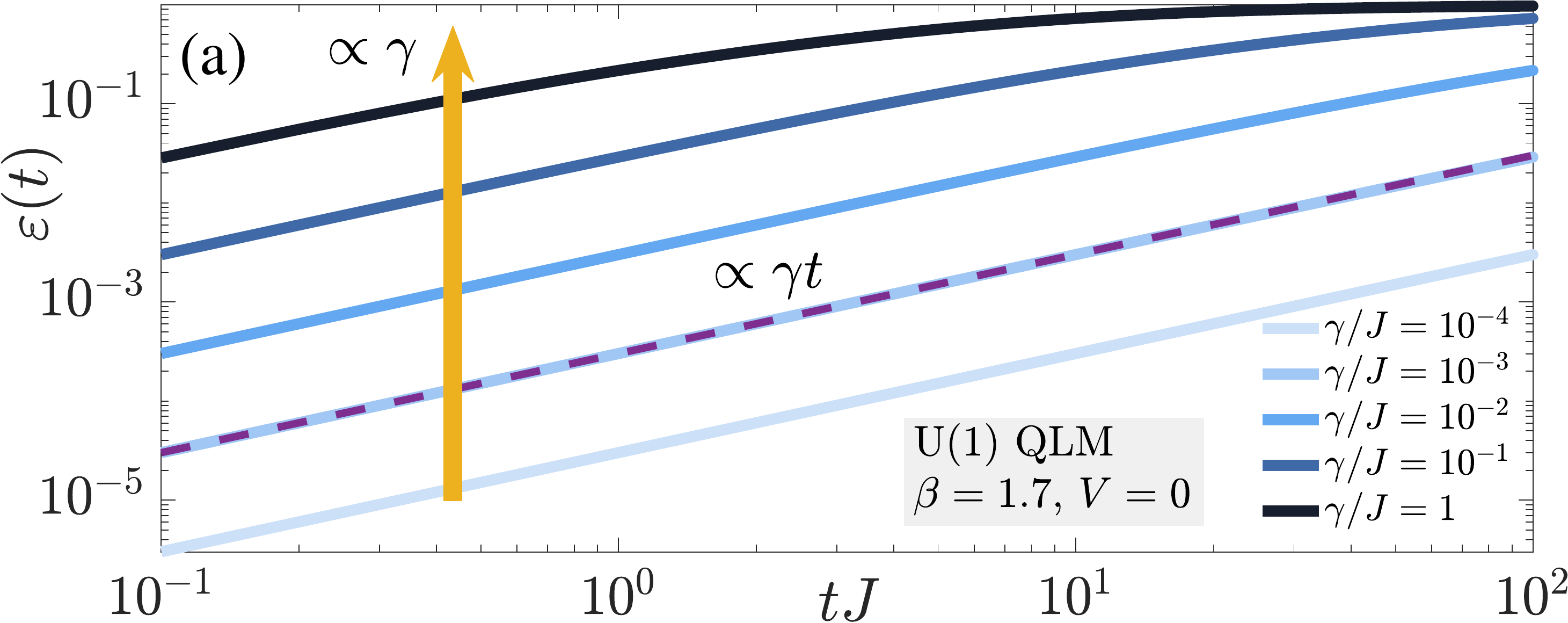}\\
	\vspace{1.1mm}
	\includegraphics[width=0.48\textwidth]{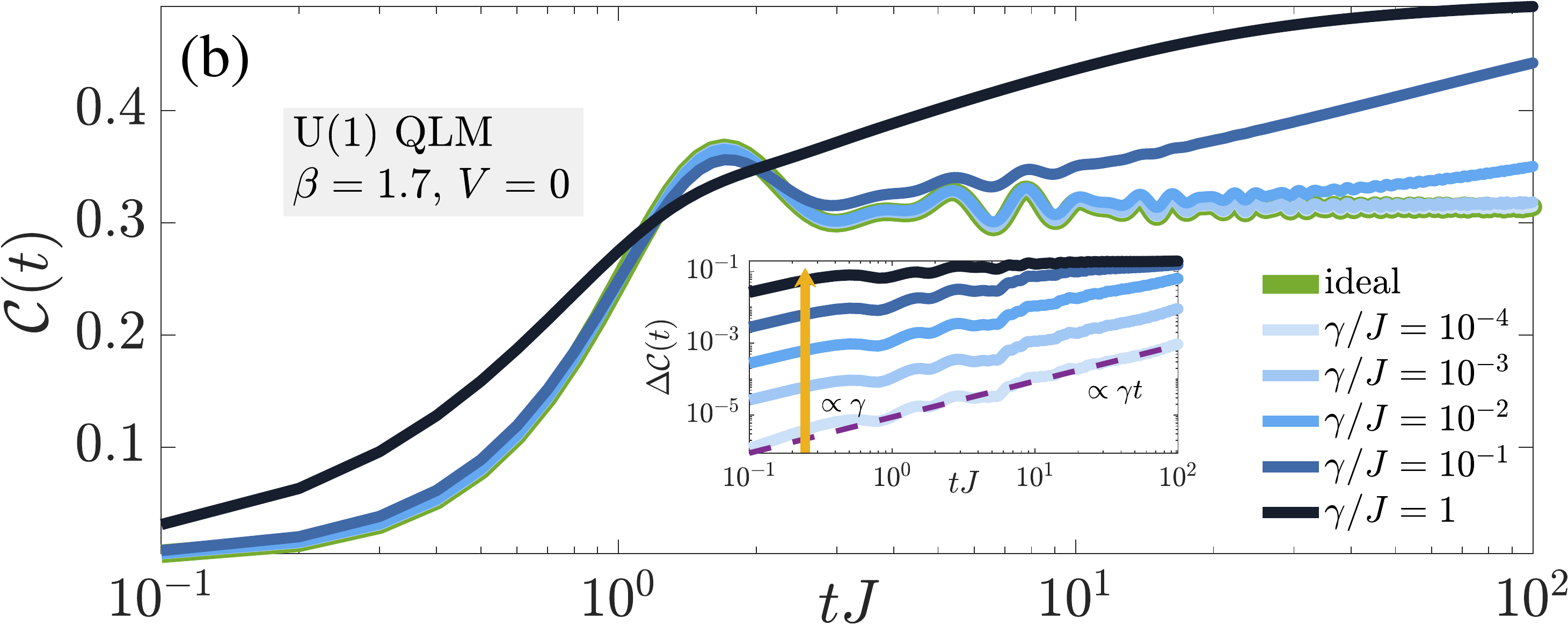}
	\caption{(Color online). Same as Fig.~\ref{fig:U1_beta1_V0} but for the noise spectral function  $S(\omega)=\gamma/|\omega|^{\beta}$ where $\beta=1.7$. The qualitative picture is identical to that of $\beta=1$ in Fig.~\ref{fig:U1_beta1_V0} for both the gauge violation and the chiral condensate, with only insignificant quantitative differences in these quantities between different values of $\beta$.}
	\label{fig:U1_beta1.7_V0} 
\end{figure}

Let us now investigate the case of a fractional coefficient $\beta$ in the spectrum $S(\omega)=\gamma/\lvert\omega\rvert^\beta$. For this purpose, we repeat the above quench protocols for $\beta=1.7$. The protection-free case is shown in Fig.~\ref{fig:U1_beta1.7_V0}. The result is qualitatively similar to that of $\beta=1$ in Fig.~\ref{fig:U1_beta1_V0}. Indeed, the gauge violation grows $\propto\gamma t$ until a timescale $\propto1/\gamma$, where it begins to settle into a maximal-violation plateau, as can be seen for large enough values of $\gamma$; see Fig.~\ref{fig:U1_beta1.7_V0}(a). This type of behavior is replicated in the chiral condensate, as depicted in Fig.~\ref{fig:U1_beta1.7_V0}(b), where the deviation from the ideal case grows $\propto\gamma t$ at short times before beginning to plateau at $t\propto1/\gamma$. We can thus conclude that the effect of $\beta$ is merely quantitative in the case of no protection.

\begin{figure}[t!]
	\centering
	\includegraphics[width=0.48\textwidth]{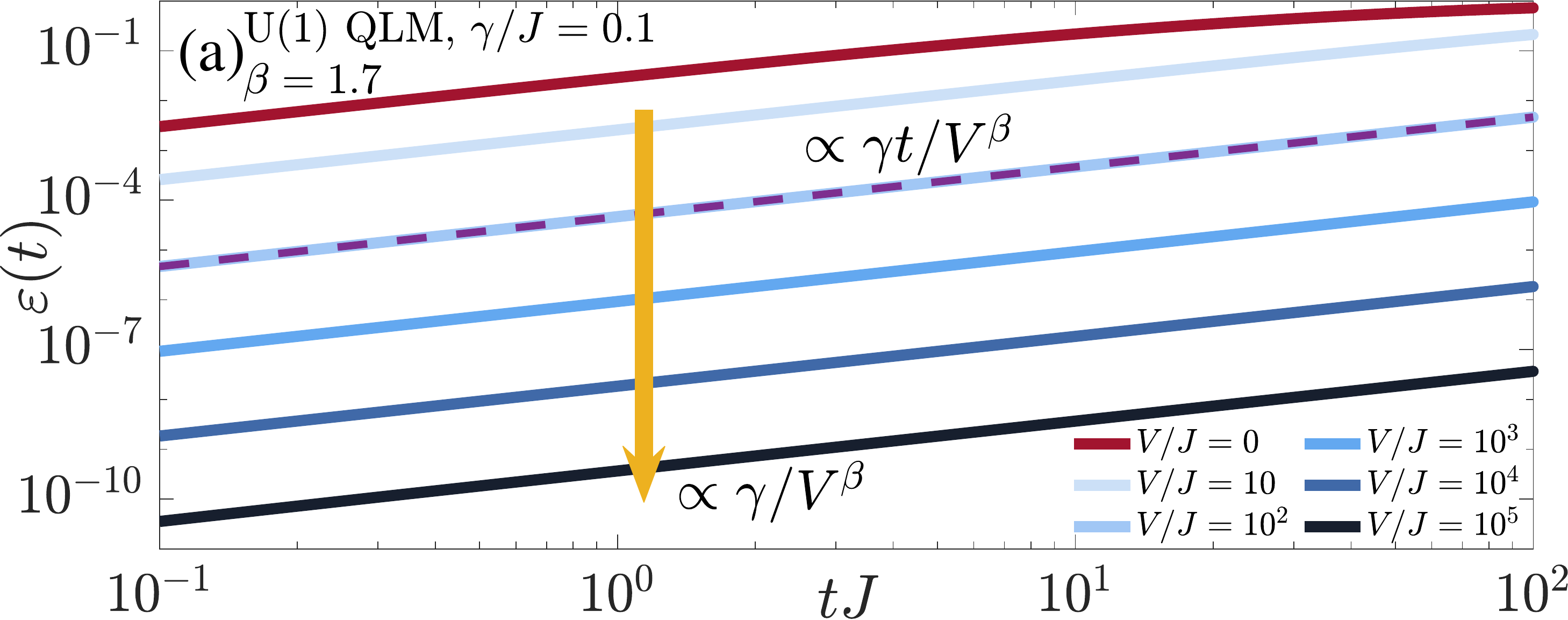}\\
	\vspace{1.1mm}
	\includegraphics[width=0.48\textwidth]{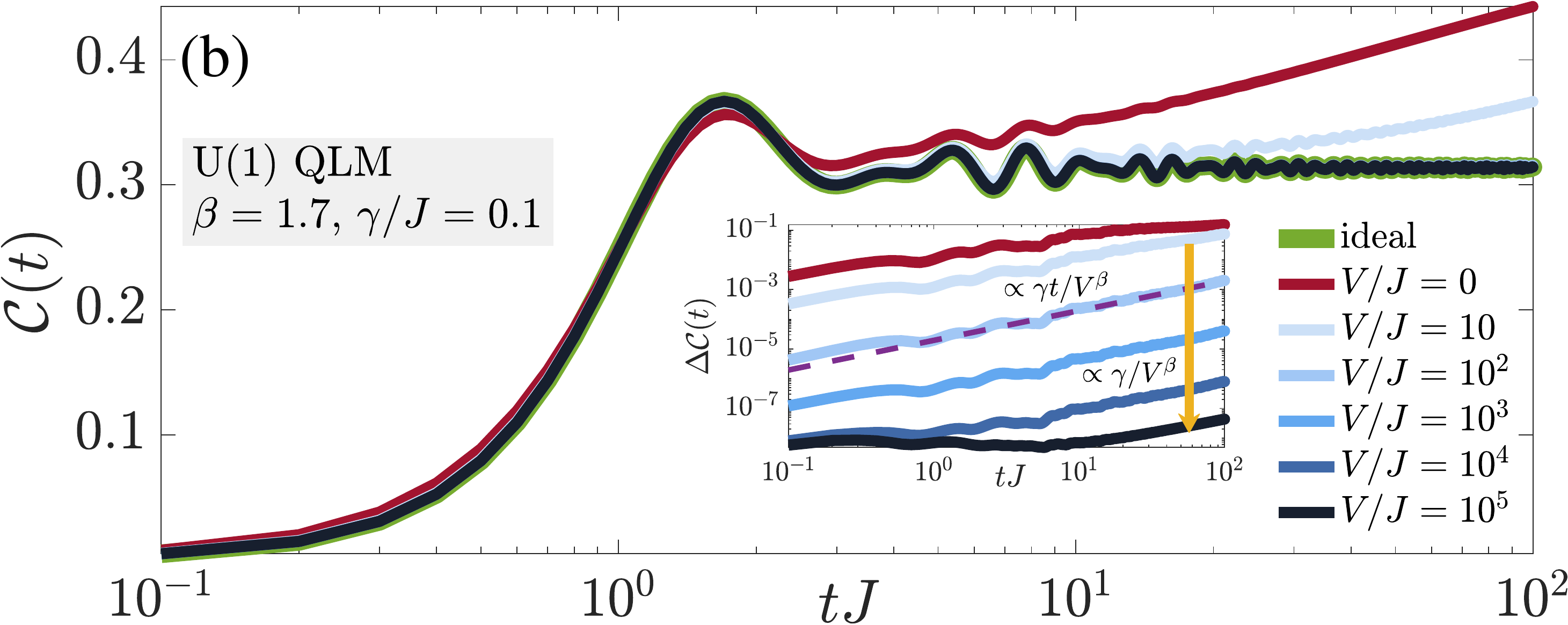}
	\caption{(Color online). Same as Fig.~\ref{fig:U1_beta1_gamma0.1}, but where $\beta=1.7$ in the noise spectral function  $S(\omega)=\gamma/|\omega|^{\beta}$. A qualitative difference arises whereby the gauge violation and the deviation of the chiral condensate from the ideal case both grow $\propto\gamma t/V^{1.7}$ instead of $\propto\gamma t/V$, showing that linear gauge protection suppresses errors more for a larger value of $\beta$.}
	\label{fig:U1_beta1.7_gamma0.1} 
\end{figure}

Upon employing gauge protection, the qualitative picture changes significantly. The gauge violation grows $\propto\gamma t/V^{1.7}$, as shown in Fig.~\ref{fig:U1_beta1.7_gamma0.1}(a) at fixed $\gamma=0.1J$. In other words, the suppression in the growth of the gauge violation directly depends on $\beta$, with greater suppression at larger $\beta$. This also happens in the case of the chiral condensate, shown in Fig.~\ref{fig:U1_beta1.7_gamma0.1}(b). We find that even though the unprotected case vastly deviates from the ideal one ($\gamma=V=0$), upon adding linear gauge protection, the chiral condensate faithfully reproduces the ideal case up to all accessible evolution times at sufficiently large $V$, with the deviation from the ideal case $\propto\gamma t/V^{1.7}$ (see inset).

This behavior can be explained in the following way. The spectral function of the considered decoherence process is $S(\omega)=\gamma/\lvert\omega\rvert^\beta$, where the relevant frequencies $\omega$ governing the system dynamics are those that create transitions between the target gauge sector and the other gauge sectors. Upon switching on the linear gauge protection, the undesired sectors are energetically separated from the target gauge sector proportionally to $V$. Hence, the relevant transition frequencies are on the order $\omega\sim V$. The strength of the spectral function thus scales as ${S}(\omega)\sim\gamma/ V^\beta$ and becomes weaker as $V$ increases.

It is worth noting that we have also checked that our conclusions hold for different jump operators, quench parameters (different values of $\mu/J$), and initial states, as well as for noncompliant sequences. See Appendix~\ref{app:supp} for supplemental numerical results.

\subsection{$\mathbb{Z}_2$ lattice gauge theory}
To check the generality of the above findings, we now turn our attention to a different model, namely a $\mathbb{Z}_2$ lattice gauge theory that has been of recent theoretical \cite{Barbiero2019,Zohar2017,Borla2019,Yang2020fragmentation,kebric2021confinement,Borla2020} and experimental relevance \cite{Goerg2019,Schweizer2019}. Its Hamiltonian reads
\begin{align}\label{eq:Z2LGT}
\hat{H}_0=J \sum_{j=1}^{L}\big(\hat{a}_j^{\dagger} \hat{\tau}_{j, j+1}^z \hat{a}_{j+1}+\text{H.c.}\big)-h \sum_{j=1}^L \hat{\tau}_{j, j+1}^x,
\end{align}
where the bosonic ladder operators $\hat{a}_j,\hat{a}^\dagger_j$ on site $j$ represent the annihilation and creation of matter, respectively. The electric (gauge) field on the link between sites $j$ and $j+1$ is represented by the Pauli operator $\hat{\tau}^x_{j,j+1}$ ($\hat{\tau}^z_{j,j+1}$), where the electric field strength is given by $h$. The overall energy scale is set by $J=1$. The generator of the $\mathbb{Z}_2$ gauge symmetry of Hamiltonian~\eqref{eq:Z2LGT} is given by
\begin{align}\label{eq:GjZ2}
    \hat{G}_j=(-1)^{\hat{a}_j^\dagger\hat{a}_j}\hat{\tau}^x_{j-1,j}\hat{\tau}^x_{j,j+1},
\end{align}
and its eigenvalues are $\pm1$, where, due to the $\mathbb{Z}_2$ gauge symmetry, $\hat{G}_j^2=\hat{\mathds{1}}_j$. Unlike the generator~\eqref{eq:GjU1} of the $\mathrm{U}(1)$ quantum link model~\eqref{eq:U1QLM}, which is composed of one-body terms, the generator~\eqref{eq:GjZ2} of the $\mathbb{Z}_2$ lattice gauge theory~\eqref{eq:Z2LGT} is a three-body term that mixes matter and gauge degrees of freedom. This renders it significantly impractical in experimental implementations. As described in Sec.~\ref{sec:linpro}, one can then utilize the concept of the local pseudogenerator \cite{Halimeh2021stabilizing}, where in this case it takes the form
\begin{align}\label{eq:Lpg}
    \hat{W}_j=\hat{\tau}^x_{j-1,j}\hat{\tau}^x_{j,j+1}+2g_j^\text{tar}\hat{a}_j^\dagger\hat{a}_j.
\end{align}
Note that even though $\big[\hat{H}_0,\hat{G}_j\big]=0,\,\forall j$, on account of the $\mathbb{Z}_2$ gauge symmetry of Hamiltonian~\eqref{eq:Z2LGT}, $\big[\hat{H}_0,\hat{W}_j\big]\neq0$. However, when working in the target sector $\mathbf{g}_\text{tar}$, then $\hat{W}_j$ and $\hat{G}_j$ are indistinguishable. It is interesting to note that the local symmetry associated with $\hat{W}_j$ contains the $\mathbb{Z}_2$ gauge symmetry generated by $\hat{G}_j$. In fact, one can prove for a given Hamiltonian $\hat{H}'$ that $\big[\hat{H}',\hat{W}_j\big]=0\Rightarrow\big[\hat{H}',\hat{G}_j\big]=0$.

\begin{figure}[t!]
	\centering
	\includegraphics[width=0.48\textwidth]{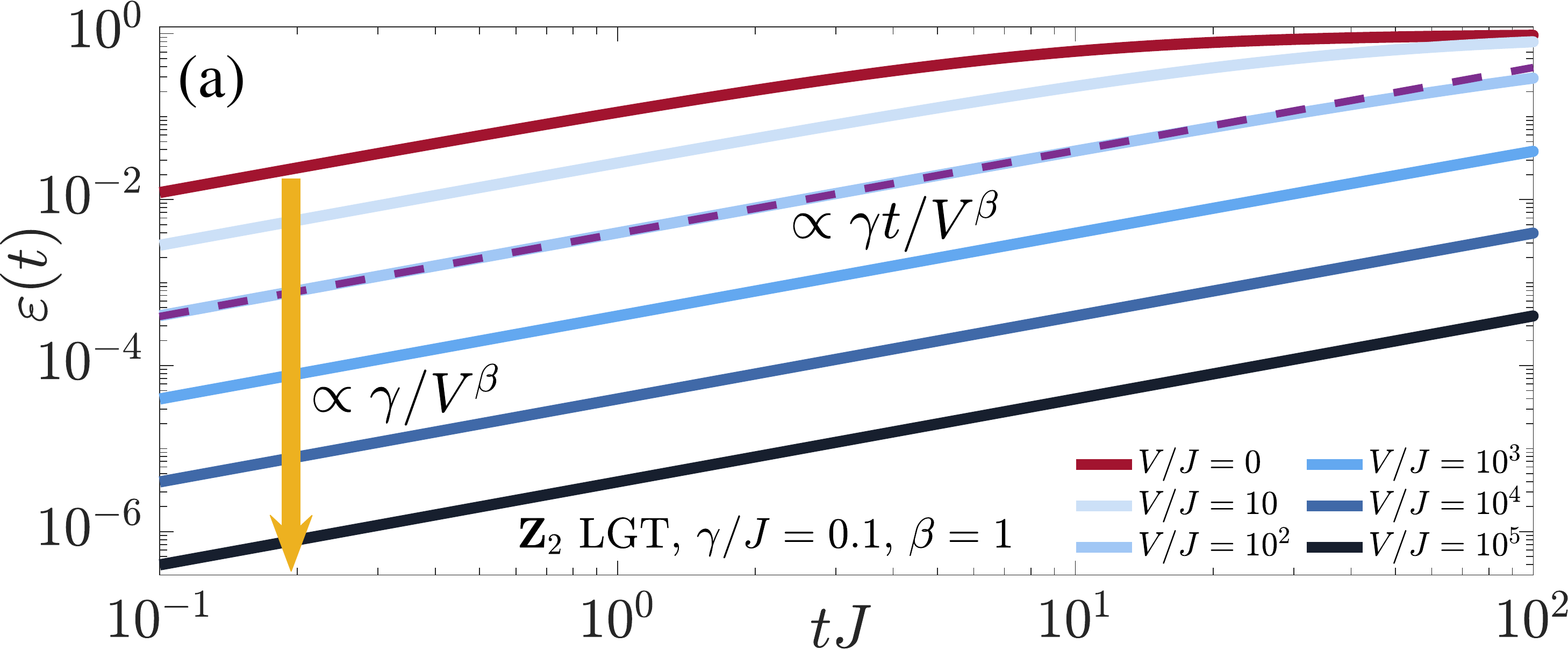}\\
	\vspace{1.1mm}
	\includegraphics[width=0.48\textwidth]{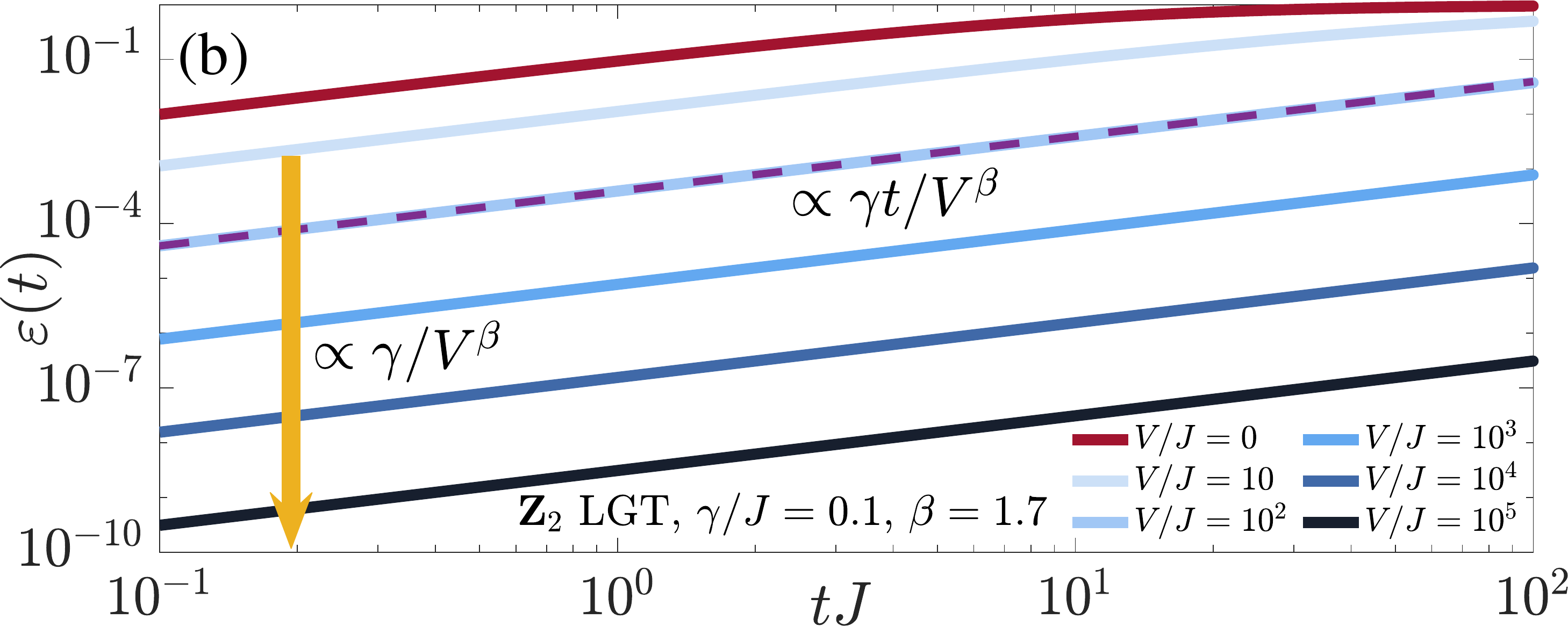}
	\caption{(Color online). Same as Figs.~\ref{fig:U1_beta1_gamma0.1}(a) and \ref{fig:U1_beta1.7_gamma0.1}(a), but for the $\mathbb{Z_2}$ lattice gauge theory~\eqref{eq:Z2LGT} and with the linear gauge protection term~\eqref{eq:HW} in the local pseudo generator, Eq.~\eqref{eq:Lpg}. The results are shown for the noncompliant sequence $[(-6)^j+5]/11$ with quantum jump operators operators $\hat{A}_j^{m}=\hat{a}_{j}+\hat{a}_{j}^\dagger$ and $\hat{A}_{j,j+1}^{g}=\hat{\tau}^{z}_{j,j+1}$. The qualitative conclusions are identical to the corresponding cases of the $\mathrm{U}(1)$ quantum link model where errors evolve $\propto\gamma t/V^\beta$, showcasing the generality of our findings.}
	\label{fig:Z2_gamma0.1} 
\end{figure}

We can now employ the concept of linear protection in terms of the local pseudogenerator according to Eq.~\eqref{eq:HW} in order to protect against $1/f$ noise in the $\mathbb{Z}_2$ lattice gauge theory. We prepare our system in a charge-density wave state in terms of the matter fields, with the electric fields aligned such that the system resides in the target sector $g_j^\text{tar}=+1,\,\forall j$. We quench this state with Hamiltonian~\eqref{eq:Z2LGT} at $h=0.54J$ in the presence of $1/f$ noise with the spectral function~\eqref{eq:spectral} and jump operators $\hat{A}_j^{m}=\hat{a}_{j}+\hat{a}_{j}^\dagger$ and $\hat{A}_{j,j+1}^{g}=\hat{\tau}^{z}_{j,j+1}$ coupling the matter and gauge fields, respectively, to the environment at a fixed value of $\gamma=0.1J$ and for several values of the protection strength $V$. The corresponding dynamics of the gauge violation is shown in Fig.~\ref{fig:Z2_gamma0.1}(a,b) for $\beta=1$ and $1.7$, respectively. The qualitative picture is identical to that of the $\mathrm{U}(1)$ quantum link model, where we find that at sufficiently large $V$ the gauge violation evolves $\propto\gamma t/V^\beta$ at short to intermediate times, before eventually plateauing at a maximal value that is delayed from a timescale $\propto 1/\gamma$ in the unprotected case to a timescale $\propto V^\beta/\gamma$ under linear gauge protection. 

We have also checked that these findings are valid for different initial states, quench parameters, and properly chosen sequenches $c_j$. As such, our conclusions are not specific to a given model, and we expect our findings to be general and applicable to any Abelian gauge theory.

\section{Conclusion and outlook}\label{sec:conc}
We have demonstrated numerically how linear gauge protection schemes based on the local full generator or on the local pseudogenerator can suppress the growth of gauge violations due to $1/f$ noise with power spectrum $S(\omega)=\gamma/\lvert\omega\rvert^\beta$ as $\varepsilon(t)\propto\gamma t/V^\beta$ in gauge-theory quantum simulations, where $V$ is the protection strength. This extends coherent lifetimes by $V^\beta$ in experiments where $1/f$ noise is the dominant source of decoherence. As examples, we have used two paradigmatic Abelian systems: the $\mathrm{U}(1)$ quantum link model and the $\mathbb{Z}_2$ lattice gauge theory. We have shown numerically, and argued analytically through time-dependent perturbation theory, that whereas without protection the gauge violation and errors in local observables evolve $\propto \gamma t$ in the presence of $1/f$ noise, under linear gauge protection this dynamics changes to $\propto\gamma t/V^\beta$.

Linear gauge protection may also help in suppressing $1/f$ noise sources in recent cold-atom experiments, where long coherent evolution times have been demonstrated \cite{Zhou2021}. This is due to the fact that the perturbative mapping of the $\mathrm{U}(1)$ quantum link model onto the Bose--Hubbard quantum simulator of Refs.~\cite{Yang2020,Zhou2021} gives rise to a leading order term that can be rearranged into a term equivalent to Eq.~\eqref{eq:HG}, with a site-dependent sequence $c_j$ \cite{Lang2022stark}.

Our findings offer the promising prospect of engineering experimentally feasible gauge protection terms that can suppress the growth of gauge violations due to $1/f$-like noise sources, and we expect our conclusions to hold in higher spatial dimensions, as well as for other generic Abelian gauge theories. An interesting avenue lies in studying how gauge violations can be further suppressed by making the time-independent sequence $c_j$ time-dependent. Indeed, for the limit of uncorrelated white noise sources it has been shown that leakage out of the target subspace can be delayed \cite{Stannigel2014}. 

\begin{acknowledgments}
We thank Zhang Jiang for stimulating discussions that inspired us to do this research, as well as Annabelle Bohrdt, Fabian Grusdt, Lukas Homeier, and Haifeng Lang for collaborations on related work. J.C.H.~acknowledges funding from the European Research Council (ERC) under the European Union’s Horizon 2020 research and innovation programm (Grant Agreement no 948141) — ERC Starting Grant SimUcQuam, and by the Deutsche Forschungsgemeinschaft (DFG, German Research Foundation) under Germany's Excellence Strategy -- EXC-2111 -- 390814868. 
P.H.~acknowledges support by the Google Research Scholar Award ProGauge, Provincia Autonoma di Trento, and Q@TN — Quantum Science and Technology in Trento.
This project has received funding from the European Research Council (ERC) under the European Union’s Horizon 2020 research and innovation programme (grant agreement No 804305). 
\end{acknowledgments}

\appendix
\section{Further details on the derivation of the Bloch--Redfield master equation}\label{app:BRE}
In this Appendix, we supply additional details for deriving Eq.~\eqref{eq:redfield}. 
Going into the interaction picture with respect to $\hat{H}_S+\hat{H}_B$ via the operators $\hat{U}_S=e^{-it\hat{H}_S}$ and $\hat{U}_B=e^{-it\hat{H}_B}$, we start by writing the von-Neumann equation
\begin{align}
d_t\hat{\tilde{\rho}}_{SB}(t)=-i\left[\hat{\tilde{H}}_{SB}(t), \hat{\tilde{\rho}}_{S B}(t)\right].    
\end{align}
After substituting the integrated solution into the equation of motion for the combined system, we can obtain the evolution of the reduced density matrix of the system in the interaction picture as
\begin{align}
d_t \hat{\tilde{\rho}}(t)=-\operatorname{Tr}_{B}\left\{\left[\hat{\tilde{H}}(t), \int_{0}^{t} d s\left[\hat{\tilde{H}}(s), \hat{\tilde{\rho}}_{S B}(s)\right]\right]\right\},
\end{align}
where $\hat{\tilde{H}}(t)=\sqrt{\gamma}\sum_{\alpha}\hat{\tilde{A}}_{\alpha}(t)\otimes {\hat{\tilde{B}}_{\alpha}(t)}$. After the change of variables $\tau=t-s$, we get
\begin{align}\label{eq:vn}
d_t \hat{\tilde{\rho}}(t)=-\operatorname{Tr}_{B}\left\{\left[\hat{\tilde{H}}(t), \int_{0}^{t} d s\left[\hat{\tilde{H}}(t-\tau), \hat{\tilde{\rho}}_{S B}(t-\tau)\right]\right]\right\}.
\end{align}

We further proceed to use a Born approximation where we assume the state of the composite system is always uncorrelated and hence can be factorized as $\hat{\tilde{\rho}}_{SB}=\hat{\tilde{\rho}}(t)\otimes{\hat{\rho}_{B}}$, also assuming the bath is much larger than the system in question.
Further, we introduce the Markov approximation, where we assume that the bath has a very short correlation time $\tau_{B}$, i.e., that the correlation function $C_{\alpha \nu}(\tau)=\gamma \operatorname{Tr}_{B}\left[\hat{\tilde{B}}_{\alpha}(t) \hat{\tilde{B}}_{\nu}(t-\tau) \hat{\rho}_{\tilde{B}}\right]=\gamma\left\langle \hat{\tilde{B}}_{\alpha}(\tau) \hat{\tilde{B}}_{\nu}(0)\right\rangle$ decays rapidly with some characteristic
timescale $\left|C_{\alpha \nu}(\tau)\right| \sim e^{-\tau / \tau_{B}}$. In the limit of $\tau_B\to0$ and replacing $\hat{\tilde{\rho}}(t-\tau)$ with $\hat{\tilde{\rho}}(t)$, which is possible due to the fact that correlation function is negligible for $\tau\gg\tau_B$, and under the assumption that $t\gg\tau_B$, one obtains a memory-less evolution of the density matrix. 
It then becomes also a good approximation to extend the integration to infinity as the integrand vanishes sufficiently fast for $\tau\gg\tau_B$, making it a fully Markovian equation. 
These approximations ensure the trace-preserving nature of the density matrix throughout the time evolution. However, the master equation that is obtained is still often times known to give rise to evolution which is not completely positive. Therefore, a secular approximation which is also known as the rotating wave approximation is then used to make the evolution of the resulting dynamical map completely positive (CPTP) \cite{Albash_2012,Davies1976,1974CMaPh..39...91D}. Writing Eq.~\eqref{eq:vn} in terms of system operators and bath correlation functions, one obtains after evaluating the partial trace
\begin{align}\nonumber
d_t\hat{\tilde{\rho}}(t)&=- \sum_{\alpha \nu} \int_{0}^{\infty} d \tau\bigg\{C_{\alpha \nu}(\tau)\Big[\hat{\tilde{A}}_{\alpha}(t) \hat{\tilde{A}}_{\nu}(t-\tau)  \hat{\tilde{\rho}}(t)-\\\nonumber
&\hat{\tilde{A}}_{\alpha}(t-\tau)  \hat{\tilde{\rho}}(t) \hat{\tilde{A}}_{\nu}(t)\Big]
+C_{\alpha \nu}(-\tau)\Big[ \hat{\tilde{\rho}}(t) \hat{\tilde{A}}_{\alpha}(t-\tau) \hat{\tilde{A}}_{\nu}(t)\\\label{eq:bre4}
& -\hat{\tilde{A}}_{\alpha}(t)  \hat{\tilde{\rho}}(t) \hat{\tilde{A}}_{\nu}(t-\tau)\Big]\bigg\}.
\end{align}
Going into the frequency domain and expanding in the eigenbasis of the system Hamiltonian $\hat{H}_S$, we can write the operators acting on the system as
\begin{align}\nonumber
\hat{\tilde{A}}_{\alpha}(t)&=\sum_{m, n} e^{-i\left(\epsilon_{m}-\epsilon_{n}\right) t}\left|\epsilon_{n}\right\rangle\left\langle\epsilon_{n}\left|\hat{A}_{\alpha}\right| \epsilon_{m}\right\rangle\left\langle\epsilon_{m}\right|\\\label{eq:Bre5}
&=\sum_{m,n} A_{mn}(\omega) e^{-i \omega_{mn} t},
\end{align}
where we have defined the transition frequencies $\omega_{mn}=\epsilon_{m}-\epsilon_{n}$. 
In the Schr\"odinger picture, we obtain the master equation in matrix form after substituting Eq.~\eqref{eq:Bre5} into Eq.~\eqref{eq:bre4} as
\begin{align}\nonumber
d_t\rho_{a b}(t) & =-i \omega_{a b} \rho_{a b}(t)- \sum_{\alpha, \nu} \sum_{c, d} \int_{0}^{\infty} d \tau  \bigg\{C_{\alpha \nu}(\tau)\Big[\delta_{b d}\\ \nonumber
&\times\sum_{n} A_{a n}^{\alpha} A_{n c}^{\nu} e^{i \omega_{c n} \tau}-A_{a c}^{\alpha} A_{d b}^{\nu} e^{i \omega_{c a} \tau}\Big]\\\nonumber
&+C_{\alpha \nu}(-\tau)\Big[\delta_{a c} \sum_{n} A_{d n}^{\alpha} A_{n b}^{\nu} e^{i \omega_{n d} \tau}\\
&-A_{a c}^{\alpha} A_{d b}^{\nu} e^{i \omega_{b d} \tau}\Big]\bigg\} \rho_{c d}(t).
\end{align}
Further substituting the expression for spectral function of Eq.~\eqref{eq:sfn} in the above equation under the assumptions of vanishing cross correlations between different environment operators acting at different particle sites, i.e, ${C}_{\nu \alpha}(\tau)={C}_{\alpha \nu}(\tau)=\delta_{\alpha \nu}{C}_{\nu}(\tau)$ we obtain Eq.~\eqref{eq:redfield} of the main text.

\begin{figure}[t!]
	\centering
	\includegraphics[width=0.48\textwidth]{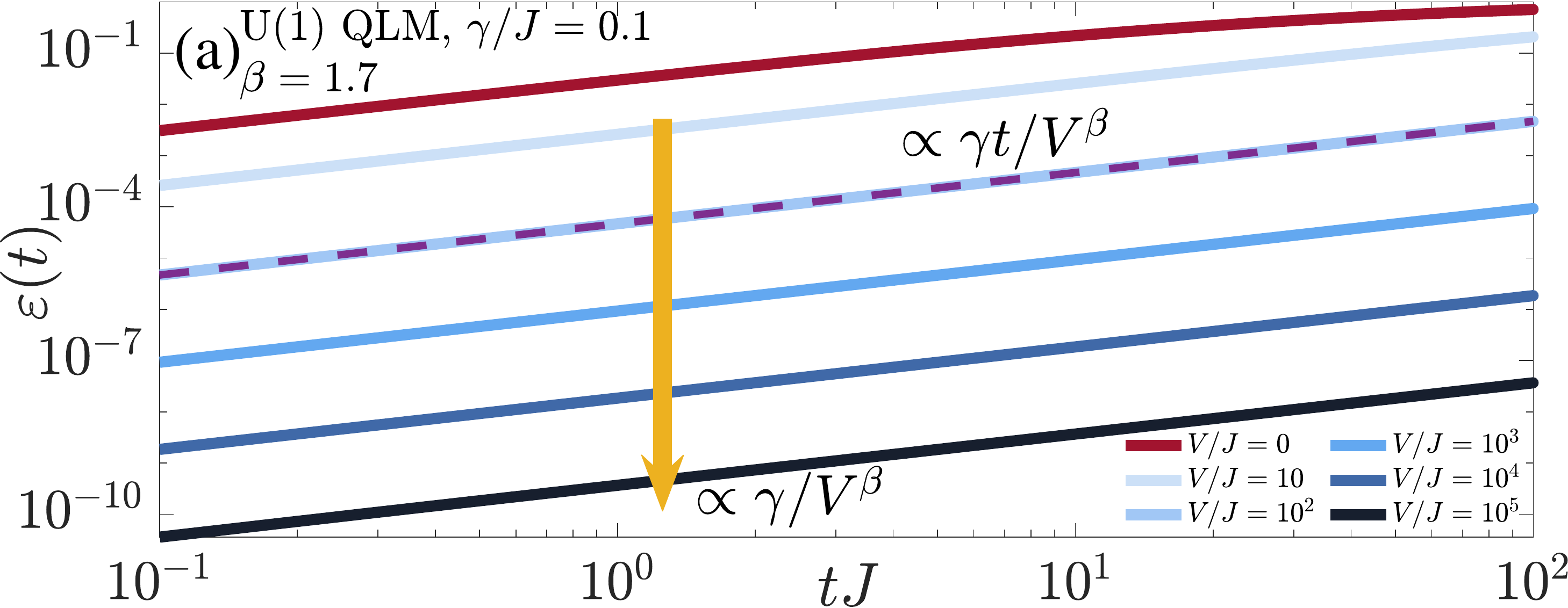}\\
	\vspace{1.1mm}
	\includegraphics[width=0.48\textwidth]{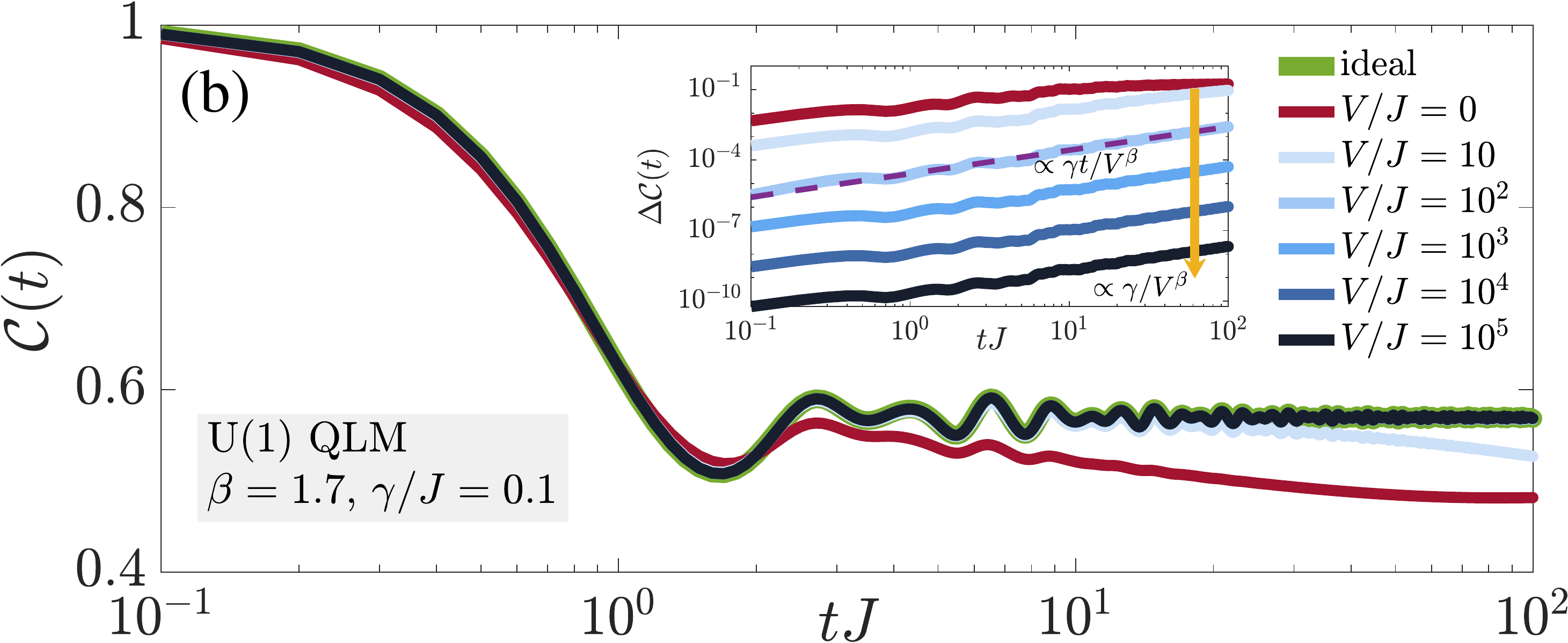}
	\caption{(Color online). Same as Fig.~\ref{fig:U1_beta1.7_gamma0.1}, but for a different gauge-invariant initial state, namely the charge-proliferated state where all matter sites are occupied and all local electric fields point down on their links. This state is also in the target sector $g_j^\text{tar}=0,\,\forall j$. The qualitative picture drawn in the main text is also valid here, where we see that the growth of gauge violation and errors in the local observables are both suppressed as $\propto{\gamma /V^{\beta}}$, indicating the independence of our conclusions from the choice of initial state.}
	\label{fig:U1_beta1.7_gamma0.1_CP} 
\end{figure}

\section{Perturbation theory}\label{app:TDPT}
We can explain the initial growth of gauge violation under $1/f$ noise in our numerical results by perturbatively expanding the Bloch--Redfield master equation. It can be shown that Eq.~\eqref{eq:vn} can be written in the familiar Lindblad form \cite{https://doi.org/10.48550/arxiv.1902.00967}, after employing the secular approximation and transforming back to the Schr\"odinger picture, as
\begin{align}\nonumber
d_t\hat{\rho}=&-i\left[\hat{H}_0+V\hat{H}_G, \hat{\rho}\right]+ \sum_{\omega}\sum_{j} S_{j}(\omega)\\
&\times\Big[\hat{A}_{j}(\omega)\hat{\rho} \hat{A}_{j}^{\dagger}(\omega)-\frac{1}{2}\big\{\hat{A}_{j}^{\dagger}(\omega)\hat{A}_{j}(\omega) ,\hat{\rho} \big\} \Big].
\end{align}
One can write the above in the concise form
\begin{align}\label{eq:a2}
d_t\hat{\rho}=(\mathcal{S}+\mathcal{D})\hat{\rho},
\end{align}
where 
\begin{subequations}
\begin{align}
\mathcal{S}[\hat{\rho}]&=-i\left[\hat{H}_0+V\hat{H}_G, \hat{\rho}\right],\\\nonumber
\mathcal{D}[\hat{\rho}]&=\sum_{\omega}\sum_{j} S_{j}(\omega)
\Big[\hat{A}_{j}(\omega)\hat{\rho} \hat{A}_{j}^{\dagger}(\omega)\\
&-\frac{1}{2}\big\{\hat{A}_{j}^{\dagger}(\omega)\hat{A}_{j}(\omega) ,\hat{\rho} \big\} \Big].
\end{align}
\end{subequations}
By Taylor expanding the solution to Eq.~\eqref{eq:a2}, we can find the leading order incoherent term to explain the growth of the gauge violation in the regimes $V=0$ and $V\gg J$. 
Choosing a target sector $g_j^\mathrm{tar}$, the gauge violation is  $\varepsilon(t)=\mathrm{Tr}\big\{\hat{\mathcal{G}}\hat{\rho}(t)\big\}$ where we have introduced the abbreviation  $\hat{\mathcal{G}}=\sum_{j}(\hat{G}_{j}-g_j^\mathrm{tar})^{2}/L$. 
The contribution of the first-order term in the absence of gauge protection (i.e., $V=0$) is
\begin{align}\nonumber
    t \mathrm{Tr}\{\hat{\mathcal{G}}\mathcal{D}\hat{\rho}_{0}\}= &t\sum_{\omega}\sum_{j} S_{j }(\omega)
\mathrm{Tr}\Big[\hat{\mathcal{G}}\hat{A}_{j}(\omega)\hat{\rho}_{0} \hat{A}_{j}^{\dagger}(\omega)\\\label{eq:A3}
&-\frac{1}{2}\big\{\hat{\mathcal{G}}\hat{A}_{j}^{\dagger}(\omega)\hat{A}_{j}(\omega) ,\hat{\rho}_{0} \big\}\Big]\sim {\gamma } t\,,
\end{align}
where we have utilized the fact that $\mathrm{Tr}\{\hat{\mathcal{G}}\mathcal{S}\hat{\rho}_{0}\}=i\mathrm{Tr}\{\left[\hat{H}_0+V\hat{H}_G, \hat{\mathcal{G}}\right]\hat{\rho}_{0}\}=0$ and the initial value $\mathrm{Tr} \{\hat{\mathcal{G}}\hat{\rho}_{0}\}=0$.

\begin{figure}[t!]
	\centering
	\includegraphics[width=0.48\textwidth]{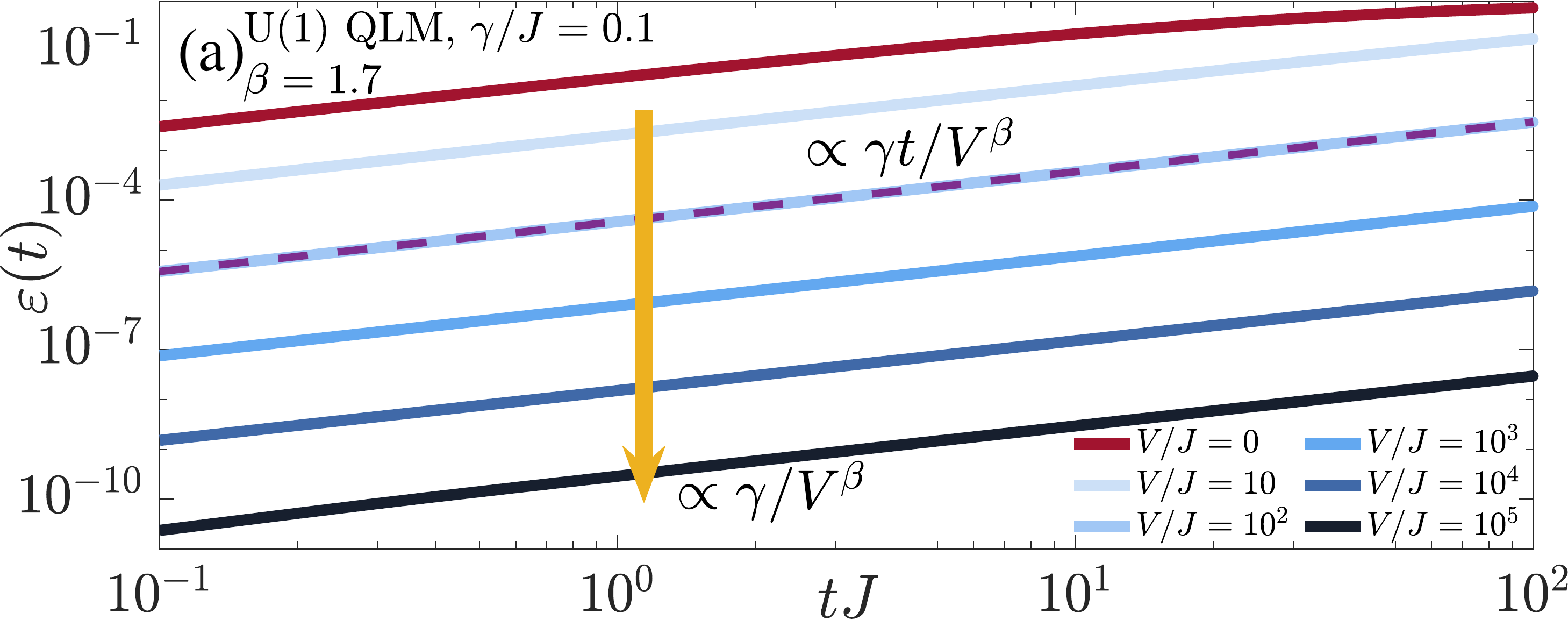}\\
	\vspace{1.1mm}
	\includegraphics[width=0.48\textwidth]{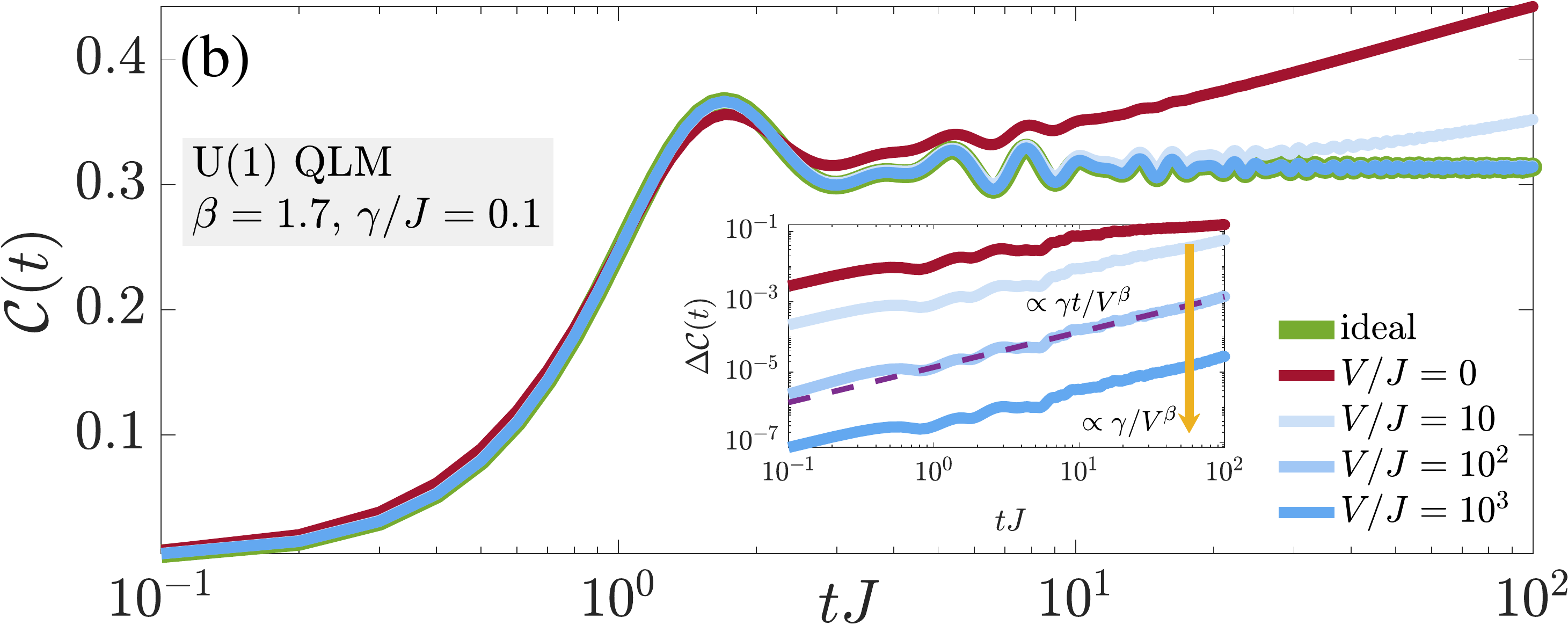}
	\caption{(Color online). Same as Fig.~\ref{fig:U1_beta1.7_gamma0.1} but for a noncompliant sequence $c_j=(-1)^j$, which is more experimentally feasible than its noncompliant counterpart. As seen in the quench dynamics of the (a) gauge violation and (b) the chiral condensate, the suppression of errors also evolves $\propto\gamma t/V^\beta$.}
	\label{fig:U1_beta1.7_gamma0.1_nc} 
\end{figure}

Once the gauge protection is switched on, in the limit $V\gg J$ the dominating coherent term is $\hat{H}_G={\sum_{m}V\epsilon_{m}^{g}|\epsilon_{m}^{g}\rangle\langle\epsilon_{m}^{g}|}$, where $\epsilon_{m}^{g}=\mathbf{c}^\intercal\mathbf{g}$, where $\mathbf{g}$ is a gauge sector. 
The relevant transition frequencies thus scale as $\omega_{mn}\sim V$. 
Taking this into account, neglecting corrections proportional to the energy scales of $\hat{H}_0$, and using the definition of the spectral function in Eq.~\eqref{eq:A3}, we obtain $\epsilon(t)\sim\gamma t/V^{\beta}$, hence explaining the corresponding scaling in the results of the main text up to first order. 
Similar results can also apply to other contexts. E.g., in applications of error correction in adiabatic quantum computing, increasing the energy gap to the excited states can suppress the transition rate out of the code space if the noise power spectrum is decreasing with frequency \cite{PhysRevA.74.052322}.

\section{Supplemental numerical results}\label{app:supp}
The linear gauge protection scheme does not depend on the initial state, and will work effectively so long as the initial state is in the correct gauge sector(s) to be protected. We demonstrate this by repeating the results of Fig.~\ref{fig:U1_beta1.7_gamma0.1} but for a charge-proliferated state, which has every site occupied with matter, and all its local electric fields pointing down. The corresponding dynamics of the gauge violation and chiral condensate are shown in Fig.~\ref{fig:U1_beta1.7_gamma0.1_CP}(a,b), respectively, and the qualitative behavior is identical to that of the vacuum initial state in Fig.~\ref{fig:U1_beta1.7_gamma0.1}, with an error $\propto\gamma t/V^\beta$ in both cases.

Due to numerical overhead, we are limited in our ED calculations to small system sizes. However, in modern cold-atom quantum simulators, much larger sizes can be attained \cite{Yang2020,Zhou2021}. This makes it difficult to construct a compliant sequence for such state-of-the-art quantum simulators, as the coefficients of the latter grow roughly exponentially with system size. However, we can use a simpler noncompliant sequence such as $c_j=(-1)^j$. We repeat the results of Fig.~\ref{fig:U1_beta1.7_gamma0.1} using such a sequence, where the corresponding dynamics is shown in Fig.~\ref{fig:U1_beta1.7_gamma0.1_nc}. We see that both the gauge violation and the chiral condensate show qualitatively identical behavior to the case of the compliant sequence of Fig.~\ref{fig:U1_beta1.7_gamma0.1_nc}, with an error $\propto\gamma t/V^\beta$ in both cases.

\bibliography{biblio}

\begin{thebibliography}{96}%
\makeatletter
\providecommand \@ifxundefined [1]{%
 \@ifx{#1\undefined}
}%
\providecommand \@ifnum [1]{%
 \ifnum #1\expandafter \@firstoftwo
 \else \expandafter \@secondoftwo
 \fi
}%
\providecommand \@ifx [1]{%
 \ifx #1\expandafter \@firstoftwo
 \else \expandafter \@secondoftwo
 \fi
}%
\providecommand \natexlab [1]{#1}%
\providecommand \enquote  [1]{``#1''}%
\providecommand \bibnamefont  [1]{#1}%
\providecommand \bibfnamefont [1]{#1}%
\providecommand \citenamefont [1]{#1}%
\providecommand \href@noop [0]{\@secondoftwo}%
\providecommand \href [0]{\begingroup \@sanitize@url \@href}%
\providecommand \@href[1]{\@@startlink{#1}\@@href}%
\providecommand \@@href[1]{\endgroup#1\@@endlink}%
\providecommand \@sanitize@url [0]{\catcode `\\12\catcode `\$12\catcode
  `\&12\catcode `\#12\catcode `\^12\catcode `\_12\catcode `\%12\relax}%
\providecommand \@@startlink[1]{}%
\providecommand \@@endlink[0]{}%
\providecommand \url  [0]{\begingroup\@sanitize@url \@url }%
\providecommand \@url [1]{\endgroup\@href {#1}{\urlprefix }}%
\providecommand \urlprefix  [0]{URL }%
\providecommand \Eprint [0]{\href }%
\providecommand \doibase [0]{http://dx.doi.org/}%
\providecommand \selectlanguage [0]{\@gobble}%
\providecommand \bibinfo  [0]{\@secondoftwo}%
\providecommand \bibfield  [0]{\@secondoftwo}%
\providecommand \translation [1]{[#1]}%
\providecommand \BibitemOpen [0]{}%
\providecommand \bibitemStop [0]{}%
\providecommand \bibitemNoStop [0]{.\EOS\space}%
\providecommand \EOS [0]{\spacefactor3000\relax}%
\providecommand \BibitemShut  [1]{\csname bibitem#1\endcsname}%
\let\auto@bib@innerbib\@empty
\bibitem [{\citenamefont {Bloch}\ \emph {et~al.}(2008)\citenamefont {Bloch},
  \citenamefont {Dalibard},\ and\ \citenamefont {Zwerger}}]{Bloch2008}%
  \BibitemOpen
  \bibfield  {author} {\bibinfo {author} {\bibfnamefont {Immanuel}\
  \bibnamefont {Bloch}}, \bibinfo {author} {\bibfnamefont {Jean}\ \bibnamefont
  {Dalibard}}, \ and\ \bibinfo {author} {\bibfnamefont {Wilhelm}\ \bibnamefont
  {Zwerger}},\ }\bibfield  {title} {\enquote {\bibinfo {title} {Many-body
  physics with ultracold gases},}\ }\href {\doibase 10.1103/RevModPhys.80.885}
  {\bibfield  {journal} {\bibinfo  {journal} {Rev. Mod. Phys.}\ }\textbf
  {\bibinfo {volume} {80}},\ \bibinfo {pages} {885--964} (\bibinfo {year}
  {2008})}\BibitemShut {NoStop}%
\bibitem [{\citenamefont {Hauke}\ \emph {et~al.}(2012)\citenamefont {Hauke},
  \citenamefont {Cucchietti}, \citenamefont {Tagliacozzo}, \citenamefont
  {Deutsch},\ and\ \citenamefont {Lewenstein}}]{Hauke2012}%
  \BibitemOpen
  \bibfield  {author} {\bibinfo {author} {\bibfnamefont {Philipp}\ \bibnamefont
  {Hauke}}, \bibinfo {author} {\bibfnamefont {Fernando~M}\ \bibnamefont
  {Cucchietti}}, \bibinfo {author} {\bibfnamefont {Luca}\ \bibnamefont
  {Tagliacozzo}}, \bibinfo {author} {\bibfnamefont {Ivan}\ \bibnamefont
  {Deutsch}}, \ and\ \bibinfo {author} {\bibfnamefont {Maciej}\ \bibnamefont
  {Lewenstein}},\ }\bibfield  {title} {\enquote {\bibinfo {title} {Can one
  trust quantum simulators?}}\ }\href {\doibase 10.1088/0034-4885/75/8/082401}
  {\bibfield  {journal} {\bibinfo  {journal} {Reports on Progress in Physics}\
  }\textbf {\bibinfo {volume} {75}},\ \bibinfo {pages} {082401} (\bibinfo
  {year} {2012})}\BibitemShut {NoStop}%
\bibitem [{\citenamefont {Georgescu}\ \emph {et~al.}(2014)\citenamefont
  {Georgescu}, \citenamefont {Ashhab},\ and\ \citenamefont
  {Nori}}]{Georgescu_review}%
  \BibitemOpen
  \bibfield  {author} {\bibinfo {author} {\bibfnamefont {I.~M.}\ \bibnamefont
  {Georgescu}}, \bibinfo {author} {\bibfnamefont {S.}~\bibnamefont {Ashhab}}, \
  and\ \bibinfo {author} {\bibfnamefont {Franco}\ \bibnamefont {Nori}},\
  }\bibfield  {title} {\enquote {\bibinfo {title} {Quantum simulation},}\
  }\href {\doibase 10.1103/RevModPhys.86.153} {\bibfield  {journal} {\bibinfo
  {journal} {Rev. Mod. Phys.}\ }\textbf {\bibinfo {volume} {86}},\ \bibinfo
  {pages} {153--185} (\bibinfo {year} {2014})}\BibitemShut {NoStop}%
\bibitem [{\citenamefont {Altman}\ \emph {et~al.}(2021)\citenamefont {Altman},
  \citenamefont {Brown}, \citenamefont {Carleo}, \citenamefont {Carr},
  \citenamefont {Demler}, \citenamefont {Chin}, \citenamefont {DeMarco},
  \citenamefont {Economou}, \citenamefont {Eriksson}, \citenamefont {Fu},
  \citenamefont {Greiner}, \citenamefont {Hazzard}, \citenamefont {Hulet},
  \citenamefont {Koll\'ar}, \citenamefont {Lev}, \citenamefont {Lukin},
  \citenamefont {Ma}, \citenamefont {Mi}, \citenamefont {Misra}, \citenamefont
  {Monroe}, \citenamefont {Murch}, \citenamefont {Nazario}, \citenamefont {Ni},
  \citenamefont {Potter}, \citenamefont {Roushan}, \citenamefont {Saffman},
  \citenamefont {Schleier-Smith}, \citenamefont {Siddiqi}, \citenamefont
  {Simmonds}, \citenamefont {Singh}, \citenamefont {Spielman}, \citenamefont
  {Temme}, \citenamefont {Weiss}, \citenamefont {Vu\ifmmode \check{c}\else
  \v{c}\fi{}kovi\ifmmode~\acute{c}\else \'{c}\fi{}}, \citenamefont
  {Vuleti\ifmmode~\acute{c}\else \'{c}\fi{}}, \citenamefont {Ye},\ and\
  \citenamefont {Zwierlein}}]{Altman_review}%
  \BibitemOpen
  \bibfield  {author} {\bibinfo {author} {\bibfnamefont {Ehud}\ \bibnamefont
  {Altman}}, \bibinfo {author} {\bibfnamefont {Kenneth~R.}\ \bibnamefont
  {Brown}}, \bibinfo {author} {\bibfnamefont {Giuseppe}\ \bibnamefont
  {Carleo}}, \bibinfo {author} {\bibfnamefont {Lincoln~D.}\ \bibnamefont
  {Carr}}, \bibinfo {author} {\bibfnamefont {Eugene}\ \bibnamefont {Demler}},
  \bibinfo {author} {\bibfnamefont {Cheng}\ \bibnamefont {Chin}}, \bibinfo
  {author} {\bibfnamefont {Brian}\ \bibnamefont {DeMarco}}, \bibinfo {author}
  {\bibfnamefont {Sophia~E.}\ \bibnamefont {Economou}}, \bibinfo {author}
  {\bibfnamefont {Mark~A.}\ \bibnamefont {Eriksson}}, \bibinfo {author}
  {\bibfnamefont {Kai-Mei~C.}\ \bibnamefont {Fu}}, \bibinfo {author}
  {\bibfnamefont {Markus}\ \bibnamefont {Greiner}}, \bibinfo {author}
  {\bibfnamefont {Kaden~R.A.}\ \bibnamefont {Hazzard}}, \bibinfo {author}
  {\bibfnamefont {Randall~G.}\ \bibnamefont {Hulet}}, \bibinfo {author}
  {\bibfnamefont {Alicia~J.}\ \bibnamefont {Koll\'ar}}, \bibinfo {author}
  {\bibfnamefont {Benjamin~L.}\ \bibnamefont {Lev}}, \bibinfo {author}
  {\bibfnamefont {Mikhail~D.}\ \bibnamefont {Lukin}}, \bibinfo {author}
  {\bibfnamefont {Ruichao}\ \bibnamefont {Ma}}, \bibinfo {author}
  {\bibfnamefont {Xiao}\ \bibnamefont {Mi}}, \bibinfo {author} {\bibfnamefont
  {Shashank}\ \bibnamefont {Misra}}, \bibinfo {author} {\bibfnamefont
  {Christopher}\ \bibnamefont {Monroe}}, \bibinfo {author} {\bibfnamefont
  {Kater}\ \bibnamefont {Murch}}, \bibinfo {author} {\bibfnamefont {Zaira}\
  \bibnamefont {Nazario}}, \bibinfo {author} {\bibfnamefont {Kang-Kuen}\
  \bibnamefont {Ni}}, \bibinfo {author} {\bibfnamefont {Andrew~C.}\
  \bibnamefont {Potter}}, \bibinfo {author} {\bibfnamefont {Pedram}\
  \bibnamefont {Roushan}}, \bibinfo {author} {\bibfnamefont {Mark}\
  \bibnamefont {Saffman}}, \bibinfo {author} {\bibfnamefont {Monika}\
  \bibnamefont {Schleier-Smith}}, \bibinfo {author} {\bibfnamefont {Irfan}\
  \bibnamefont {Siddiqi}}, \bibinfo {author} {\bibfnamefont {Raymond}\
  \bibnamefont {Simmonds}}, \bibinfo {author} {\bibfnamefont {Meenakshi}\
  \bibnamefont {Singh}}, \bibinfo {author} {\bibfnamefont {I.B.}\ \bibnamefont
  {Spielman}}, \bibinfo {author} {\bibfnamefont {Kristan}\ \bibnamefont
  {Temme}}, \bibinfo {author} {\bibfnamefont {David~S.}\ \bibnamefont {Weiss}},
  \bibinfo {author} {\bibfnamefont {Jelena}\ \bibnamefont {Vu\ifmmode
  \check{c}\else \v{c}\fi{}kovi\ifmmode~\acute{c}\else \'{c}\fi{}}}, \bibinfo
  {author} {\bibfnamefont {Vladan}\ \bibnamefont {Vuleti\ifmmode~\acute{c}\else
  \'{c}\fi{}}}, \bibinfo {author} {\bibfnamefont {Jun}\ \bibnamefont {Ye}}, \
  and\ \bibinfo {author} {\bibfnamefont {Martin}\ \bibnamefont {Zwierlein}},\
  }\bibfield  {title} {\enquote {\bibinfo {title} {Quantum simulators:
  Architectures and opportunities},}\ }\href {\doibase
  10.1103/PRXQuantum.2.017003} {\bibfield  {journal} {\bibinfo  {journal} {PRX
  Quantum}\ }\textbf {\bibinfo {volume} {2}},\ \bibinfo {pages} {017003}
  (\bibinfo {year} {2021})}\BibitemShut {NoStop}%
\bibitem [{\citenamefont {Rothe}(2005)}]{Rothe_book}%
  \BibitemOpen
  \bibfield  {author} {\bibinfo {author} {\bibfnamefont {H.J.}\ \bibnamefont
  {Rothe}},\ }\href {https://books.google.de/books?id=U1hBLG-\_WxAC} {\emph
  {\bibinfo {title} {Lattice Gauge Theories: An Introduction}}},\ EBSCO ebook
  academic collection\ (\bibinfo  {publisher} {World Scientific},\ \bibinfo
  {year} {2005})\BibitemShut {NoStop}%
\bibitem [{\citenamefont {Alexeev}\ \emph {et~al.}(2021)\citenamefont
  {Alexeev}, \citenamefont {Bacon}, \citenamefont {Brown}, \citenamefont
  {Calderbank}, \citenamefont {Carr}, \citenamefont {Chong}, \citenamefont
  {DeMarco}, \citenamefont {Englund}, \citenamefont {Farhi}, \citenamefont
  {Fefferman}, \citenamefont {Gorshkov}, \citenamefont {Houck}, \citenamefont
  {Kim}, \citenamefont {Kimmel}, \citenamefont {Lange}, \citenamefont {Lloyd},
  \citenamefont {Lukin}, \citenamefont {Maslov}, \citenamefont {Maunz},
  \citenamefont {Monroe}, \citenamefont {Preskill}, \citenamefont {Roetteler},
  \citenamefont {Savage},\ and\ \citenamefont {Thompson}}]{Alexeev_review}%
  \BibitemOpen
  \bibfield  {author} {\bibinfo {author} {\bibfnamefont {Yuri}\ \bibnamefont
  {Alexeev}}, \bibinfo {author} {\bibfnamefont {Dave}\ \bibnamefont {Bacon}},
  \bibinfo {author} {\bibfnamefont {Kenneth~R.}\ \bibnamefont {Brown}},
  \bibinfo {author} {\bibfnamefont {Robert}\ \bibnamefont {Calderbank}},
  \bibinfo {author} {\bibfnamefont {Lincoln~D.}\ \bibnamefont {Carr}}, \bibinfo
  {author} {\bibfnamefont {Frederic~T.}\ \bibnamefont {Chong}}, \bibinfo
  {author} {\bibfnamefont {Brian}\ \bibnamefont {DeMarco}}, \bibinfo {author}
  {\bibfnamefont {Dirk}\ \bibnamefont {Englund}}, \bibinfo {author}
  {\bibfnamefont {Edward}\ \bibnamefont {Farhi}}, \bibinfo {author}
  {\bibfnamefont {Bill}\ \bibnamefont {Fefferman}}, \bibinfo {author}
  {\bibfnamefont {Alexey~V.}\ \bibnamefont {Gorshkov}}, \bibinfo {author}
  {\bibfnamefont {Andrew}\ \bibnamefont {Houck}}, \bibinfo {author}
  {\bibfnamefont {Jungsang}\ \bibnamefont {Kim}}, \bibinfo {author}
  {\bibfnamefont {Shelby}\ \bibnamefont {Kimmel}}, \bibinfo {author}
  {\bibfnamefont {Michael}\ \bibnamefont {Lange}}, \bibinfo {author}
  {\bibfnamefont {Seth}\ \bibnamefont {Lloyd}}, \bibinfo {author}
  {\bibfnamefont {Mikhail~D.}\ \bibnamefont {Lukin}}, \bibinfo {author}
  {\bibfnamefont {Dmitri}\ \bibnamefont {Maslov}}, \bibinfo {author}
  {\bibfnamefont {Peter}\ \bibnamefont {Maunz}}, \bibinfo {author}
  {\bibfnamefont {Christopher}\ \bibnamefont {Monroe}}, \bibinfo {author}
  {\bibfnamefont {John}\ \bibnamefont {Preskill}}, \bibinfo {author}
  {\bibfnamefont {Martin}\ \bibnamefont {Roetteler}}, \bibinfo {author}
  {\bibfnamefont {Martin~J.}\ \bibnamefont {Savage}}, \ and\ \bibinfo {author}
  {\bibfnamefont {Jeff}\ \bibnamefont {Thompson}},\ }\href {\doibase
  10.1103/PRXQuantum.2.017001} {\enquote {\bibinfo {title} {Quantum computer
  systems for scientific discovery},}\ } (\bibinfo {year} {2021})\BibitemShut
  {NoStop}%
\bibitem [{\citenamefont {Klco}\ \emph {et~al.}(2022)\citenamefont {Klco},
  \citenamefont {Roggero},\ and\ \citenamefont {Savage}}]{klco2021standard}%
  \BibitemOpen
  \bibfield  {author} {\bibinfo {author} {\bibfnamefont {Natalie}\ \bibnamefont
  {Klco}}, \bibinfo {author} {\bibfnamefont {Alessandro}\ \bibnamefont
  {Roggero}}, \ and\ \bibinfo {author} {\bibfnamefont {Martin~J}\ \bibnamefont
  {Savage}},\ }\bibfield  {title} {\enquote {\bibinfo {title} {Standard model
  physics and the digital quantum revolution: thoughts about the interface},}\
  }\href {\doibase 10.1088/1361-6633/ac58a4} {\bibfield  {journal} {\bibinfo
  {journal} {Reports on Progress in Physics}\ }\textbf {\bibinfo {volume}
  {85}},\ \bibinfo {pages} {064301} (\bibinfo {year} {2022})}\BibitemShut
  {NoStop}%
\bibitem [{\citenamefont {Dalmonte}\ and\ \citenamefont
  {Montangero}(2016)}]{Dalmonte_review}%
  \BibitemOpen
  \bibfield  {author} {\bibinfo {author} {\bibfnamefont {M.}~\bibnamefont
  {Dalmonte}}\ and\ \bibinfo {author} {\bibfnamefont {S.}~\bibnamefont
  {Montangero}},\ }\bibfield  {title} {\enquote {\bibinfo {title} {Lattice
  gauge theory simulations in the quantum information era},}\ }\href {\doibase
  10.1080/00107514.2016.1151199} {\bibfield  {journal} {\bibinfo  {journal}
  {Contemporary Physics}\ }\textbf {\bibinfo {volume} {57}},\ \bibinfo {pages}
  {388--412} (\bibinfo {year} {2016})},\ \Eprint
  {http://arxiv.org/abs/https://doi.org/10.1080/00107514.2016.1151199}
  {https://doi.org/10.1080/00107514.2016.1151199} \BibitemShut {NoStop}%
\bibitem [{\citenamefont {Zohar}\ \emph {et~al.}(2015)\citenamefont {Zohar},
  \citenamefont {Cirac},\ and\ \citenamefont {Reznik}}]{Zohar_review}%
  \BibitemOpen
  \bibfield  {author} {\bibinfo {author} {\bibfnamefont {Erez}\ \bibnamefont
  {Zohar}}, \bibinfo {author} {\bibfnamefont {J~Ignacio}\ \bibnamefont
  {Cirac}}, \ and\ \bibinfo {author} {\bibfnamefont {Benni}\ \bibnamefont
  {Reznik}},\ }\bibfield  {title} {\enquote {\bibinfo {title} {Quantum
  simulations of lattice gauge theories using ultracold atoms in optical
  lattices},}\ }\href {\doibase 10.1088/0034-4885/79/1/014401} {\bibfield
  {journal} {\bibinfo  {journal} {Reports on Progress in Physics}\ }\textbf
  {\bibinfo {volume} {79}},\ \bibinfo {pages} {014401} (\bibinfo {year}
  {2015})}\BibitemShut {NoStop}%
\bibitem [{\citenamefont {Aidelsburger}\ \emph {et~al.}(2022)\citenamefont
  {Aidelsburger}, \citenamefont {Barbiero}, \citenamefont {Bermudez},
  \citenamefont {Chanda}, \citenamefont {Dauphin}, \citenamefont
  {González-Cuadra}, \citenamefont {Grzybowski}, \citenamefont {Hands},
  \citenamefont {Jendrzejewski}, \citenamefont {Jünemann}, \citenamefont
  {Juzeliūnas}, \citenamefont {Kasper}, \citenamefont {Piga}, \citenamefont
  {Ran}, \citenamefont {Rizzi}, \citenamefont {Sierra}, \citenamefont
  {Tagliacozzo}, \citenamefont {Tirrito}, \citenamefont {Zache}, \citenamefont
  {Zakrzewski}, \citenamefont {Zohar},\ and\ \citenamefont
  {Lewenstein}}]{aidelsburger2021cold}%
  \BibitemOpen
  \bibfield  {author} {\bibinfo {author} {\bibfnamefont {Monika}\ \bibnamefont
  {Aidelsburger}}, \bibinfo {author} {\bibfnamefont {Luca}\ \bibnamefont
  {Barbiero}}, \bibinfo {author} {\bibfnamefont {Alejandro}\ \bibnamefont
  {Bermudez}}, \bibinfo {author} {\bibfnamefont {Titas}\ \bibnamefont
  {Chanda}}, \bibinfo {author} {\bibfnamefont {Alexandre}\ \bibnamefont
  {Dauphin}}, \bibinfo {author} {\bibfnamefont {Daniel}\ \bibnamefont
  {González-Cuadra}}, \bibinfo {author} {\bibfnamefont {Przemysław~R.}\
  \bibnamefont {Grzybowski}}, \bibinfo {author} {\bibfnamefont {Simon}\
  \bibnamefont {Hands}}, \bibinfo {author} {\bibfnamefont {Fred}\ \bibnamefont
  {Jendrzejewski}}, \bibinfo {author} {\bibfnamefont {Johannes}\ \bibnamefont
  {Jünemann}}, \bibinfo {author} {\bibfnamefont {Gediminas}\ \bibnamefont
  {Juzeliūnas}}, \bibinfo {author} {\bibfnamefont {Valentin}\ \bibnamefont
  {Kasper}}, \bibinfo {author} {\bibfnamefont {Angelo}\ \bibnamefont {Piga}},
  \bibinfo {author} {\bibfnamefont {Shi-Ju}\ \bibnamefont {Ran}}, \bibinfo
  {author} {\bibfnamefont {Matteo}\ \bibnamefont {Rizzi}}, \bibinfo {author}
  {\bibfnamefont {Germán}\ \bibnamefont {Sierra}}, \bibinfo {author}
  {\bibfnamefont {Luca}\ \bibnamefont {Tagliacozzo}}, \bibinfo {author}
  {\bibfnamefont {Emanuele}\ \bibnamefont {Tirrito}}, \bibinfo {author}
  {\bibfnamefont {Torsten~V.}\ \bibnamefont {Zache}}, \bibinfo {author}
  {\bibfnamefont {Jakub}\ \bibnamefont {Zakrzewski}}, \bibinfo {author}
  {\bibfnamefont {Erez}\ \bibnamefont {Zohar}}, \ and\ \bibinfo {author}
  {\bibfnamefont {Maciej}\ \bibnamefont {Lewenstein}},\ }\bibfield  {title}
  {\enquote {\bibinfo {title} {Cold atoms meet lattice gauge theory},}\ }\href
  {\doibase 10.1098/rsta.2021.0064} {\bibfield  {journal} {\bibinfo  {journal}
  {Philosophical Transactions of the Royal Society A: Mathematical, Physical
  and Engineering Sciences}\ }\textbf {\bibinfo {volume} {380}},\ \bibinfo
  {pages} {20210064} (\bibinfo {year} {2022})},\ \Eprint
  {http://arxiv.org/abs/https://royalsocietypublishing.org/doi/pdf/10.1098/rsta.2021.0064}
  {https://royalsocietypublishing.org/doi/pdf/10.1098/rsta.2021.0064}
  \BibitemShut {NoStop}%
\bibitem [{\citenamefont {{Zohar}}(2022)}]{Zohar_NewReview}%
  \BibitemOpen
  \bibfield  {author} {\bibinfo {author} {\bibfnamefont {Erez}\ \bibnamefont
  {{Zohar}}},\ }\bibfield  {title} {\enquote {\bibinfo {title} {{Quantum
  simulation of lattice gauge theories in more than one space
  dimension{\textemdash}requirements, challenges and methods}},}\ }\href
  {\doibase 10.1098/rsta.2021.0069} {\bibfield  {journal} {\bibinfo  {journal}
  {Philosophical Transactions of the Royal Society of London Series A}\
  }\textbf {\bibinfo {volume} {380}},\ \bibinfo {eid} {20210069} (\bibinfo
  {year} {2022})},\ \Eprint {http://arxiv.org/abs/2106.04609} {arXiv:2106.04609
  [quant-ph]} \BibitemShut {NoStop}%
\bibitem [{\citenamefont {Davoudi}\ \emph {et~al.}(2022)\citenamefont
  {Davoudi}, \citenamefont {Balantekin}, \citenamefont {Bhattacharya},
  \citenamefont {Carena}, \citenamefont {de~Jong}, \citenamefont {Draper},
  \citenamefont {El-Khadra}, \citenamefont {Gemelke}, \citenamefont {Hanada},
  \citenamefont {Kharzeev}, \citenamefont {Lamm}, \citenamefont {Li},
  \citenamefont {Liu}, \citenamefont {Lukin}, \citenamefont {Meurice},
  \citenamefont {Monroe}, \citenamefont {Nachman}, \citenamefont {Pagano},
  \citenamefont {Preskill}, \citenamefont {Rinaldi}, \citenamefont {Roggero},
  \citenamefont {Santiago}, \citenamefont {Savage}, \citenamefont {Siddiqi},
  \citenamefont {Siopsis}, \citenamefont {Van~Zanten}, \citenamefont {Wiebe},
  \citenamefont {Yamauchi}, \citenamefont {Yeter-Aydeniz},\ and\ \citenamefont
  {Zorzetti}}]{Bauer_review}%
  \BibitemOpen
  \bibfield  {author} {\bibinfo {author} {\bibfnamefont {Christian W.
  Bauer.~Zohreh}\ \bibnamefont {Davoudi}}, \bibinfo {author} {\bibfnamefont
  {A.~Baha}\ \bibnamefont {Balantekin}}, \bibinfo {author} {\bibfnamefont
  {Tanmoy}\ \bibnamefont {Bhattacharya}}, \bibinfo {author} {\bibfnamefont
  {Marcela}\ \bibnamefont {Carena}}, \bibinfo {author} {\bibfnamefont
  {Wibe~A.}\ \bibnamefont {de~Jong}}, \bibinfo {author} {\bibfnamefont
  {Patrick}\ \bibnamefont {Draper}}, \bibinfo {author} {\bibfnamefont {Aida}\
  \bibnamefont {El-Khadra}}, \bibinfo {author} {\bibfnamefont {Nate}\
  \bibnamefont {Gemelke}}, \bibinfo {author} {\bibfnamefont {Masanori}\
  \bibnamefont {Hanada}}, \bibinfo {author} {\bibfnamefont {Dmitri}\
  \bibnamefont {Kharzeev}}, \bibinfo {author} {\bibfnamefont {Henry}\
  \bibnamefont {Lamm}}, \bibinfo {author} {\bibfnamefont {Ying-Ying}\
  \bibnamefont {Li}}, \bibinfo {author} {\bibfnamefont {Junyu}\ \bibnamefont
  {Liu}}, \bibinfo {author} {\bibfnamefont {Mikhail}\ \bibnamefont {Lukin}},
  \bibinfo {author} {\bibfnamefont {Yannick}\ \bibnamefont {Meurice}}, \bibinfo
  {author} {\bibfnamefont {Christopher}\ \bibnamefont {Monroe}}, \bibinfo
  {author} {\bibfnamefont {Benjamin}\ \bibnamefont {Nachman}}, \bibinfo
  {author} {\bibfnamefont {Guido}\ \bibnamefont {Pagano}}, \bibinfo {author}
  {\bibfnamefont {John}\ \bibnamefont {Preskill}}, \bibinfo {author}
  {\bibfnamefont {Enrico}\ \bibnamefont {Rinaldi}}, \bibinfo {author}
  {\bibfnamefont {Alessandro}\ \bibnamefont {Roggero}}, \bibinfo {author}
  {\bibfnamefont {David~I.}\ \bibnamefont {Santiago}}, \bibinfo {author}
  {\bibfnamefont {Martin~J.}\ \bibnamefont {Savage}}, \bibinfo {author}
  {\bibfnamefont {Irfan}\ \bibnamefont {Siddiqi}}, \bibinfo {author}
  {\bibfnamefont {George}\ \bibnamefont {Siopsis}}, \bibinfo {author}
  {\bibfnamefont {David}\ \bibnamefont {Van~Zanten}}, \bibinfo {author}
  {\bibfnamefont {Nathan}\ \bibnamefont {Wiebe}}, \bibinfo {author}
  {\bibfnamefont {Yukari}\ \bibnamefont {Yamauchi}}, \bibinfo {author}
  {\bibfnamefont {Kübra}\ \bibnamefont {Yeter-Aydeniz}}, \ and\ \bibinfo
  {author} {\bibfnamefont {Silvia}\ \bibnamefont {Zorzetti}},\ }\bibfield
  {title} {\enquote {\bibinfo {title} {Quantum simulation for high energy
  physics},}\ }\href {\doibase 10.48550/ARXIV.2204.03381} {\  (\bibinfo {year}
  {2022}),\ 10.48550/ARXIV.2204.03381}\BibitemShut {NoStop}%
\bibitem [{\citenamefont {Catterall}\ \emph {et~al.}(2022)\citenamefont
  {Catterall}, \citenamefont {Harnik}, \citenamefont {Hubeny}, \citenamefont
  {Bauer}, \citenamefont {Berlin}, \citenamefont {Davoudi}, \citenamefont
  {Faulkner}, \citenamefont {Hartman}, \citenamefont {Headrick}, \citenamefont
  {Kahn}, \citenamefont {Lamm}, \citenamefont {Meurice}, \citenamefont
  {Rajendran}, \citenamefont {Rangamani},\ and\ \citenamefont
  {Swingle}}]{Catterall2022}%
  \BibitemOpen
  \bibfield  {author} {\bibinfo {author} {\bibfnamefont {Simon}\ \bibnamefont
  {Catterall}}, \bibinfo {author} {\bibfnamefont {Roni}\ \bibnamefont
  {Harnik}}, \bibinfo {author} {\bibfnamefont {Veronika~E.}\ \bibnamefont
  {Hubeny}}, \bibinfo {author} {\bibfnamefont {Christian~W.}\ \bibnamefont
  {Bauer}}, \bibinfo {author} {\bibfnamefont {Asher}\ \bibnamefont {Berlin}},
  \bibinfo {author} {\bibfnamefont {Zohreh}\ \bibnamefont {Davoudi}}, \bibinfo
  {author} {\bibfnamefont {Thomas}\ \bibnamefont {Faulkner}}, \bibinfo {author}
  {\bibfnamefont {Thomas}\ \bibnamefont {Hartman}}, \bibinfo {author}
  {\bibfnamefont {Matthew}\ \bibnamefont {Headrick}}, \bibinfo {author}
  {\bibfnamefont {Yonatan~F.}\ \bibnamefont {Kahn}}, \bibinfo {author}
  {\bibfnamefont {Henry}\ \bibnamefont {Lamm}}, \bibinfo {author}
  {\bibfnamefont {Yannick}\ \bibnamefont {Meurice}}, \bibinfo {author}
  {\bibfnamefont {Surjeet}\ \bibnamefont {Rajendran}}, \bibinfo {author}
  {\bibfnamefont {Mukund}\ \bibnamefont {Rangamani}}, \ and\ \bibinfo {author}
  {\bibfnamefont {Brian}\ \bibnamefont {Swingle}},\ }\bibfield  {title}
  {\enquote {\bibinfo {title} {Report of the snowmass 2021 theory frontier
  topical group on quantum information science},}\ }\href {\doibase
  10.48550/ARXIV.2209.14839} {\  (\bibinfo {year} {2022}),\
  10.48550/ARXIV.2209.14839}\BibitemShut {NoStop}%
\bibitem [{\citenamefont {Martinez}\ \emph {et~al.}(2016)\citenamefont
  {Martinez}, \citenamefont {Muschik}, \citenamefont {Schindler}, \citenamefont
  {Nigg}, \citenamefont {Erhard}, \citenamefont {Heyl}, \citenamefont {Hauke},
  \citenamefont {Dalmonte}, \citenamefont {Monz}, \citenamefont {Zoller},\ and\
  \citenamefont {Blatt}}]{Martinez2016}%
  \BibitemOpen
  \bibfield  {author} {\bibinfo {author} {\bibfnamefont {Esteban~A.}\
  \bibnamefont {Martinez}}, \bibinfo {author} {\bibfnamefont {Christine~A.}\
  \bibnamefont {Muschik}}, \bibinfo {author} {\bibfnamefont {Philipp}\
  \bibnamefont {Schindler}}, \bibinfo {author} {\bibfnamefont {Daniel}\
  \bibnamefont {Nigg}}, \bibinfo {author} {\bibfnamefont {Alexander}\
  \bibnamefont {Erhard}}, \bibinfo {author} {\bibfnamefont {Markus}\
  \bibnamefont {Heyl}}, \bibinfo {author} {\bibfnamefont {Philipp}\
  \bibnamefont {Hauke}}, \bibinfo {author} {\bibfnamefont {Marcello}\
  \bibnamefont {Dalmonte}}, \bibinfo {author} {\bibfnamefont {Thomas}\
  \bibnamefont {Monz}}, \bibinfo {author} {\bibfnamefont {Peter}\ \bibnamefont
  {Zoller}}, \ and\ \bibinfo {author} {\bibfnamefont {Rainer}\ \bibnamefont
  {Blatt}},\ }\bibfield  {title} {\enquote {\bibinfo {title} {Real-time
  dynamics of lattice gauge theories with a few-qubit quantum computer},}\
  }\href {\doibase 10.1038/nature18318} {\bibfield  {journal} {\bibinfo
  {journal} {Nature}\ }\textbf {\bibinfo {volume} {534}},\ \bibinfo {pages}
  {516--519} (\bibinfo {year} {2016})}\BibitemShut {NoStop}%
\bibitem [{\citenamefont {Muschik}\ \emph {et~al.}(2017)\citenamefont
  {Muschik}, \citenamefont {Heyl}, \citenamefont {Martinez}, \citenamefont
  {Monz}, \citenamefont {Schindler}, \citenamefont {Vogell}, \citenamefont
  {Dalmonte}, \citenamefont {Hauke}, \citenamefont {Blatt},\ and\ \citenamefont
  {Zoller}}]{Muschik2017}%
  \BibitemOpen
  \bibfield  {author} {\bibinfo {author} {\bibfnamefont {Christine}\
  \bibnamefont {Muschik}}, \bibinfo {author} {\bibfnamefont {Markus}\
  \bibnamefont {Heyl}}, \bibinfo {author} {\bibfnamefont {Esteban}\
  \bibnamefont {Martinez}}, \bibinfo {author} {\bibfnamefont {Thomas}\
  \bibnamefont {Monz}}, \bibinfo {author} {\bibfnamefont {Philipp}\
  \bibnamefont {Schindler}}, \bibinfo {author} {\bibfnamefont {Berit}\
  \bibnamefont {Vogell}}, \bibinfo {author} {\bibfnamefont {Marcello}\
  \bibnamefont {Dalmonte}}, \bibinfo {author} {\bibfnamefont {Philipp}\
  \bibnamefont {Hauke}}, \bibinfo {author} {\bibfnamefont {Rainer}\
  \bibnamefont {Blatt}}, \ and\ \bibinfo {author} {\bibfnamefont {Peter}\
  \bibnamefont {Zoller}},\ }\bibfield  {title} {\enquote {\bibinfo {title}
  {U(1) wilson lattice gauge theories in digital quantum simulators},}\ }\href
  {\doibase 10.1088/1367-2630/aa89ab} {\bibfield  {journal} {\bibinfo
  {journal} {New Journal of Physics}\ }\textbf {\bibinfo {volume} {19}},\
  \bibinfo {pages} {103020} (\bibinfo {year} {2017})}\BibitemShut {NoStop}%
\bibitem [{\citenamefont {Bernien}\ \emph {et~al.}(2017)\citenamefont
  {Bernien}, \citenamefont {Schwartz}, \citenamefont {Keesling}, \citenamefont
  {Levine}, \citenamefont {Omran}, \citenamefont {Pichler}, \citenamefont
  {Choi}, \citenamefont {Zibrov}, \citenamefont {Endres}, \citenamefont
  {Greiner}, \citenamefont {Vuleti{\'c}},\ and\ \citenamefont
  {Lukin}}]{Bernien2017}%
  \BibitemOpen
  \bibfield  {author} {\bibinfo {author} {\bibfnamefont {Hannes}\ \bibnamefont
  {Bernien}}, \bibinfo {author} {\bibfnamefont {Sylvain}\ \bibnamefont
  {Schwartz}}, \bibinfo {author} {\bibfnamefont {Alexander}\ \bibnamefont
  {Keesling}}, \bibinfo {author} {\bibfnamefont {Harry}\ \bibnamefont
  {Levine}}, \bibinfo {author} {\bibfnamefont {Ahmed}\ \bibnamefont {Omran}},
  \bibinfo {author} {\bibfnamefont {Hannes}\ \bibnamefont {Pichler}}, \bibinfo
  {author} {\bibfnamefont {Soonwon}\ \bibnamefont {Choi}}, \bibinfo {author}
  {\bibfnamefont {Alexander~S.}\ \bibnamefont {Zibrov}}, \bibinfo {author}
  {\bibfnamefont {Manuel}\ \bibnamefont {Endres}}, \bibinfo {author}
  {\bibfnamefont {Markus}\ \bibnamefont {Greiner}}, \bibinfo {author}
  {\bibfnamefont {Vladan}\ \bibnamefont {Vuleti{\'c}}}, \ and\ \bibinfo
  {author} {\bibfnamefont {Mikhail~D.}\ \bibnamefont {Lukin}},\ }\bibfield
  {title} {\enquote {\bibinfo {title} {Probing many-body dynamics on a 51-atom
  quantum simulator},}\ }\href {\doibase 10.1038/nature24622} {\bibfield
  {journal} {\bibinfo  {journal} {Nature}\ }\textbf {\bibinfo {volume} {551}},\
  \bibinfo {pages} {579--584} (\bibinfo {year} {2017})}\BibitemShut {NoStop}%
\bibitem [{\citenamefont {Klco}\ \emph {et~al.}(2018)\citenamefont {Klco},
  \citenamefont {Dumitrescu}, \citenamefont {McCaskey}, \citenamefont {Morris},
  \citenamefont {Pooser}, \citenamefont {Sanz}, \citenamefont {Solano},
  \citenamefont {Lougovski},\ and\ \citenamefont {Savage}}]{Klco2018}%
  \BibitemOpen
  \bibfield  {author} {\bibinfo {author} {\bibfnamefont {N.}~\bibnamefont
  {Klco}}, \bibinfo {author} {\bibfnamefont {E.~F.}\ \bibnamefont
  {Dumitrescu}}, \bibinfo {author} {\bibfnamefont {A.~J.}\ \bibnamefont
  {McCaskey}}, \bibinfo {author} {\bibfnamefont {T.~D.}\ \bibnamefont
  {Morris}}, \bibinfo {author} {\bibfnamefont {R.~C.}\ \bibnamefont {Pooser}},
  \bibinfo {author} {\bibfnamefont {M.}~\bibnamefont {Sanz}}, \bibinfo {author}
  {\bibfnamefont {E.}~\bibnamefont {Solano}}, \bibinfo {author} {\bibfnamefont
  {P.}~\bibnamefont {Lougovski}}, \ and\ \bibinfo {author} {\bibfnamefont
  {M.~J.}\ \bibnamefont {Savage}},\ }\bibfield  {title} {\enquote {\bibinfo
  {title} {Quantum-classical computation of schwinger model dynamics using
  quantum computers},}\ }\href {\doibase 10.1103/PhysRevA.98.032331} {\bibfield
   {journal} {\bibinfo  {journal} {Phys. Rev. A}\ }\textbf {\bibinfo {volume}
  {98}},\ \bibinfo {pages} {032331} (\bibinfo {year} {2018})}\BibitemShut
  {NoStop}%
\bibitem [{\citenamefont {Kokail}\ \emph {et~al.}(2019)\citenamefont {Kokail},
  \citenamefont {Maier}, \citenamefont {van Bijnen}, \citenamefont {Brydges},
  \citenamefont {Joshi}, \citenamefont {Jurcevic}, \citenamefont {Muschik},
  \citenamefont {Silvi}, \citenamefont {Blatt}, \citenamefont {Roos},\ and\
  \citenamefont {Zoller}}]{Kokail2019}%
  \BibitemOpen
  \bibfield  {author} {\bibinfo {author} {\bibfnamefont {C.}~\bibnamefont
  {Kokail}}, \bibinfo {author} {\bibfnamefont {C.}~\bibnamefont {Maier}},
  \bibinfo {author} {\bibfnamefont {R.}~\bibnamefont {van Bijnen}}, \bibinfo
  {author} {\bibfnamefont {T.}~\bibnamefont {Brydges}}, \bibinfo {author}
  {\bibfnamefont {M.~K.}\ \bibnamefont {Joshi}}, \bibinfo {author}
  {\bibfnamefont {P.}~\bibnamefont {Jurcevic}}, \bibinfo {author}
  {\bibfnamefont {C.~A.}\ \bibnamefont {Muschik}}, \bibinfo {author}
  {\bibfnamefont {P.}~\bibnamefont {Silvi}}, \bibinfo {author} {\bibfnamefont
  {R.}~\bibnamefont {Blatt}}, \bibinfo {author} {\bibfnamefont {C.~F.}\
  \bibnamefont {Roos}}, \ and\ \bibinfo {author} {\bibfnamefont
  {P.}~\bibnamefont {Zoller}},\ }\bibfield  {title} {\enquote {\bibinfo {title}
  {Self-verifying variational quantum simulation of lattice models},}\ }\href
  {\doibase 10.1038/s41586-019-1177-4} {\bibfield  {journal} {\bibinfo
  {journal} {Nature}\ }\textbf {\bibinfo {volume} {569}},\ \bibinfo {pages}
  {355--360} (\bibinfo {year} {2019})}\BibitemShut {NoStop}%
\bibitem [{\citenamefont {Schweizer}\ \emph {et~al.}(2019)\citenamefont
  {Schweizer}, \citenamefont {Grusdt}, \citenamefont {Berngruber},
  \citenamefont {Barbiero}, \citenamefont {Demler}, \citenamefont {Goldman},
  \citenamefont {Bloch},\ and\ \citenamefont {Aidelsburger}}]{Schweizer2019}%
  \BibitemOpen
  \bibfield  {author} {\bibinfo {author} {\bibfnamefont {Christian}\
  \bibnamefont {Schweizer}}, \bibinfo {author} {\bibfnamefont {Fabian}\
  \bibnamefont {Grusdt}}, \bibinfo {author} {\bibfnamefont {Moritz}\
  \bibnamefont {Berngruber}}, \bibinfo {author} {\bibfnamefont {Luca}\
  \bibnamefont {Barbiero}}, \bibinfo {author} {\bibfnamefont {Eugene}\
  \bibnamefont {Demler}}, \bibinfo {author} {\bibfnamefont {Nathan}\
  \bibnamefont {Goldman}}, \bibinfo {author} {\bibfnamefont {Immanuel}\
  \bibnamefont {Bloch}}, \ and\ \bibinfo {author} {\bibfnamefont {Monika}\
  \bibnamefont {Aidelsburger}},\ }\bibfield  {title} {\enquote {\bibinfo
  {title} {Floquet approach to $\mathbb{Z}$2 lattice gauge theories with
  ultracold atoms in optical lattices},}\ }\href {\doibase
  10.1038/s41567-019-0649-7} {\bibfield  {journal} {\bibinfo  {journal} {Nature
  Physics}\ }\textbf {\bibinfo {volume} {15}},\ \bibinfo {pages} {1168--1173}
  (\bibinfo {year} {2019})}\BibitemShut {NoStop}%
\bibitem [{\citenamefont {G{\"o}rg}\ \emph {et~al.}(2019)\citenamefont
  {G{\"o}rg}, \citenamefont {Sandholzer}, \citenamefont {Minguzzi},
  \citenamefont {Desbuquois}, \citenamefont {Messer},\ and\ \citenamefont
  {Esslinger}}]{Goerg2019}%
  \BibitemOpen
  \bibfield  {author} {\bibinfo {author} {\bibfnamefont {Frederik}\
  \bibnamefont {G{\"o}rg}}, \bibinfo {author} {\bibfnamefont {Kilian}\
  \bibnamefont {Sandholzer}}, \bibinfo {author} {\bibfnamefont {Joaqu{\'\i}n}\
  \bibnamefont {Minguzzi}}, \bibinfo {author} {\bibfnamefont {R{\'e}mi}\
  \bibnamefont {Desbuquois}}, \bibinfo {author} {\bibfnamefont {Michael}\
  \bibnamefont {Messer}}, \ and\ \bibinfo {author} {\bibfnamefont {Tilman}\
  \bibnamefont {Esslinger}},\ }\bibfield  {title} {\enquote {\bibinfo {title}
  {Realization of density-dependent peierls phases to engineer quantized gauge
  fields coupled to ultracold matter},}\ }\href {\doibase
  10.1038/s41567-019-0615-4} {\bibfield  {journal} {\bibinfo  {journal} {Nature
  Physics}\ }\textbf {\bibinfo {volume} {15}},\ \bibinfo {pages} {1161--1167}
  (\bibinfo {year} {2019})}\BibitemShut {NoStop}%
\bibitem [{\citenamefont {Mil}\ \emph {et~al.}(2020)\citenamefont {Mil},
  \citenamefont {Zache}, \citenamefont {Hegde}, \citenamefont {Xia},
  \citenamefont {Bhatt}, \citenamefont {Oberthaler}, \citenamefont {Hauke},
  \citenamefont {Berges},\ and\ \citenamefont {Jendrzejewski}}]{Mil2020}%
  \BibitemOpen
  \bibfield  {author} {\bibinfo {author} {\bibfnamefont {Alexander}\
  \bibnamefont {Mil}}, \bibinfo {author} {\bibfnamefont {Torsten~V.}\
  \bibnamefont {Zache}}, \bibinfo {author} {\bibfnamefont {Apoorva}\
  \bibnamefont {Hegde}}, \bibinfo {author} {\bibfnamefont {Andy}\ \bibnamefont
  {Xia}}, \bibinfo {author} {\bibfnamefont {Rohit~P.}\ \bibnamefont {Bhatt}},
  \bibinfo {author} {\bibfnamefont {Markus~K.}\ \bibnamefont {Oberthaler}},
  \bibinfo {author} {\bibfnamefont {Philipp}\ \bibnamefont {Hauke}}, \bibinfo
  {author} {\bibfnamefont {J{\"u}rgen}\ \bibnamefont {Berges}}, \ and\ \bibinfo
  {author} {\bibfnamefont {Fred}\ \bibnamefont {Jendrzejewski}},\ }\bibfield
  {title} {\enquote {\bibinfo {title} {A scalable realization of local u(1)
  gauge invariance in cold atomic mixtures},}\ }\href {\doibase
  10.1126/science.aaz5312} {\bibfield  {journal} {\bibinfo  {journal}
  {Science}\ }\textbf {\bibinfo {volume} {367}},\ \bibinfo {pages} {1128--1130}
  (\bibinfo {year} {2020})}\BibitemShut {NoStop}%
\bibitem [{\citenamefont {Klco}\ \emph {et~al.}(2020)\citenamefont {Klco},
  \citenamefont {Savage},\ and\ \citenamefont {Stryker}}]{Klco2020}%
  \BibitemOpen
  \bibfield  {author} {\bibinfo {author} {\bibfnamefont {Natalie}\ \bibnamefont
  {Klco}}, \bibinfo {author} {\bibfnamefont {Martin~J.}\ \bibnamefont
  {Savage}}, \ and\ \bibinfo {author} {\bibfnamefont {Jesse~R.}\ \bibnamefont
  {Stryker}},\ }\bibfield  {title} {\enquote {\bibinfo {title} {Su(2)
  non-abelian gauge field theory in one dimension on digital quantum
  computers},}\ }\href {\doibase 10.1103/PhysRevD.101.074512} {\bibfield
  {journal} {\bibinfo  {journal} {Phys. Rev. D}\ }\textbf {\bibinfo {volume}
  {101}},\ \bibinfo {pages} {074512} (\bibinfo {year} {2020})}\BibitemShut
  {NoStop}%
\bibitem [{\citenamefont {Yang}\ \emph
  {et~al.}(2020{\natexlab{a}})\citenamefont {Yang}, \citenamefont {Sun},
  \citenamefont {Ott}, \citenamefont {Wang}, \citenamefont {Zache},
  \citenamefont {Halimeh}, \citenamefont {Yuan}, \citenamefont {Hauke},\ and\
  \citenamefont {Pan}}]{Yang2020}%
  \BibitemOpen
  \bibfield  {author} {\bibinfo {author} {\bibfnamefont {Bing}\ \bibnamefont
  {Yang}}, \bibinfo {author} {\bibfnamefont {Hui}\ \bibnamefont {Sun}},
  \bibinfo {author} {\bibfnamefont {Robert}\ \bibnamefont {Ott}}, \bibinfo
  {author} {\bibfnamefont {Han-Yi}\ \bibnamefont {Wang}}, \bibinfo {author}
  {\bibfnamefont {Torsten~V.}\ \bibnamefont {Zache}}, \bibinfo {author}
  {\bibfnamefont {Jad~C.}\ \bibnamefont {Halimeh}}, \bibinfo {author}
  {\bibfnamefont {Zhen-Sheng}\ \bibnamefont {Yuan}}, \bibinfo {author}
  {\bibfnamefont {Philipp}\ \bibnamefont {Hauke}}, \ and\ \bibinfo {author}
  {\bibfnamefont {Jian-Wei}\ \bibnamefont {Pan}},\ }\bibfield  {title}
  {\enquote {\bibinfo {title} {Observation of gauge invariance in a 71-site
  bose--hubbard quantum simulator},}\ }\href {\doibase
  10.1038/s41586-020-2910-8} {\bibfield  {journal} {\bibinfo  {journal}
  {Nature}\ }\textbf {\bibinfo {volume} {587}},\ \bibinfo {pages} {392--396}
  (\bibinfo {year} {2020}{\natexlab{a}})}\BibitemShut {NoStop}%
\bibitem [{\citenamefont {Zhou}\ \emph {et~al.}(2022)\citenamefont {Zhou},
  \citenamefont {Su}, \citenamefont {Halimeh}, \citenamefont {Ott},
  \citenamefont {Sun}, \citenamefont {Hauke}, \citenamefont {Yang},
  \citenamefont {Yuan}, \citenamefont {Berges},\ and\ \citenamefont
  {Pan}}]{Zhou2021}%
  \BibitemOpen
  \bibfield  {author} {\bibinfo {author} {\bibfnamefont {Zhao-Yu}\ \bibnamefont
  {Zhou}}, \bibinfo {author} {\bibfnamefont {Guo-Xian}\ \bibnamefont {Su}},
  \bibinfo {author} {\bibfnamefont {Jad~C.}\ \bibnamefont {Halimeh}}, \bibinfo
  {author} {\bibfnamefont {Robert}\ \bibnamefont {Ott}}, \bibinfo {author}
  {\bibfnamefont {Hui}\ \bibnamefont {Sun}}, \bibinfo {author} {\bibfnamefont
  {Philipp}\ \bibnamefont {Hauke}}, \bibinfo {author} {\bibfnamefont {Bing}\
  \bibnamefont {Yang}}, \bibinfo {author} {\bibfnamefont {Zhen-Sheng}\
  \bibnamefont {Yuan}}, \bibinfo {author} {\bibfnamefont {Jürgen}\
  \bibnamefont {Berges}}, \ and\ \bibinfo {author} {\bibfnamefont {Jian-Wei}\
  \bibnamefont {Pan}},\ }\bibfield  {title} {\enquote {\bibinfo {title}
  {Thermalization dynamics of a gauge theory on a quantum simulator},}\ }\href
  {\doibase 10.1126/science.abl6277} {\bibfield  {journal} {\bibinfo  {journal}
  {Science}\ }\textbf {\bibinfo {volume} {377}},\ \bibinfo {pages} {311--314}
  (\bibinfo {year} {2022})},\ \Eprint
  {http://arxiv.org/abs/https://www.science.org/doi/pdf/10.1126/science.abl6277}
  {https://www.science.org/doi/pdf/10.1126/science.abl6277} \BibitemShut
  {NoStop}%
\bibitem [{\citenamefont {Nguyen}\ \emph {et~al.}(2021)\citenamefont {Nguyen},
  \citenamefont {Tran}, \citenamefont {Zhu}, \citenamefont {Green},
  \citenamefont {Alderete}, \citenamefont {Davoudi},\ and\ \citenamefont
  {Linke}}]{Nguyen2021}%
  \BibitemOpen
  \bibfield  {author} {\bibinfo {author} {\bibfnamefont {Nhung~H.}\
  \bibnamefont {Nguyen}}, \bibinfo {author} {\bibfnamefont {Minh~C.}\
  \bibnamefont {Tran}}, \bibinfo {author} {\bibfnamefont {Yingyue}\
  \bibnamefont {Zhu}}, \bibinfo {author} {\bibfnamefont {Alaina~M.}\
  \bibnamefont {Green}}, \bibinfo {author} {\bibfnamefont {C.~Huerta}\
  \bibnamefont {Alderete}}, \bibinfo {author} {\bibfnamefont {Zohreh}\
  \bibnamefont {Davoudi}}, \ and\ \bibinfo {author} {\bibfnamefont
  {Norbert~M.}\ \bibnamefont {Linke}},\ }\bibfield  {title} {\enquote {\bibinfo
  {title} {Digital quantum simulation of the schwinger model and symmetry
  protection with trapped ions},}\ }\href {\doibase 10.48550/ARXIV.2112.14262}
  {\  (\bibinfo {year} {2021}),\ 10.48550/ARXIV.2112.14262}\BibitemShut
  {NoStop}%
\bibitem [{\citenamefont {Wang}\ \emph {et~al.}(2022)\citenamefont {Wang},
  \citenamefont {Ge}, \citenamefont {Xiang}, \citenamefont {Song},
  \citenamefont {Huang}, \citenamefont {Song}, \citenamefont {Guo},
  \citenamefont {Su}, \citenamefont {Xu}, \citenamefont {Zheng},\ and\
  \citenamefont {Fan}}]{Wang2022}%
  \BibitemOpen
  \bibfield  {author} {\bibinfo {author} {\bibfnamefont {Zhan}\ \bibnamefont
  {Wang}}, \bibinfo {author} {\bibfnamefont {Zi-Yong}\ \bibnamefont {Ge}},
  \bibinfo {author} {\bibfnamefont {Zhongcheng}\ \bibnamefont {Xiang}},
  \bibinfo {author} {\bibfnamefont {Xiaohui}\ \bibnamefont {Song}}, \bibinfo
  {author} {\bibfnamefont {Rui-Zhen}\ \bibnamefont {Huang}}, \bibinfo {author}
  {\bibfnamefont {Pengtao}\ \bibnamefont {Song}}, \bibinfo {author}
  {\bibfnamefont {Xue-Yi}\ \bibnamefont {Guo}}, \bibinfo {author}
  {\bibfnamefont {Luhong}\ \bibnamefont {Su}}, \bibinfo {author} {\bibfnamefont
  {Kai}\ \bibnamefont {Xu}}, \bibinfo {author} {\bibfnamefont {Dongning}\
  \bibnamefont {Zheng}}, \ and\ \bibinfo {author} {\bibfnamefont {Heng}\
  \bibnamefont {Fan}},\ }\bibfield  {title} {\enquote {\bibinfo {title}
  {Observation of emergent $\mathbb{Z}_2$ gauge invariance in a superconducting
  circuit},}\ }\href {\doibase 10.1103/PhysRevResearch.4.L022060} {\bibfield
  {journal} {\bibinfo  {journal} {Phys. Rev. Research}\ }\textbf {\bibinfo
  {volume} {4}},\ \bibinfo {pages} {L022060} (\bibinfo {year}
  {2022})}\BibitemShut {NoStop}%
\bibitem [{\citenamefont {{Mildenberger}}\ \emph {et~al.}(2022)\citenamefont
  {{Mildenberger}}, \citenamefont {{Mruczkiewicz}}, \citenamefont {{Halimeh}},
  \citenamefont {{Jiang}},\ and\ \citenamefont {{Hauke}}}]{Mildenberger2022}%
  \BibitemOpen
  \bibfield  {author} {\bibinfo {author} {\bibfnamefont {Julius}\ \bibnamefont
  {{Mildenberger}}}, \bibinfo {author} {\bibfnamefont {Wojciech}\ \bibnamefont
  {{Mruczkiewicz}}}, \bibinfo {author} {\bibfnamefont {Jad~C.}\ \bibnamefont
  {{Halimeh}}}, \bibinfo {author} {\bibfnamefont {Zhang}\ \bibnamefont
  {{Jiang}}}, \ and\ \bibinfo {author} {\bibfnamefont {Philipp}\ \bibnamefont
  {{Hauke}}},\ }\bibfield  {title} {\enquote {\bibinfo {title} {{Probing
  confinement in a $\mathbb{Z}_2$ lattice gauge theory on a quantum
  computer}},}\ }\href@noop {} {\bibfield  {journal} {\bibinfo  {journal}
  {arXiv e-prints}\ ,\ \bibinfo {eid} {arXiv:2203.08905}} (\bibinfo {year}
  {2022})},\ \Eprint {http://arxiv.org/abs/2203.08905} {arXiv:2203.08905
  [quant-ph]} \BibitemShut {NoStop}%
\bibitem [{\citenamefont {Weinberg}(1995)}]{Weinberg_book}%
  \BibitemOpen
  \bibfield  {author} {\bibinfo {author} {\bibfnamefont {S.}~\bibnamefont
  {Weinberg}},\ }\href {https://books.google.de/books?id=doeDB3\_WLvwC} {\emph
  {\bibinfo {title} {The Quantum Theory of Fields}}},\ Vol. 2: Modern
  Applications\ (\bibinfo  {publisher} {Cambridge University Press},\ \bibinfo
  {year} {1995})\BibitemShut {NoStop}%
\bibitem [{\citenamefont {Gattringer}\ and\ \citenamefont
  {Lang}(2009)}]{Gattringer_book}%
  \BibitemOpen
  \bibfield  {author} {\bibinfo {author} {\bibfnamefont {C.}~\bibnamefont
  {Gattringer}}\ and\ \bibinfo {author} {\bibfnamefont {C.}~\bibnamefont
  {Lang}},\ }\href {https://books.google.de/books?id=l2hZKnlYDxoC} {\emph
  {\bibinfo {title} {Quantum Chromodynamics on the Lattice: An Introductory
  Presentation}}},\ Lecture Notes in Physics\ (\bibinfo  {publisher} {Springer
  Berlin Heidelberg},\ \bibinfo {year} {2009})\BibitemShut {NoStop}%
\bibitem [{\citenamefont {Zee}(2003)}]{Zee_book}%
  \BibitemOpen
  \bibfield  {author} {\bibinfo {author} {\bibfnamefont {A.}~\bibnamefont
  {Zee}},\ }\href {https://books.google.de/books?id=85G9QgAACAAJ} {\emph
  {\bibinfo {title} {Quantum Field Theory in a Nutshell}}}\ (\bibinfo
  {publisher} {Princeton University Press},\ \bibinfo {year}
  {2003})\BibitemShut {NoStop}%
\bibitem [{\citenamefont {Halimeh}\ and\ \citenamefont
  {Hauke}(2020{\natexlab{a}})}]{Halimeh2020a}%
  \BibitemOpen
  \bibfield  {author} {\bibinfo {author} {\bibfnamefont {Jad~C.}\ \bibnamefont
  {Halimeh}}\ and\ \bibinfo {author} {\bibfnamefont {Philipp}\ \bibnamefont
  {Hauke}},\ }\bibfield  {title} {\enquote {\bibinfo {title} {Reliability of
  lattice gauge theories},}\ }\href {\doibase 10.1103/PhysRevLett.125.030503}
  {\bibfield  {journal} {\bibinfo  {journal} {Phys. Rev. Lett.}\ }\textbf
  {\bibinfo {volume} {125}},\ \bibinfo {pages} {030503} (\bibinfo {year}
  {2020}{\natexlab{a}})}\BibitemShut {NoStop}%
\bibitem [{\citenamefont {Halimeh}\ and\ \citenamefont
  {Hauke}(2020{\natexlab{b}})}]{Halimeh2020b}%
  \BibitemOpen
  \bibfield  {author} {\bibinfo {author} {\bibfnamefont {Jad~C.}\ \bibnamefont
  {Halimeh}}\ and\ \bibinfo {author} {\bibfnamefont {Philipp}\ \bibnamefont
  {Hauke}},\ }\bibfield  {title} {\enquote {\bibinfo {title} {Staircase
  prethermalization and constrained dynamics in lattice gauge theories},}\
  }\href@noop {} {\  (\bibinfo {year} {2020}{\natexlab{b}})},\ \Eprint
  {http://arxiv.org/abs/2004.07248} {arXiv:2004.07248 [cond-mat.quant-gas]}
  \BibitemShut {NoStop}%
\bibitem [{\citenamefont {Halimeh}\ and\ \citenamefont
  {Hauke}(2020{\natexlab{c}})}]{Halimeh2020c}%
  \BibitemOpen
  \bibfield  {author} {\bibinfo {author} {\bibfnamefont {Jad~C.}\ \bibnamefont
  {Halimeh}}\ and\ \bibinfo {author} {\bibfnamefont {Philipp}\ \bibnamefont
  {Hauke}},\ }\bibfield  {title} {\enquote {\bibinfo {title} {Origin of
  staircase prethermalization in lattice gauge theories},}\ }\href@noop {} {\
  (\bibinfo {year} {2020}{\natexlab{c}})},\ \Eprint
  {http://arxiv.org/abs/2004.07254} {arXiv:2004.07254 [cond-mat.str-el]}
  \BibitemShut {NoStop}%
\bibitem [{\citenamefont {Halimeh}\ \emph {et~al.}(2020)\citenamefont
  {Halimeh}, \citenamefont {Kasper},\ and\ \citenamefont
  {Hauke}}]{Halimeh2020f}%
  \BibitemOpen
  \bibfield  {author} {\bibinfo {author} {\bibfnamefont {Jad~C.}\ \bibnamefont
  {Halimeh}}, \bibinfo {author} {\bibfnamefont {Valentin}\ \bibnamefont
  {Kasper}}, \ and\ \bibinfo {author} {\bibfnamefont {Philipp}\ \bibnamefont
  {Hauke}},\ }\bibfield  {title} {\enquote {\bibinfo {title} {Fate of lattice
  gauge theories under decoherence},}\ }\href@noop {} {\  (\bibinfo {year}
  {2020})},\ \Eprint {http://arxiv.org/abs/2009.07848} {arXiv:2009.07848
  [cond-mat.quant-gas]} \BibitemShut {NoStop}%
\bibitem [{\citenamefont {Zohar}\ and\ \citenamefont
  {Reznik}(2011)}]{Zohar2011}%
  \BibitemOpen
  \bibfield  {author} {\bibinfo {author} {\bibfnamefont {Erez}\ \bibnamefont
  {Zohar}}\ and\ \bibinfo {author} {\bibfnamefont {Benni}\ \bibnamefont
  {Reznik}},\ }\bibfield  {title} {\enquote {\bibinfo {title} {Confinement and
  lattice quantum-electrodynamic electric flux tubes simulated with ultracold
  atoms},}\ }\href {\doibase 10.1103/PhysRevLett.107.275301} {\bibfield
  {journal} {\bibinfo  {journal} {Phys. Rev. Lett.}\ }\textbf {\bibinfo
  {volume} {107}},\ \bibinfo {pages} {275301} (\bibinfo {year}
  {2011})}\BibitemShut {NoStop}%
\bibitem [{\citenamefont {Zohar}\ \emph {et~al.}(2012)\citenamefont {Zohar},
  \citenamefont {Cirac},\ and\ \citenamefont {Reznik}}]{Zohar2012}%
  \BibitemOpen
  \bibfield  {author} {\bibinfo {author} {\bibfnamefont {Erez}\ \bibnamefont
  {Zohar}}, \bibinfo {author} {\bibfnamefont {J.~Ignacio}\ \bibnamefont
  {Cirac}}, \ and\ \bibinfo {author} {\bibfnamefont {Benni}\ \bibnamefont
  {Reznik}},\ }\bibfield  {title} {\enquote {\bibinfo {title} {Simulating
  compact quantum electrodynamics with ultracold atoms: Probing confinement and
  nonperturbative effects},}\ }\href {\doibase 10.1103/PhysRevLett.109.125302}
  {\bibfield  {journal} {\bibinfo  {journal} {Phys. Rev. Lett.}\ }\textbf
  {\bibinfo {volume} {109}},\ \bibinfo {pages} {125302} (\bibinfo {year}
  {2012})}\BibitemShut {NoStop}%
\bibitem [{\citenamefont {Banerjee}\ \emph {et~al.}(2012)\citenamefont
  {Banerjee}, \citenamefont {Dalmonte}, \citenamefont {M\"uller}, \citenamefont
  {Rico}, \citenamefont {Stebler}, \citenamefont {Wiese},\ and\ \citenamefont
  {Zoller}}]{Banerjee2012}%
  \BibitemOpen
  \bibfield  {author} {\bibinfo {author} {\bibfnamefont {D.}~\bibnamefont
  {Banerjee}}, \bibinfo {author} {\bibfnamefont {M.}~\bibnamefont {Dalmonte}},
  \bibinfo {author} {\bibfnamefont {M.}~\bibnamefont {M\"uller}}, \bibinfo
  {author} {\bibfnamefont {E.}~\bibnamefont {Rico}}, \bibinfo {author}
  {\bibfnamefont {P.}~\bibnamefont {Stebler}}, \bibinfo {author} {\bibfnamefont
  {U.-J.}\ \bibnamefont {Wiese}}, \ and\ \bibinfo {author} {\bibfnamefont
  {P.}~\bibnamefont {Zoller}},\ }\bibfield  {title} {\enquote {\bibinfo {title}
  {Atomic quantum simulation of dynamical gauge fields coupled to fermionic
  matter: From string breaking to evolution after a quench},}\ }\href {\doibase
  10.1103/PhysRevLett.109.175302} {\bibfield  {journal} {\bibinfo  {journal}
  {Phys. Rev. Lett.}\ }\textbf {\bibinfo {volume} {109}},\ \bibinfo {pages}
  {175302} (\bibinfo {year} {2012})}\BibitemShut {NoStop}%
\bibitem [{\citenamefont {Zohar}\ \emph {et~al.}(2013)\citenamefont {Zohar},
  \citenamefont {Cirac},\ and\ \citenamefont {Reznik}}]{Zohar2013}%
  \BibitemOpen
  \bibfield  {author} {\bibinfo {author} {\bibfnamefont {Erez}\ \bibnamefont
  {Zohar}}, \bibinfo {author} {\bibfnamefont {J.~Ignacio}\ \bibnamefont
  {Cirac}}, \ and\ \bibinfo {author} {\bibfnamefont {Benni}\ \bibnamefont
  {Reznik}},\ }\bibfield  {title} {\enquote {\bibinfo {title} {Simulating
  ($2+1$)-dimensional lattice qed with dynamical matter using ultracold
  atoms},}\ }\href {\doibase 10.1103/PhysRevLett.110.055302} {\bibfield
  {journal} {\bibinfo  {journal} {Phys. Rev. Lett.}\ }\textbf {\bibinfo
  {volume} {110}},\ \bibinfo {pages} {055302} (\bibinfo {year}
  {2013})}\BibitemShut {NoStop}%
\bibitem [{\citenamefont {Banerjee}\ \emph {et~al.}(2013)\citenamefont
  {Banerjee}, \citenamefont {B\"ogli}, \citenamefont {Dalmonte}, \citenamefont
  {Rico}, \citenamefont {Stebler}, \citenamefont {Wiese},\ and\ \citenamefont
  {Zoller}}]{Banerjee2013}%
  \BibitemOpen
  \bibfield  {author} {\bibinfo {author} {\bibfnamefont {D.}~\bibnamefont
  {Banerjee}}, \bibinfo {author} {\bibfnamefont {M.}~\bibnamefont {B\"ogli}},
  \bibinfo {author} {\bibfnamefont {M.}~\bibnamefont {Dalmonte}}, \bibinfo
  {author} {\bibfnamefont {E.}~\bibnamefont {Rico}}, \bibinfo {author}
  {\bibfnamefont {P.}~\bibnamefont {Stebler}}, \bibinfo {author} {\bibfnamefont
  {U.-J.}\ \bibnamefont {Wiese}}, \ and\ \bibinfo {author} {\bibfnamefont
  {P.}~\bibnamefont {Zoller}},\ }\bibfield  {title} {\enquote {\bibinfo {title}
  {Atomic quantum simulation of $\mathbf{U}(n)$ and $\mathrm{SU}(n)$
  non-abelian lattice gauge theories},}\ }\href {\doibase
  10.1103/PhysRevLett.110.125303} {\bibfield  {journal} {\bibinfo  {journal}
  {Phys. Rev. Lett.}\ }\textbf {\bibinfo {volume} {110}},\ \bibinfo {pages}
  {125303} (\bibinfo {year} {2013})}\BibitemShut {NoStop}%
\bibitem [{\citenamefont {Hauke}\ \emph {et~al.}(2013)\citenamefont {Hauke},
  \citenamefont {Marcos}, \citenamefont {Dalmonte},\ and\ \citenamefont
  {Zoller}}]{Hauke2013}%
  \BibitemOpen
  \bibfield  {author} {\bibinfo {author} {\bibfnamefont {P.}~\bibnamefont
  {Hauke}}, \bibinfo {author} {\bibfnamefont {D.}~\bibnamefont {Marcos}},
  \bibinfo {author} {\bibfnamefont {M.}~\bibnamefont {Dalmonte}}, \ and\
  \bibinfo {author} {\bibfnamefont {P.}~\bibnamefont {Zoller}},\ }\bibfield
  {title} {\enquote {\bibinfo {title} {Quantum simulation of a lattice
  schwinger model in a chain of trapped ions},}\ }\href {\doibase
  10.1103/PhysRevX.3.041018} {\bibfield  {journal} {\bibinfo  {journal} {Phys.
  Rev. X}\ }\textbf {\bibinfo {volume} {3}},\ \bibinfo {pages} {041018}
  (\bibinfo {year} {2013})}\BibitemShut {NoStop}%
\bibitem [{\citenamefont {Stannigel}\ \emph {et~al.}(2014)\citenamefont
  {Stannigel}, \citenamefont {Hauke}, \citenamefont {Marcos}, \citenamefont
  {Hafezi}, \citenamefont {Diehl}, \citenamefont {Dalmonte},\ and\
  \citenamefont {Zoller}}]{Stannigel2014}%
  \BibitemOpen
  \bibfield  {author} {\bibinfo {author} {\bibfnamefont {K.}~\bibnamefont
  {Stannigel}}, \bibinfo {author} {\bibfnamefont {P.}~\bibnamefont {Hauke}},
  \bibinfo {author} {\bibfnamefont {D.}~\bibnamefont {Marcos}}, \bibinfo
  {author} {\bibfnamefont {M.}~\bibnamefont {Hafezi}}, \bibinfo {author}
  {\bibfnamefont {S.}~\bibnamefont {Diehl}}, \bibinfo {author} {\bibfnamefont
  {M.}~\bibnamefont {Dalmonte}}, \ and\ \bibinfo {author} {\bibfnamefont
  {P.}~\bibnamefont {Zoller}},\ }\bibfield  {title} {\enquote {\bibinfo {title}
  {Constrained dynamics via the zeno effect in quantum simulation: Implementing
  non-abelian lattice gauge theories with cold atoms},}\ }\href {\doibase
  10.1103/PhysRevLett.112.120406} {\bibfield  {journal} {\bibinfo  {journal}
  {Phys. Rev. Lett.}\ }\textbf {\bibinfo {volume} {112}},\ \bibinfo {pages}
  {120406} (\bibinfo {year} {2014})}\BibitemShut {NoStop}%
\bibitem [{\citenamefont {K\"uhn}\ \emph {et~al.}(2014)\citenamefont {K\"uhn},
  \citenamefont {Cirac},\ and\ \citenamefont {Ba\~nuls}}]{Kuehn2014}%
  \BibitemOpen
  \bibfield  {author} {\bibinfo {author} {\bibfnamefont {Stefan}\ \bibnamefont
  {K\"uhn}}, \bibinfo {author} {\bibfnamefont {J.~Ignacio}\ \bibnamefont
  {Cirac}}, \ and\ \bibinfo {author} {\bibfnamefont {Mari-Carmen}\ \bibnamefont
  {Ba\~nuls}},\ }\bibfield  {title} {\enquote {\bibinfo {title} {Quantum
  simulation of the schwinger model: A study of feasibility},}\ }\href
  {\doibase 10.1103/PhysRevA.90.042305} {\bibfield  {journal} {\bibinfo
  {journal} {Phys. Rev. A}\ }\textbf {\bibinfo {volume} {90}},\ \bibinfo
  {pages} {042305} (\bibinfo {year} {2014})}\BibitemShut {NoStop}%
\bibitem [{\citenamefont {Kuno}\ \emph {et~al.}(2015)\citenamefont {Kuno},
  \citenamefont {Kasamatsu}, \citenamefont {Takahashi}, \citenamefont
  {Ichinose},\ and\ \citenamefont {Matsui}}]{Kuno2015}%
  \BibitemOpen
  \bibfield  {author} {\bibinfo {author} {\bibfnamefont {Yoshihito}\
  \bibnamefont {Kuno}}, \bibinfo {author} {\bibfnamefont {Kenichi}\
  \bibnamefont {Kasamatsu}}, \bibinfo {author} {\bibfnamefont {Yoshiro}\
  \bibnamefont {Takahashi}}, \bibinfo {author} {\bibfnamefont {Ikuo}\
  \bibnamefont {Ichinose}}, \ and\ \bibinfo {author} {\bibfnamefont {Tetsuo}\
  \bibnamefont {Matsui}},\ }\bibfield  {title} {\enquote {\bibinfo {title}
  {Real-time dynamics and proposal for feasible experiments of lattice
  gauge{\textendash}higgs model simulated by cold atoms},}\ }\href {\doibase
  10.1088/1367-2630/17/6/063005} {\bibfield  {journal} {\bibinfo  {journal}
  {New Journal of Physics}\ }\textbf {\bibinfo {volume} {17}},\ \bibinfo
  {pages} {063005} (\bibinfo {year} {2015})}\BibitemShut {NoStop}%
\bibitem [{\citenamefont {Yang}\ \emph {et~al.}(2016)\citenamefont {Yang},
  \citenamefont {Giri}, \citenamefont {Johanning}, \citenamefont {Wunderlich},
  \citenamefont {Zoller},\ and\ \citenamefont {Hauke}}]{Yang2016}%
  \BibitemOpen
  \bibfield  {author} {\bibinfo {author} {\bibfnamefont {Dayou}\ \bibnamefont
  {Yang}}, \bibinfo {author} {\bibfnamefont {Gouri~Shankar}\ \bibnamefont
  {Giri}}, \bibinfo {author} {\bibfnamefont {Michael}\ \bibnamefont
  {Johanning}}, \bibinfo {author} {\bibfnamefont {Christof}\ \bibnamefont
  {Wunderlich}}, \bibinfo {author} {\bibfnamefont {Peter}\ \bibnamefont
  {Zoller}}, \ and\ \bibinfo {author} {\bibfnamefont {Philipp}\ \bibnamefont
  {Hauke}},\ }\bibfield  {title} {\enquote {\bibinfo {title} {Analog quantum
  simulation of $(1+1)$-dimensional lattice qed with trapped ions},}\ }\href
  {\doibase 10.1103/PhysRevA.94.052321} {\bibfield  {journal} {\bibinfo
  {journal} {Phys. Rev. A}\ }\textbf {\bibinfo {volume} {94}},\ \bibinfo
  {pages} {052321} (\bibinfo {year} {2016})}\BibitemShut {NoStop}%
\bibitem [{\citenamefont {Kuno}\ \emph {et~al.}(2017)\citenamefont {Kuno},
  \citenamefont {Sakane}, \citenamefont {Kasamatsu}, \citenamefont {Ichinose},\
  and\ \citenamefont {Matsui}}]{Kuno2017}%
  \BibitemOpen
  \bibfield  {author} {\bibinfo {author} {\bibfnamefont {Yoshihito}\
  \bibnamefont {Kuno}}, \bibinfo {author} {\bibfnamefont {Shinya}\ \bibnamefont
  {Sakane}}, \bibinfo {author} {\bibfnamefont {Kenichi}\ \bibnamefont
  {Kasamatsu}}, \bibinfo {author} {\bibfnamefont {Ikuo}\ \bibnamefont
  {Ichinose}}, \ and\ \bibinfo {author} {\bibfnamefont {Tetsuo}\ \bibnamefont
  {Matsui}},\ }\bibfield  {title} {\enquote {\bibinfo {title} {Quantum
  simulation of ($1+1$)-dimensional u(1) gauge-higgs model on a lattice by cold
  bose gases},}\ }\href {\doibase 10.1103/PhysRevD.95.094507} {\bibfield
  {journal} {\bibinfo  {journal} {Phys. Rev. D}\ }\textbf {\bibinfo {volume}
  {95}},\ \bibinfo {pages} {094507} (\bibinfo {year} {2017})}\BibitemShut
  {NoStop}%
\bibitem [{\citenamefont {Dehkharghani}\ \emph {et~al.}(2017)\citenamefont
  {Dehkharghani}, \citenamefont {Rico}, \citenamefont {Zinner},\ and\
  \citenamefont {Negretti}}]{Negretti2017}%
  \BibitemOpen
  \bibfield  {author} {\bibinfo {author} {\bibfnamefont {A.~S.}\ \bibnamefont
  {Dehkharghani}}, \bibinfo {author} {\bibfnamefont {E.}~\bibnamefont {Rico}},
  \bibinfo {author} {\bibfnamefont {N.~T.}\ \bibnamefont {Zinner}}, \ and\
  \bibinfo {author} {\bibfnamefont {A.}~\bibnamefont {Negretti}},\ }\bibfield
  {title} {\enquote {\bibinfo {title} {Quantum simulation of abelian lattice
  gauge theories via state-dependent hopping},}\ }\href {\doibase
  10.1103/PhysRevA.96.043611} {\bibfield  {journal} {\bibinfo  {journal} {Phys.
  Rev. A}\ }\textbf {\bibinfo {volume} {96}},\ \bibinfo {pages} {043611}
  (\bibinfo {year} {2017})}\BibitemShut {NoStop}%
\bibitem [{\citenamefont {Dutta}\ \emph {et~al.}(2017)\citenamefont {Dutta},
  \citenamefont {Tagliacozzo}, \citenamefont {Lewenstein},\ and\ \citenamefont
  {Zakrzewski}}]{Dutta2017}%
  \BibitemOpen
  \bibfield  {author} {\bibinfo {author} {\bibfnamefont {Omjyoti}\ \bibnamefont
  {Dutta}}, \bibinfo {author} {\bibfnamefont {Luca}\ \bibnamefont
  {Tagliacozzo}}, \bibinfo {author} {\bibfnamefont {Maciej}\ \bibnamefont
  {Lewenstein}}, \ and\ \bibinfo {author} {\bibfnamefont {Jakub}\ \bibnamefont
  {Zakrzewski}},\ }\bibfield  {title} {\enquote {\bibinfo {title} {Toolbox for
  abelian lattice gauge theories with synthetic matter},}\ }\href {\doibase
  10.1103/PhysRevA.95.053608} {\bibfield  {journal} {\bibinfo  {journal} {Phys.
  Rev. A}\ }\textbf {\bibinfo {volume} {95}},\ \bibinfo {pages} {053608}
  (\bibinfo {year} {2017})}\BibitemShut {NoStop}%
\bibitem [{\citenamefont {Pinto~Barros}\ \emph {et~al.}(2020)\citenamefont
  {Pinto~Barros}, \citenamefont {Burrello},\ and\ \citenamefont
  {Trombettoni}}]{Barros2019}%
  \BibitemOpen
  \bibfield  {author} {\bibinfo {author} {\bibfnamefont {Jo{\~a}o~C.}\
  \bibnamefont {Pinto~Barros}}, \bibinfo {author} {\bibfnamefont {Michele}\
  \bibnamefont {Burrello}}, \ and\ \bibinfo {author} {\bibfnamefont {Andrea}\
  \bibnamefont {Trombettoni}},\ }\bibfield  {title} {\enquote {\bibinfo {title}
  {Gauge theories with ultracold atoms},}\ }in\ \href@noop {} {\emph {\bibinfo
  {booktitle} {Strongly Coupled Field Theories for Condensed Matter and Quantum
  Information Theory}}},\ \bibinfo {editor} {edited by\ \bibinfo {editor}
  {\bibfnamefont {Alvaro}\ \bibnamefont {Ferraz}}, \bibinfo {editor}
  {\bibfnamefont {Kumar~S.}\ \bibnamefont {Gupta}}, \bibinfo {editor}
  {\bibfnamefont {Gordon~Walter}\ \bibnamefont {Semenoff}}, \ and\ \bibinfo
  {editor} {\bibfnamefont {Pasquale}\ \bibnamefont {Sodano}}}\ (\bibinfo
  {publisher} {Springer International Publishing},\ \bibinfo {address} {Cham},\
  \bibinfo {year} {2020})\ pp.\ \bibinfo {pages} {217--245}\BibitemShut
  {NoStop}%
\bibitem [{\citenamefont {Kasper}\ \emph {et~al.}(2021)\citenamefont {Kasper},
  \citenamefont {Zache}, \citenamefont {Jendrzejewski}, \citenamefont
  {Lewenstein},\ and\ \citenamefont {Zohar}}]{Kasper2021nonabelian}%
  \BibitemOpen
  \bibfield  {author} {\bibinfo {author} {\bibfnamefont {Valentin}\
  \bibnamefont {Kasper}}, \bibinfo {author} {\bibfnamefont {Torsten~V.}\
  \bibnamefont {Zache}}, \bibinfo {author} {\bibfnamefont {Fred}\ \bibnamefont
  {Jendrzejewski}}, \bibinfo {author} {\bibfnamefont {Maciej}\ \bibnamefont
  {Lewenstein}}, \ and\ \bibinfo {author} {\bibfnamefont {Erez}\ \bibnamefont
  {Zohar}},\ }\bibfield  {title} {\enquote {\bibinfo {title} {Non-abelian gauge
  invariance from dynamical decoupling},}\ }\href@noop {} {\  (\bibinfo {year}
  {2021})},\ \Eprint {http://arxiv.org/abs/2012.08620} {arXiv:2012.08620
  [quant-ph]} \BibitemShut {NoStop}%
\bibitem [{\citenamefont {Lamm}\ \emph {et~al.}(2020)\citenamefont {Lamm},
  \citenamefont {Lawrence},\ and\ \citenamefont {Yamauchi}}]{Lamm2020}%
  \BibitemOpen
  \bibfield  {author} {\bibinfo {author} {\bibfnamefont {Henry}\ \bibnamefont
  {Lamm}}, \bibinfo {author} {\bibfnamefont {Scott}\ \bibnamefont {Lawrence}},
  \ and\ \bibinfo {author} {\bibfnamefont {Yukari}\ \bibnamefont {Yamauchi}},\
  }\bibfield  {title} {\enquote {\bibinfo {title} {Suppressing coherent gauge
  drift in quantum simulations},}\ }\href {https://arxiv.org/abs/2005.12688} {\
   (\bibinfo {year} {2020})},\ \Eprint {http://arxiv.org/abs/2005.12688}
  {arXiv:2005.12688 [quant-ph]} \BibitemShut {NoStop}%
\bibitem [{\citenamefont {Halimeh}\ \emph {et~al.}(2021)\citenamefont
  {Halimeh}, \citenamefont {Lang}, \citenamefont {Mildenberger}, \citenamefont
  {Jiang},\ and\ \citenamefont {Hauke}}]{Halimeh2020e}%
  \BibitemOpen
  \bibfield  {author} {\bibinfo {author} {\bibfnamefont {Jad~C.}\ \bibnamefont
  {Halimeh}}, \bibinfo {author} {\bibfnamefont {Haifeng}\ \bibnamefont {Lang}},
  \bibinfo {author} {\bibfnamefont {Julius}\ \bibnamefont {Mildenberger}},
  \bibinfo {author} {\bibfnamefont {Zhang}\ \bibnamefont {Jiang}}, \ and\
  \bibinfo {author} {\bibfnamefont {Philipp}\ \bibnamefont {Hauke}},\
  }\bibfield  {title} {\enquote {\bibinfo {title} {Gauge-symmetry protection
  using single-body terms},}\ }\href {\doibase 10.1103/PRXQuantum.2.040311}
  {\bibfield  {journal} {\bibinfo  {journal} {PRX Quantum}\ }\textbf {\bibinfo
  {volume} {2}},\ \bibinfo {pages} {040311} (\bibinfo {year}
  {2021})}\BibitemShut {NoStop}%
\bibitem [{\citenamefont {Halimeh}\ \emph
  {et~al.}(2022{\natexlab{a}})\citenamefont {Halimeh}, \citenamefont {Homeier},
  \citenamefont {Schweizer}, \citenamefont {Aidelsburger}, \citenamefont
  {Hauke},\ and\ \citenamefont {Grusdt}}]{Halimeh2021stabilizing}%
  \BibitemOpen
  \bibfield  {author} {\bibinfo {author} {\bibfnamefont {Jad~C.}\ \bibnamefont
  {Halimeh}}, \bibinfo {author} {\bibfnamefont {Lukas}\ \bibnamefont
  {Homeier}}, \bibinfo {author} {\bibfnamefont {Christian}\ \bibnamefont
  {Schweizer}}, \bibinfo {author} {\bibfnamefont {Monika}\ \bibnamefont
  {Aidelsburger}}, \bibinfo {author} {\bibfnamefont {Philipp}\ \bibnamefont
  {Hauke}}, \ and\ \bibinfo {author} {\bibfnamefont {Fabian}\ \bibnamefont
  {Grusdt}},\ }\bibfield  {title} {\enquote {\bibinfo {title} {Stabilizing
  lattice gauge theories through simplified local pseudogenerators},}\ }\href
  {\doibase 10.1103/PhysRevResearch.4.033120} {\bibfield  {journal} {\bibinfo
  {journal} {Phys. Rev. Research}\ }\textbf {\bibinfo {volume} {4}},\ \bibinfo
  {pages} {033120} (\bibinfo {year} {2022}{\natexlab{a}})}\BibitemShut
  {NoStop}%
\bibitem [{\citenamefont {Halimeh}\ \emph
  {et~al.}(2022{\natexlab{b}})\citenamefont {Halimeh}, \citenamefont {Lang},\
  and\ \citenamefont {Hauke}}]{Halimeh2021gauge}%
  \BibitemOpen
  \bibfield  {author} {\bibinfo {author} {\bibfnamefont {Jad~C.}\ \bibnamefont
  {Halimeh}}, \bibinfo {author} {\bibfnamefont {Haifeng}\ \bibnamefont {Lang}},
  \ and\ \bibinfo {author} {\bibfnamefont {Philipp}\ \bibnamefont {Hauke}},\
  }\bibfield  {title} {\enquote {\bibinfo {title} {Gauge protection in
  non-abelian lattice gauge theories},}\ }\href
  {http://iopscience.iop.org/article/10.1088/1367-2630/ac5564} {\bibfield
  {journal} {\bibinfo  {journal} {New Journal of Physics}\ } (\bibinfo {year}
  {2022}{\natexlab{b}})}\BibitemShut {NoStop}%
\bibitem [{\citenamefont {Halimeh}\ and\ \citenamefont
  {Hauke}(2022)}]{Halimeh_BriefReview}%
  \BibitemOpen
  \bibfield  {author} {\bibinfo {author} {\bibfnamefont {Jad~C.}\ \bibnamefont
  {Halimeh}}\ and\ \bibinfo {author} {\bibfnamefont {Philipp}\ \bibnamefont
  {Hauke}},\ }\bibfield  {title} {\enquote {\bibinfo {title} {Stabilizing gauge
  theories in quantum simulators: A brief review},}\ }\href {\doibase
  10.48550/ARXIV.2204.13709} {\  (\bibinfo {year} {2022}),\
  10.48550/ARXIV.2204.13709}\BibitemShut {NoStop}%
\bibitem [{\citenamefont {Halimeh}\ and\ \citenamefont
  {Hauke}(2020{\natexlab{d}})}]{Halimeh2020g}%
  \BibitemOpen
  \bibfield  {author} {\bibinfo {author} {\bibfnamefont {Jad~C.}\ \bibnamefont
  {Halimeh}}\ and\ \bibinfo {author} {\bibfnamefont {Philipp}\ \bibnamefont
  {Hauke}},\ }\bibfield  {title} {\enquote {\bibinfo {title}
  {Diffusive-to-ballistic crossover of symmetry violation in open many-body
  systems},}\ }\href@noop {} {\  (\bibinfo {year} {2020}{\natexlab{d}})},\
  \Eprint {http://arxiv.org/abs/2010.00009} {arXiv:2010.00009
  [cond-mat.quant-gas]} \BibitemShut {NoStop}%
\bibitem [{\citenamefont {Zeh}(1970)}]{Zeh1970}%
  \BibitemOpen
  \bibfield  {author} {\bibinfo {author} {\bibfnamefont {H.~D.}\ \bibnamefont
  {Zeh}},\ }\bibfield  {title} {\enquote {\bibinfo {title} {On the
  interpretation of measurement in quantum theory},}\ }\href {\doibase
  10.1007/BF00708656} {\bibfield  {journal} {\bibinfo  {journal} {Foundations
  of Physics}\ }\textbf {\bibinfo {volume} {1}},\ \bibinfo {pages} {69--76}
  (\bibinfo {year} {1970})}\BibitemShut {NoStop}%
\bibitem [{\citenamefont {Schlosshauer}(2005)}]{Schlosshauer2005}%
  \BibitemOpen
  \bibfield  {author} {\bibinfo {author} {\bibfnamefont {Maximilian}\
  \bibnamefont {Schlosshauer}},\ }\bibfield  {title} {\enquote {\bibinfo
  {title} {Decoherence, the measurement problem, and interpretations of quantum
  mechanics},}\ }\href {\doibase 10.1103/RevModPhys.76.1267} {\bibfield
  {journal} {\bibinfo  {journal} {Rev. Mod. Phys.}\ }\textbf {\bibinfo {volume}
  {76}},\ \bibinfo {pages} {1267--1305} (\bibinfo {year} {2005})}\BibitemShut
  {NoStop}%
\bibitem [{\citenamefont {Yoshihara}\ \emph {et~al.}(2006)\citenamefont
  {Yoshihara}, \citenamefont {Harrabi}, \citenamefont {Niskanen}, \citenamefont
  {Nakamura},\ and\ \citenamefont {Tsai}}]{Yoshihara2006}%
  \BibitemOpen
  \bibfield  {author} {\bibinfo {author} {\bibfnamefont {F.}~\bibnamefont
  {Yoshihara}}, \bibinfo {author} {\bibfnamefont {K.}~\bibnamefont {Harrabi}},
  \bibinfo {author} {\bibfnamefont {A.~O.}\ \bibnamefont {Niskanen}}, \bibinfo
  {author} {\bibfnamefont {Y.}~\bibnamefont {Nakamura}}, \ and\ \bibinfo
  {author} {\bibfnamefont {J.~S.}\ \bibnamefont {Tsai}},\ }\bibfield  {title}
  {\enquote {\bibinfo {title} {Decoherence of flux qubits due to $1/f$ flux
  noise},}\ }\href {\doibase 10.1103/PhysRevLett.97.167001} {\bibfield
  {journal} {\bibinfo  {journal} {Phys. Rev. Lett.}\ }\textbf {\bibinfo
  {volume} {97}},\ \bibinfo {pages} {167001} (\bibinfo {year}
  {2006})}\BibitemShut {NoStop}%
\bibitem [{\citenamefont {Kakuyanagi}\ \emph {et~al.}(2007)\citenamefont
  {Kakuyanagi}, \citenamefont {Meno}, \citenamefont {Saito}, \citenamefont
  {Nakano}, \citenamefont {Semba}, \citenamefont {Takayanagi}, \citenamefont
  {Deppe},\ and\ \citenamefont {Shnirman}}]{Kakuyanagi2007}%
  \BibitemOpen
  \bibfield  {author} {\bibinfo {author} {\bibfnamefont {K.}~\bibnamefont
  {Kakuyanagi}}, \bibinfo {author} {\bibfnamefont {T.}~\bibnamefont {Meno}},
  \bibinfo {author} {\bibfnamefont {S.}~\bibnamefont {Saito}}, \bibinfo
  {author} {\bibfnamefont {H.}~\bibnamefont {Nakano}}, \bibinfo {author}
  {\bibfnamefont {K.}~\bibnamefont {Semba}}, \bibinfo {author} {\bibfnamefont
  {H.}~\bibnamefont {Takayanagi}}, \bibinfo {author} {\bibfnamefont
  {F.}~\bibnamefont {Deppe}}, \ and\ \bibinfo {author} {\bibfnamefont
  {A.}~\bibnamefont {Shnirman}},\ }\bibfield  {title} {\enquote {\bibinfo
  {title} {Dephasing of a superconducting flux qubit},}\ }\href {\doibase
  10.1103/PhysRevLett.98.047004} {\bibfield  {journal} {\bibinfo  {journal}
  {Phys. Rev. Lett.}\ }\textbf {\bibinfo {volume} {98}},\ \bibinfo {pages}
  {047004} (\bibinfo {year} {2007})}\BibitemShut {NoStop}%
\bibitem [{\citenamefont {Bialczak}\ \emph {et~al.}(2007)\citenamefont
  {Bialczak}, \citenamefont {McDermott}, \citenamefont {Ansmann}, \citenamefont
  {Hofheinz}, \citenamefont {Katz}, \citenamefont {Lucero}, \citenamefont
  {Neeley}, \citenamefont {O'Connell}, \citenamefont {Wang}, \citenamefont
  {Cleland},\ and\ \citenamefont {Martinis}}]{Bialczak2007}%
  \BibitemOpen
  \bibfield  {author} {\bibinfo {author} {\bibfnamefont {Radoslaw~C.}\
  \bibnamefont {Bialczak}}, \bibinfo {author} {\bibfnamefont {R.}~\bibnamefont
  {McDermott}}, \bibinfo {author} {\bibfnamefont {M.}~\bibnamefont {Ansmann}},
  \bibinfo {author} {\bibfnamefont {M.}~\bibnamefont {Hofheinz}}, \bibinfo
  {author} {\bibfnamefont {N.}~\bibnamefont {Katz}}, \bibinfo {author}
  {\bibfnamefont {Erik}\ \bibnamefont {Lucero}}, \bibinfo {author}
  {\bibfnamefont {Matthew}\ \bibnamefont {Neeley}}, \bibinfo {author}
  {\bibfnamefont {A.~D.}\ \bibnamefont {O'Connell}}, \bibinfo {author}
  {\bibfnamefont {H.}~\bibnamefont {Wang}}, \bibinfo {author} {\bibfnamefont
  {A.~N.}\ \bibnamefont {Cleland}}, \ and\ \bibinfo {author} {\bibfnamefont
  {John~M.}\ \bibnamefont {Martinis}},\ }\bibfield  {title} {\enquote {\bibinfo
  {title} {$1/f$ flux noise in josephson phase qubits},}\ }\href {\doibase
  10.1103/PhysRevLett.99.187006} {\bibfield  {journal} {\bibinfo  {journal}
  {Phys. Rev. Lett.}\ }\textbf {\bibinfo {volume} {99}},\ \bibinfo {pages}
  {187006} (\bibinfo {year} {2007})}\BibitemShut {NoStop}%
\bibitem [{\citenamefont {Bylander}\ \emph {et~al.}(2011)\citenamefont
  {Bylander}, \citenamefont {Gustavsson}, \citenamefont {Yan}, \citenamefont
  {Yoshihara}, \citenamefont {Harrabi}, \citenamefont {Fitch}, \citenamefont
  {Cory}, \citenamefont {Nakamura}, \citenamefont {Tsai},\ and\ \citenamefont
  {Oliver}}]{Bylander2011}%
  \BibitemOpen
  \bibfield  {author} {\bibinfo {author} {\bibfnamefont {Jonas}\ \bibnamefont
  {Bylander}}, \bibinfo {author} {\bibfnamefont {Simon}\ \bibnamefont
  {Gustavsson}}, \bibinfo {author} {\bibfnamefont {Fei}\ \bibnamefont {Yan}},
  \bibinfo {author} {\bibfnamefont {Fumiki}\ \bibnamefont {Yoshihara}},
  \bibinfo {author} {\bibfnamefont {Khalil}\ \bibnamefont {Harrabi}}, \bibinfo
  {author} {\bibfnamefont {George}\ \bibnamefont {Fitch}}, \bibinfo {author}
  {\bibfnamefont {David~G.}\ \bibnamefont {Cory}}, \bibinfo {author}
  {\bibfnamefont {Yasunobu}\ \bibnamefont {Nakamura}}, \bibinfo {author}
  {\bibfnamefont {Jaw-Shen}\ \bibnamefont {Tsai}}, \ and\ \bibinfo {author}
  {\bibfnamefont {William~D.}\ \bibnamefont {Oliver}},\ }\bibfield  {title}
  {\enquote {\bibinfo {title} {Noise spectroscopy through dynamical decoupling
  with a superconducting flux qubit},}\ }\href {\doibase 10.1038/nphys1994}
  {\bibfield  {journal} {\bibinfo  {journal} {Nature Physics}\ }\textbf
  {\bibinfo {volume} {7}},\ \bibinfo {pages} {565--570} (\bibinfo {year}
  {2011})}\BibitemShut {NoStop}%
\bibitem [{\citenamefont {Wang}\ \emph {et~al.}(2015)\citenamefont {Wang},
  \citenamefont {Shi}, \citenamefont {Hu}, \citenamefont {Han}, \citenamefont
  {Yu},\ and\ \citenamefont {Wu}}]{Wang2015}%
  \BibitemOpen
  \bibfield  {author} {\bibinfo {author} {\bibfnamefont {Hui}\ \bibnamefont
  {Wang}}, \bibinfo {author} {\bibfnamefont {Chuntai}\ \bibnamefont {Shi}},
  \bibinfo {author} {\bibfnamefont {Jun}\ \bibnamefont {Hu}}, \bibinfo {author}
  {\bibfnamefont {Sungho}\ \bibnamefont {Han}}, \bibinfo {author}
  {\bibfnamefont {Clare~C.}\ \bibnamefont {Yu}}, \ and\ \bibinfo {author}
  {\bibfnamefont {R.~Q.}\ \bibnamefont {Wu}},\ }\bibfield  {title} {\enquote
  {\bibinfo {title} {Candidate source of flux noise in squids: Adsorbed oxygen
  molecules},}\ }\href {\doibase 10.1103/PhysRevLett.115.077002} {\bibfield
  {journal} {\bibinfo  {journal} {Phys. Rev. Lett.}\ }\textbf {\bibinfo
  {volume} {115}},\ \bibinfo {pages} {077002} (\bibinfo {year}
  {2015})}\BibitemShut {NoStop}%
\bibitem [{\citenamefont {Kumar}\ \emph {et~al.}(2016)\citenamefont {Kumar},
  \citenamefont {Sendelbach}, \citenamefont {Beck}, \citenamefont {Freeland},
  \citenamefont {Wang}, \citenamefont {Wang}, \citenamefont {Yu}, \citenamefont
  {Wu}, \citenamefont {Pappas},\ and\ \citenamefont {McDermott}}]{Kumar2016}%
  \BibitemOpen
  \bibfield  {author} {\bibinfo {author} {\bibfnamefont {P.}~\bibnamefont
  {Kumar}}, \bibinfo {author} {\bibfnamefont {S.}~\bibnamefont {Sendelbach}},
  \bibinfo {author} {\bibfnamefont {M.~A.}\ \bibnamefont {Beck}}, \bibinfo
  {author} {\bibfnamefont {J.~W.}\ \bibnamefont {Freeland}}, \bibinfo {author}
  {\bibfnamefont {Zhe}\ \bibnamefont {Wang}}, \bibinfo {author} {\bibfnamefont
  {Hui}\ \bibnamefont {Wang}}, \bibinfo {author} {\bibfnamefont {Clare~C.}\
  \bibnamefont {Yu}}, \bibinfo {author} {\bibfnamefont {R.~Q.}\ \bibnamefont
  {Wu}}, \bibinfo {author} {\bibfnamefont {D.~P.}\ \bibnamefont {Pappas}}, \
  and\ \bibinfo {author} {\bibfnamefont {R.}~\bibnamefont {McDermott}},\
  }\bibfield  {title} {\enquote {\bibinfo {title} {Origin and reduction of
  $1/f$ magnetic flux noise in superconducting devices},}\ }\href {\doibase
  10.1103/PhysRevApplied.6.041001} {\bibfield  {journal} {\bibinfo  {journal}
  {Phys. Rev. Applied}\ }\textbf {\bibinfo {volume} {6}},\ \bibinfo {pages}
  {041001} (\bibinfo {year} {2016})}\BibitemShut {NoStop}%
\bibitem [{\citenamefont {Abanin}\ \emph {et~al.}(2017)\citenamefont {Abanin},
  \citenamefont {De~Roeck}, \citenamefont {Ho},\ and\ \citenamefont
  {Huveneers}}]{ARHH2017}%
  \BibitemOpen
  \bibfield  {author} {\bibinfo {author} {\bibfnamefont {Dmitry}\ \bibnamefont
  {Abanin}}, \bibinfo {author} {\bibfnamefont {Wojciech}\ \bibnamefont
  {De~Roeck}}, \bibinfo {author} {\bibfnamefont {Wen~Wei}\ \bibnamefont {Ho}},
  \ and\ \bibinfo {author} {\bibfnamefont {Fran{\c c}ois}\ \bibnamefont
  {Huveneers}},\ }\bibfield  {title} {\enquote {\bibinfo {title} {A rigorous
  theory of many-body prethermalization for periodically driven and closed
  quantum systems},}\ }\href {\doibase 10.1007/s00220-017-2930-x} {\bibfield
  {journal} {\bibinfo  {journal} {Communications in Mathematical Physics}\
  }\textbf {\bibinfo {volume} {354}},\ \bibinfo {pages} {809--827} (\bibinfo
  {year} {2017})}\BibitemShut {NoStop}%
\bibitem [{\citenamefont {Damme}\ \emph
  {et~al.}(2021{\natexlab{a}})\citenamefont {Damme}, \citenamefont {Lang},
  \citenamefont {Hauke},\ and\ \citenamefont
  {Halimeh}}]{vandamme2021reliability}%
  \BibitemOpen
  \bibfield  {author} {\bibinfo {author} {\bibfnamefont {Maarten~Van}\
  \bibnamefont {Damme}}, \bibinfo {author} {\bibfnamefont {Haifeng}\
  \bibnamefont {Lang}}, \bibinfo {author} {\bibfnamefont {Philipp}\
  \bibnamefont {Hauke}}, \ and\ \bibinfo {author} {\bibfnamefont {Jad~C.}\
  \bibnamefont {Halimeh}},\ }\bibfield  {title} {\enquote {\bibinfo {title}
  {Reliability of lattice gauge theories in the thermodynamic limit},}\
  }\href@noop {} {\  (\bibinfo {year} {2021}{\natexlab{a}})},\ \Eprint
  {http://arxiv.org/abs/2104.07040} {arXiv:2104.07040 [cond-mat.quant-gas]}
  \BibitemShut {NoStop}%
\bibitem [{\citenamefont {Facchi}\ and\ \citenamefont
  {Pascazio}(2002)}]{facchi2002quantum}%
  \BibitemOpen
  \bibfield  {author} {\bibinfo {author} {\bibfnamefont {P.}~\bibnamefont
  {Facchi}}\ and\ \bibinfo {author} {\bibfnamefont {S.}~\bibnamefont
  {Pascazio}},\ }\bibfield  {title} {\enquote {\bibinfo {title} {Quantum zeno
  subspaces},}\ }\href {\doibase 10.1103/PhysRevLett.89.080401} {\bibfield
  {journal} {\bibinfo  {journal} {Phys. Rev. Lett.}\ }\textbf {\bibinfo
  {volume} {89}},\ \bibinfo {pages} {080401} (\bibinfo {year}
  {2002})}\BibitemShut {NoStop}%
\bibitem [{\citenamefont {Facchi}\ \emph {et~al.}(2004)\citenamefont {Facchi},
  \citenamefont {Lidar},\ and\ \citenamefont
  {Pascazio}}]{facchi2004unification}%
  \BibitemOpen
  \bibfield  {author} {\bibinfo {author} {\bibfnamefont {P.}~\bibnamefont
  {Facchi}}, \bibinfo {author} {\bibfnamefont {D.~A.}\ \bibnamefont {Lidar}}, \
  and\ \bibinfo {author} {\bibfnamefont {S.}~\bibnamefont {Pascazio}},\
  }\bibfield  {title} {\enquote {\bibinfo {title} {Unification of dynamical
  decoupling and the quantum zeno effect},}\ }\href {\doibase
  10.1103/PhysRevA.69.032314} {\bibfield  {journal} {\bibinfo  {journal} {Phys.
  Rev. A}\ }\textbf {\bibinfo {volume} {69}},\ \bibinfo {pages} {032314}
  (\bibinfo {year} {2004})}\BibitemShut {NoStop}%
\bibitem [{\citenamefont {Facchi}\ \emph {et~al.}(2009)\citenamefont {Facchi},
  \citenamefont {Marmo},\ and\ \citenamefont {Pascazio}}]{facchi2009quantum}%
  \BibitemOpen
  \bibfield  {author} {\bibinfo {author} {\bibfnamefont {Paolo}\ \bibnamefont
  {Facchi}}, \bibinfo {author} {\bibfnamefont {Giuseppe}\ \bibnamefont
  {Marmo}}, \ and\ \bibinfo {author} {\bibfnamefont {Saverio}\ \bibnamefont
  {Pascazio}},\ }\bibfield  {title} {\enquote {\bibinfo {title} {Quantum zeno
  dynamics and quantum zeno subspaces},}\ }\href {\doibase
  10.1088/1742-6596/196/1/012017} {\ \textbf {\bibinfo {volume} {196}},\
  \bibinfo {pages} {012017} (\bibinfo {year} {2009})}\BibitemShut {NoStop}%
\bibitem [{\citenamefont {Burgarth}\ \emph {et~al.}(2019)\citenamefont
  {Burgarth}, \citenamefont {Facchi}, \citenamefont {Nakazato}, \citenamefont
  {Pascazio},\ and\ \citenamefont {Yuasa}}]{burgarth2019generalized}%
  \BibitemOpen
  \bibfield  {author} {\bibinfo {author} {\bibfnamefont {Daniel}\ \bibnamefont
  {Burgarth}}, \bibinfo {author} {\bibfnamefont {Paolo}\ \bibnamefont
  {Facchi}}, \bibinfo {author} {\bibfnamefont {Hiromichi}\ \bibnamefont
  {Nakazato}}, \bibinfo {author} {\bibfnamefont {Saverio}\ \bibnamefont
  {Pascazio}}, \ and\ \bibinfo {author} {\bibfnamefont {Kazuya}\ \bibnamefont
  {Yuasa}},\ }\bibfield  {title} {\enquote {\bibinfo {title} {Generalized
  {A}diabatic {T}heorem and {S}trong-{C}oupling {L}imits},}\ }\href {\doibase
  10.22331/q-2019-06-12-152} {\bibfield  {journal} {\bibinfo  {journal}
  {{Quantum}}\ }\textbf {\bibinfo {volume} {3}},\ \bibinfo {pages} {152}
  (\bibinfo {year} {2019})}\BibitemShut {NoStop}%
\bibitem [{\citenamefont {Halimeh}\ \emph
  {et~al.}(2022{\natexlab{c}})\citenamefont {Halimeh}, \citenamefont {Homeier},
  \citenamefont {Zhao}, \citenamefont {Bohrdt}, \citenamefont {Grusdt},
  \citenamefont {Hauke},\ and\ \citenamefont {Knolle}}]{Halimeh2021enhancing}%
  \BibitemOpen
  \bibfield  {author} {\bibinfo {author} {\bibfnamefont {Jad~C.}\ \bibnamefont
  {Halimeh}}, \bibinfo {author} {\bibfnamefont {Lukas}\ \bibnamefont
  {Homeier}}, \bibinfo {author} {\bibfnamefont {Hongzheng}\ \bibnamefont
  {Zhao}}, \bibinfo {author} {\bibfnamefont {Annabelle}\ \bibnamefont
  {Bohrdt}}, \bibinfo {author} {\bibfnamefont {Fabian}\ \bibnamefont {Grusdt}},
  \bibinfo {author} {\bibfnamefont {Philipp}\ \bibnamefont {Hauke}}, \ and\
  \bibinfo {author} {\bibfnamefont {Johannes}\ \bibnamefont {Knolle}},\
  }\bibfield  {title} {\enquote {\bibinfo {title} {Enhancing disorder-free
  localization through dynamically emergent local symmetries},}\ }\href
  {\doibase 10.1103/PRXQuantum.3.020345} {\bibfield  {journal} {\bibinfo
  {journal} {PRX Quantum}\ }\textbf {\bibinfo {volume} {3}},\ \bibinfo {pages}
  {020345} (\bibinfo {year} {2022}{\natexlab{c}})}\BibitemShut {NoStop}%
\bibitem [{\citenamefont {Haegeman}\ \emph {et~al.}(2011)\citenamefont
  {Haegeman}, \citenamefont {Cirac}, \citenamefont {Osborne}, \citenamefont
  {Pi\ifmmode~\check{z}\else \v{z}\fi{}orn}, \citenamefont {Verschelde},\ and\
  \citenamefont {Verstraete}}]{Haegeman2011}%
  \BibitemOpen
  \bibfield  {author} {\bibinfo {author} {\bibfnamefont {Jutho}\ \bibnamefont
  {Haegeman}}, \bibinfo {author} {\bibfnamefont {J.~Ignacio}\ \bibnamefont
  {Cirac}}, \bibinfo {author} {\bibfnamefont {Tobias~J.}\ \bibnamefont
  {Osborne}}, \bibinfo {author} {\bibfnamefont {Iztok}\ \bibnamefont
  {Pi\ifmmode~\check{z}\else \v{z}\fi{}orn}}, \bibinfo {author} {\bibfnamefont
  {Henri}\ \bibnamefont {Verschelde}}, \ and\ \bibinfo {author} {\bibfnamefont
  {Frank}\ \bibnamefont {Verstraete}},\ }\bibfield  {title} {\enquote {\bibinfo
  {title} {Time-dependent variational principle for quantum lattices},}\ }\href
  {\doibase 10.1103/PhysRevLett.107.070601} {\bibfield  {journal} {\bibinfo
  {journal} {Phys. Rev. Lett.}\ }\textbf {\bibinfo {volume} {107}},\ \bibinfo
  {pages} {070601} (\bibinfo {year} {2011})}\BibitemShut {NoStop}%
\bibitem [{\citenamefont {Haegeman}\ \emph {et~al.}(2013)\citenamefont
  {Haegeman}, \citenamefont {Osborne},\ and\ \citenamefont
  {Verstraete}}]{Haegeman2013}%
  \BibitemOpen
  \bibfield  {author} {\bibinfo {author} {\bibfnamefont {Jutho}\ \bibnamefont
  {Haegeman}}, \bibinfo {author} {\bibfnamefont {Tobias~J.}\ \bibnamefont
  {Osborne}}, \ and\ \bibinfo {author} {\bibfnamefont {Frank}\ \bibnamefont
  {Verstraete}},\ }\bibfield  {title} {\enquote {\bibinfo {title} {Post-matrix
  product state methods: To tangent space and beyond},}\ }\href {\doibase
  10.1103/PhysRevB.88.075133} {\bibfield  {journal} {\bibinfo  {journal} {Phys.
  Rev. B}\ }\textbf {\bibinfo {volume} {88}},\ \bibinfo {pages} {075133}
  (\bibinfo {year} {2013})}\BibitemShut {NoStop}%
\bibitem [{\citenamefont {Haegeman}\ \emph {et~al.}(2016)\citenamefont
  {Haegeman}, \citenamefont {Lubich}, \citenamefont {Oseledets}, \citenamefont
  {Vandereycken},\ and\ \citenamefont {Verstraete}}]{Haegeman2016}%
  \BibitemOpen
  \bibfield  {author} {\bibinfo {author} {\bibfnamefont {Jutho}\ \bibnamefont
  {Haegeman}}, \bibinfo {author} {\bibfnamefont {Christian}\ \bibnamefont
  {Lubich}}, \bibinfo {author} {\bibfnamefont {Ivan}\ \bibnamefont
  {Oseledets}}, \bibinfo {author} {\bibfnamefont {Bart}\ \bibnamefont
  {Vandereycken}}, \ and\ \bibinfo {author} {\bibfnamefont {Frank}\
  \bibnamefont {Verstraete}},\ }\bibfield  {title} {\enquote {\bibinfo {title}
  {Unifying time evolution and optimization with matrix product states},}\
  }\href {\doibase 10.1103/PhysRevB.94.165116} {\bibfield  {journal} {\bibinfo
  {journal} {Phys. Rev. B}\ }\textbf {\bibinfo {volume} {94}},\ \bibinfo
  {pages} {165116} (\bibinfo {year} {2016})}\BibitemShut {NoStop}%
\bibitem [{\citenamefont {Damme}\ \emph
  {et~al.}(2021{\natexlab{b}})\citenamefont {Damme}, \citenamefont
  {Mildenberger}, \citenamefont {Grusdt}, \citenamefont {Hauke},\ and\
  \citenamefont {Halimeh}}]{vandamme2021suppressing}%
  \BibitemOpen
  \bibfield  {author} {\bibinfo {author} {\bibfnamefont {Maarten~Van}\
  \bibnamefont {Damme}}, \bibinfo {author} {\bibfnamefont {Julius}\
  \bibnamefont {Mildenberger}}, \bibinfo {author} {\bibfnamefont {Fabian}\
  \bibnamefont {Grusdt}}, \bibinfo {author} {\bibfnamefont {Philipp}\
  \bibnamefont {Hauke}}, \ and\ \bibinfo {author} {\bibfnamefont {Jad~C.}\
  \bibnamefont {Halimeh}},\ }\bibfield  {title} {\enquote {\bibinfo {title}
  {Suppressing nonperturbative gauge errors in the thermodynamic limit using
  local pseudogenerators},}\ }\href@noop {} {\  (\bibinfo {year}
  {2021}{\natexlab{b}})},\ \Eprint {http://arxiv.org/abs/2110.08041}
  {arXiv:2110.08041 [quant-ph]} \BibitemShut {NoStop}%
\bibitem [{\citenamefont {{Press}}(1978)}]{Press1978}%
  \BibitemOpen
  \bibfield  {author} {\bibinfo {author} {\bibfnamefont {W.~H.}\ \bibnamefont
  {{Press}}},\ }\bibfield  {title} {\enquote {\bibinfo {title} {{Flicker noises
  in astronomy and elsewhere.}}}\ }\href@noop {} {\bibfield  {journal}
  {\bibinfo  {journal} {Comments on Astrophysics}\ }\textbf {\bibinfo {volume}
  {7}},\ \bibinfo {pages} {103--119} (\bibinfo {year} {1978})}\BibitemShut
  {NoStop}%
\bibitem [{\citenamefont {Kogan}\ and\ \citenamefont
  {Kogan}(1996)}]{kogan1996electronic}%
  \BibitemOpen
  \bibfield  {author} {\bibinfo {author} {\bibfnamefont {S.}~\bibnamefont
  {Kogan}}\ and\ \bibinfo {author} {\bibfnamefont {S.}~\bibnamefont {Kogan}},\
  }\href {https://books.google.de/books?id=s5tupGCMzBYC} {\emph {\bibinfo
  {title} {Electronic Noise and Fluctuations in Solids}}}\ (\bibinfo
  {publisher} {Cambridge University Press},\ \bibinfo {year}
  {1996})\BibitemShut {NoStop}%
\bibitem [{\citenamefont {Cohen-Tannoudji}\ \emph {et~al.}(1992)\citenamefont
  {Cohen-Tannoudji}, \citenamefont {Dupont-Roc},\ and\ \citenamefont
  {Grynberg}}]{cohen1992atom}%
  \BibitemOpen
  \bibfield  {author} {\bibinfo {author} {\bibfnamefont {C.}~\bibnamefont
  {Cohen-Tannoudji}}, \bibinfo {author} {\bibfnamefont {J.}~\bibnamefont
  {Dupont-Roc}}, \ and\ \bibinfo {author} {\bibfnamefont {G.}~\bibnamefont
  {Grynberg}},\ }\href {https://books.google.de/books?id=m7gPAQAAMAAJ} {\emph
  {\bibinfo {title} {Atom-photon interactions: basic processes and
  applications}}},\ Wiley-Interscience publication\ (\bibinfo  {publisher} {J.
  Wiley},\ \bibinfo {year} {1992})\BibitemShut {NoStop}%
\bibitem [{\citenamefont {Breuer}\ \emph {et~al.}(2002)\citenamefont {Breuer},
  \citenamefont {Petruccione},\ and\ \citenamefont
  {Petruccione}}]{breuer2002theory}%
  \BibitemOpen
  \bibfield  {author} {\bibinfo {author} {\bibfnamefont {H.P.}\ \bibnamefont
  {Breuer}}, \bibinfo {author} {\bibfnamefont {F.}~\bibnamefont {Petruccione}},
  \ and\ \bibinfo {author} {\bibfnamefont {S.P.A.P.F.}\ \bibnamefont
  {Petruccione}},\ }\href {https://books.google.de/books?id=0Yx5VzaMYm8C}
  {\emph {\bibinfo {title} {The Theory of Open Quantum Systems}}}\ (\bibinfo
  {publisher} {Oxford University Press},\ \bibinfo {year} {2002})\BibitemShut
  {NoStop}%
\bibitem [{\citenamefont {Amin}\ \emph {et~al.}(2009)\citenamefont {Amin},
  \citenamefont {Truncik},\ and\ \citenamefont {Averin}}]{PhysRevA.80.022303}%
  \BibitemOpen
  \bibfield  {author} {\bibinfo {author} {\bibfnamefont {M.~H.~S.}\
  \bibnamefont {Amin}}, \bibinfo {author} {\bibfnamefont {C.~J.~S.}\
  \bibnamefont {Truncik}}, \ and\ \bibinfo {author} {\bibfnamefont {D.~V.}\
  \bibnamefont {Averin}},\ }\bibfield  {title} {\enquote {\bibinfo {title}
  {Role of single-qubit decoherence time in adiabatic quantum computation},}\
  }\href {\doibase 10.1103/PhysRevA.80.022303} {\bibfield  {journal} {\bibinfo
  {journal} {Phys. Rev. A}\ }\textbf {\bibinfo {volume} {80}},\ \bibinfo
  {pages} {022303} (\bibinfo {year} {2009})}\BibitemShut {NoStop}%
\bibitem [{\citenamefont {Johansson}\ \emph {et~al.}(2012)\citenamefont
  {Johansson}, \citenamefont {Nation},\ and\ \citenamefont
  {Nori}}]{Johansson2012}%
  \BibitemOpen
  \bibfield  {author} {\bibinfo {author} {\bibfnamefont {J.R.}\ \bibnamefont
  {Johansson}}, \bibinfo {author} {\bibfnamefont {P.D.}\ \bibnamefont
  {Nation}}, \ and\ \bibinfo {author} {\bibfnamefont {Franco}\ \bibnamefont
  {Nori}},\ }\bibfield  {title} {\enquote {\bibinfo {title} {Qutip: An
  open-source python framework for the dynamics of open quantum systems},}\
  }\href {\doibase https://doi.org/10.1016/j.cpc.2012.02.021} {\bibfield
  {journal} {\bibinfo  {journal} {Computer Physics Communications}\ }\textbf
  {\bibinfo {volume} {183}},\ \bibinfo {pages} {1760 -- 1772} (\bibinfo {year}
  {2012})}\BibitemShut {NoStop}%
\bibitem [{\citenamefont {Johansson}\ \emph {et~al.}(2013)\citenamefont
  {Johansson}, \citenamefont {Nation},\ and\ \citenamefont
  {Nori}}]{Johansson2013}%
  \BibitemOpen
  \bibfield  {author} {\bibinfo {author} {\bibfnamefont {J.R.}\ \bibnamefont
  {Johansson}}, \bibinfo {author} {\bibfnamefont {P.D.}\ \bibnamefont
  {Nation}}, \ and\ \bibinfo {author} {\bibfnamefont {Franco}\ \bibnamefont
  {Nori}},\ }\bibfield  {title} {\enquote {\bibinfo {title} {Qutip 2: A python
  framework for the dynamics of open quantum systems},}\ }\href {\doibase
  https://doi.org/10.1016/j.cpc.2012.11.019} {\bibfield  {journal} {\bibinfo
  {journal} {Computer Physics Communications}\ }\textbf {\bibinfo {volume}
  {184}},\ \bibinfo {pages} {1234 -- 1240} (\bibinfo {year}
  {2013})}\BibitemShut {NoStop}%
\bibitem [{\citenamefont {Chandrasekharan}\ and\ \citenamefont
  {Wiese}(1997)}]{Chandrasekharan1997}%
  \BibitemOpen
  \bibfield  {author} {\bibinfo {author} {\bibfnamefont {S}~\bibnamefont
  {Chandrasekharan}}\ and\ \bibinfo {author} {\bibfnamefont {U.-J}\
  \bibnamefont {Wiese}},\ }\bibfield  {title} {\enquote {\bibinfo {title}
  {Quantum link models: A discrete approach to gauge theories},}\ }\href
  {\doibase https://doi.org/10.1016/S0550-3213(97)80041-7} {\bibfield
  {journal} {\bibinfo  {journal} {Nuclear Physics B}\ }\textbf {\bibinfo
  {volume} {492}},\ \bibinfo {pages} {455 -- 471} (\bibinfo {year}
  {1997})}\BibitemShut {NoStop}%
\bibitem [{\citenamefont {Wiese}(2013)}]{Wiese_review}%
  \BibitemOpen
  \bibfield  {author} {\bibinfo {author} {\bibfnamefont {U.-J.}\ \bibnamefont
  {Wiese}},\ }\bibfield  {title} {\enquote {\bibinfo {title} {Ultracold quantum
  gases and lattice systems: quantum simulation of lattice gauge theories},}\
  }\href {\doibase 10.1002/andp.201300104} {\bibfield  {journal} {\bibinfo
  {journal} {Annalen der Physik}\ }\textbf {\bibinfo {volume} {525}},\ \bibinfo
  {pages} {777--796} (\bibinfo {year} {2013})}\BibitemShut {NoStop}%
\bibitem [{\citenamefont {Kasper}\ \emph {et~al.}(2017)\citenamefont {Kasper},
  \citenamefont {Hebenstreit}, \citenamefont {Jendrzejewski}, \citenamefont
  {Oberthaler},\ and\ \citenamefont {Berges}}]{Kasper2017}%
  \BibitemOpen
  \bibfield  {author} {\bibinfo {author} {\bibfnamefont {V}~\bibnamefont
  {Kasper}}, \bibinfo {author} {\bibfnamefont {F}~\bibnamefont {Hebenstreit}},
  \bibinfo {author} {\bibfnamefont {F}~\bibnamefont {Jendrzejewski}}, \bibinfo
  {author} {\bibfnamefont {M~K}\ \bibnamefont {Oberthaler}}, \ and\ \bibinfo
  {author} {\bibfnamefont {J}~\bibnamefont {Berges}},\ }\bibfield  {title}
  {\enquote {\bibinfo {title} {Implementing quantum electrodynamics with
  ultracold atomic systems},}\ }\href {\doibase 10.1088/1367-2630/aa54e0}
  {\bibfield  {journal} {\bibinfo  {journal} {New Journal of Physics}\ }\textbf
  {\bibinfo {volume} {19}},\ \bibinfo {pages} {023030} (\bibinfo {year}
  {2017})}\BibitemShut {NoStop}%
\bibitem [{\citenamefont {Barbiero}\ \emph {et~al.}(2019)\citenamefont
  {Barbiero}, \citenamefont {Schweizer}, \citenamefont {Aidelsburger},
  \citenamefont {Demler}, \citenamefont {Goldman},\ and\ \citenamefont
  {Grusdt}}]{Barbiero2019}%
  \BibitemOpen
  \bibfield  {author} {\bibinfo {author} {\bibfnamefont {Luca}\ \bibnamefont
  {Barbiero}}, \bibinfo {author} {\bibfnamefont {Christian}\ \bibnamefont
  {Schweizer}}, \bibinfo {author} {\bibfnamefont {Monika}\ \bibnamefont
  {Aidelsburger}}, \bibinfo {author} {\bibfnamefont {Eugene}\ \bibnamefont
  {Demler}}, \bibinfo {author} {\bibfnamefont {Nathan}\ \bibnamefont
  {Goldman}}, \ and\ \bibinfo {author} {\bibfnamefont {Fabian}\ \bibnamefont
  {Grusdt}},\ }\bibfield  {title} {\enquote {\bibinfo {title} {Coupling
  ultracold matter to dynamical gauge fields in optical lattices: From flux
  attachment to $\mathbb{Z}_2$ lattice gauge theories},}\ }\href {\doibase
  10.1126/sciadv.aav7444} {\bibfield  {journal} {\bibinfo  {journal} {Science
  Advances}\ }\textbf {\bibinfo {volume} {5}} (\bibinfo {year} {2019}),\
  10.1126/sciadv.aav7444}\BibitemShut {NoStop}%
\bibitem [{\citenamefont {Zohar}\ \emph {et~al.}(2017)\citenamefont {Zohar},
  \citenamefont {Farace}, \citenamefont {Reznik},\ and\ \citenamefont
  {Cirac}}]{Zohar2017}%
  \BibitemOpen
  \bibfield  {author} {\bibinfo {author} {\bibfnamefont {Erez}\ \bibnamefont
  {Zohar}}, \bibinfo {author} {\bibfnamefont {Alessandro}\ \bibnamefont
  {Farace}}, \bibinfo {author} {\bibfnamefont {Benni}\ \bibnamefont {Reznik}},
  \ and\ \bibinfo {author} {\bibfnamefont {J.~Ignacio}\ \bibnamefont {Cirac}},\
  }\bibfield  {title} {\enquote {\bibinfo {title} {Digital quantum simulation
  of ${\mathbb{z}}_{2}$ lattice gauge theories with dynamical fermionic
  matter},}\ }\href {\doibase 10.1103/PhysRevLett.118.070501} {\bibfield
  {journal} {\bibinfo  {journal} {Phys. Rev. Lett.}\ }\textbf {\bibinfo
  {volume} {118}},\ \bibinfo {pages} {070501} (\bibinfo {year}
  {2017})}\BibitemShut {NoStop}%
\bibitem [{\citenamefont {Borla}\ \emph {et~al.}(2020)\citenamefont {Borla},
  \citenamefont {Verresen}, \citenamefont {Grusdt},\ and\ \citenamefont
  {Moroz}}]{Borla2019}%
  \BibitemOpen
  \bibfield  {author} {\bibinfo {author} {\bibfnamefont {Umberto}\ \bibnamefont
  {Borla}}, \bibinfo {author} {\bibfnamefont {Ruben}\ \bibnamefont {Verresen}},
  \bibinfo {author} {\bibfnamefont {Fabian}\ \bibnamefont {Grusdt}}, \ and\
  \bibinfo {author} {\bibfnamefont {Sergej}\ \bibnamefont {Moroz}},\ }\bibfield
   {title} {\enquote {\bibinfo {title} {Confined phases of one-dimensional
  spinless fermions coupled to ${Z}_{2}$ gauge theory},}\ }\href {\doibase
  10.1103/PhysRevLett.124.120503} {\bibfield  {journal} {\bibinfo  {journal}
  {Phys. Rev. Lett.}\ }\textbf {\bibinfo {volume} {124}},\ \bibinfo {pages}
  {120503} (\bibinfo {year} {2020})}\BibitemShut {NoStop}%
\bibitem [{\citenamefont {Yang}\ \emph
  {et~al.}(2020{\natexlab{b}})\citenamefont {Yang}, \citenamefont {Liu},
  \citenamefont {Gorshkov},\ and\ \citenamefont
  {Iadecola}}]{Yang2020fragmentation}%
  \BibitemOpen
  \bibfield  {author} {\bibinfo {author} {\bibfnamefont {Zhi-Cheng}\
  \bibnamefont {Yang}}, \bibinfo {author} {\bibfnamefont {Fangli}\ \bibnamefont
  {Liu}}, \bibinfo {author} {\bibfnamefont {Alexey~V.}\ \bibnamefont
  {Gorshkov}}, \ and\ \bibinfo {author} {\bibfnamefont {Thomas}\ \bibnamefont
  {Iadecola}},\ }\bibfield  {title} {\enquote {\bibinfo {title} {Hilbert-space
  fragmentation from strict confinement},}\ }\href {\doibase
  10.1103/PhysRevLett.124.207602} {\bibfield  {journal} {\bibinfo  {journal}
  {Phys. Rev. Lett.}\ }\textbf {\bibinfo {volume} {124}},\ \bibinfo {pages}
  {207602} (\bibinfo {year} {2020}{\natexlab{b}})}\BibitemShut {NoStop}%
\bibitem [{\citenamefont {Kebri\ifmmode~\check{c}\else \v{c}\fi{}}\ \emph
  {et~al.}(2021)\citenamefont {Kebri\ifmmode~\check{c}\else \v{c}\fi{}},
  \citenamefont {Barbiero}, \citenamefont {Reinmoser}, \citenamefont
  {Schollw\"ock},\ and\ \citenamefont {Grusdt}}]{kebric2021confinement}%
  \BibitemOpen
  \bibfield  {author} {\bibinfo {author} {\bibfnamefont {Matja\ifmmode
  \check{z}\else~\v{z}\fi{}}\ \bibnamefont {Kebri\ifmmode~\check{c}\else
  \v{c}\fi{}}}, \bibinfo {author} {\bibfnamefont {Luca}\ \bibnamefont
  {Barbiero}}, \bibinfo {author} {\bibfnamefont {Christian}\ \bibnamefont
  {Reinmoser}}, \bibinfo {author} {\bibfnamefont {Ulrich}\ \bibnamefont
  {Schollw\"ock}}, \ and\ \bibinfo {author} {\bibfnamefont {Fabian}\
  \bibnamefont {Grusdt}},\ }\bibfield  {title} {\enquote {\bibinfo {title}
  {Confinement and mott transitions of dynamical charges in one-dimensional
  lattice gauge theories},}\ }\href {\doibase 10.1103/PhysRevLett.127.167203}
  {\bibfield  {journal} {\bibinfo  {journal} {Phys. Rev. Lett.}\ }\textbf
  {\bibinfo {volume} {127}},\ \bibinfo {pages} {167203} (\bibinfo {year}
  {2021})}\BibitemShut {NoStop}%
\bibitem [{\citenamefont {Borla}\ \emph {et~al.}(2021)\citenamefont {Borla},
  \citenamefont {Verresen}, \citenamefont {Shah},\ and\ \citenamefont
  {Moroz}}]{Borla2020}%
  \BibitemOpen
  \bibfield  {author} {\bibinfo {author} {\bibfnamefont {Umberto}\ \bibnamefont
  {Borla}}, \bibinfo {author} {\bibfnamefont {Ruben}\ \bibnamefont {Verresen}},
  \bibinfo {author} {\bibfnamefont {Jeet}\ \bibnamefont {Shah}}, \ and\
  \bibinfo {author} {\bibfnamefont {Sergej}\ \bibnamefont {Moroz}},\
  }\href@noop {} {\enquote {\bibinfo {title} {Gauging the kitaev chain},}\ }
  (\bibinfo {year} {2021}),\ \Eprint {http://arxiv.org/abs/2010.00607}
  {arXiv:2010.00607 [cond-mat.str-el]} \BibitemShut {NoStop}%
\bibitem [{\citenamefont {{Lang}}\ \emph {et~al.}(2022)\citenamefont {{Lang}},
  \citenamefont {{Hauke}}, \citenamefont {{Knolle}}, \citenamefont {{Grusdt}},\
  and\ \citenamefont {{Halimeh}}}]{Lang2022stark}%
  \BibitemOpen
  \bibfield  {author} {\bibinfo {author} {\bibfnamefont {Haifeng}\ \bibnamefont
  {{Lang}}}, \bibinfo {author} {\bibfnamefont {Philipp}\ \bibnamefont
  {{Hauke}}}, \bibinfo {author} {\bibfnamefont {Johannes}\ \bibnamefont
  {{Knolle}}}, \bibinfo {author} {\bibfnamefont {Fabian}\ \bibnamefont
  {{Grusdt}}}, \ and\ \bibinfo {author} {\bibfnamefont {Jad~C.}\ \bibnamefont
  {{Halimeh}}},\ }\bibfield  {title} {\enquote {\bibinfo {title}
  {{Disorder-free localization with Stark gauge protection}},}\ }\href@noop {}
  {\bibfield  {journal} {\bibinfo  {journal} {arXiv e-prints}\ ,\ \bibinfo
  {eid} {arXiv:2203.01338}} (\bibinfo {year} {2022})},\ \Eprint
  {http://arxiv.org/abs/2203.01338} {arXiv:2203.01338 [cond-mat.quant-gas]}
  \BibitemShut {NoStop}%
\bibitem [{\citenamefont {Albash}\ \emph {et~al.}(2012)\citenamefont {Albash},
  \citenamefont {Boixo}, \citenamefont {Lidar},\ and\ \citenamefont
  {Zanardi}}]{Albash_2012}%
  \BibitemOpen
  \bibfield  {author} {\bibinfo {author} {\bibfnamefont {Tameem}\ \bibnamefont
  {Albash}}, \bibinfo {author} {\bibfnamefont {Sergio}\ \bibnamefont {Boixo}},
  \bibinfo {author} {\bibfnamefont {Daniel~A}\ \bibnamefont {Lidar}}, \ and\
  \bibinfo {author} {\bibfnamefont {Paolo}\ \bibnamefont {Zanardi}},\
  }\bibfield  {title} {\enquote {\bibinfo {title} {Quantum adiabatic markovian
  master equations},}\ }\href {\doibase 10.1088/1367-2630/14/12/123016}
  {\bibfield  {journal} {\bibinfo  {journal} {New Journal of Physics}\ }\textbf
  {\bibinfo {volume} {14}},\ \bibinfo {pages} {123016} (\bibinfo {year}
  {2012})}\BibitemShut {NoStop}%
\bibitem [{\citenamefont {Davies}(1976)}]{Davies1976}%
  \BibitemOpen
  \bibfield  {author} {\bibinfo {author} {\bibfnamefont {E.B.}\ \bibnamefont
  {Davies}},\ }\bibfield  {title} {\enquote {\bibinfo {title} {Markovian master
  equations. ii.}}\ }\href {http://eudml.org/doc/182764} {\bibfield  {journal}
  {\bibinfo  {journal} {Mathematische Annalen}\ }\textbf {\bibinfo {volume}
  {219}},\ \bibinfo {pages} {147--158} (\bibinfo {year} {1976})}\BibitemShut
  {NoStop}%
\bibitem [{\citenamefont {{Davies}}(1974)}]{1974CMaPh..39...91D}%
  \BibitemOpen
  \bibfield  {author} {\bibinfo {author} {\bibfnamefont {E.~B.}\ \bibnamefont
  {{Davies}}},\ }\bibfield  {title} {\enquote {\bibinfo {title} {{Markovian
  master equations}},}\ }\href {\doibase 10.1007/BF01608389} {\bibfield
  {journal} {\bibinfo  {journal} {Communications in Mathematical Physics}\
  }\textbf {\bibinfo {volume} {39}},\ \bibinfo {pages} {91--110} (\bibinfo
  {year} {1974})}\BibitemShut {NoStop}%
\bibitem [{\citenamefont
  {Lidar}(2019)}]{https://doi.org/10.48550/arxiv.1902.00967}%
  \BibitemOpen
  \bibfield  {author} {\bibinfo {author} {\bibfnamefont {Daniel~A.}\
  \bibnamefont {Lidar}},\ }\href {\doibase 10.48550/ARXIV.1902.00967} {\enquote
  {\bibinfo {title} {Lecture notes on the theory of open quantum systems},}\ }
  (\bibinfo {year} {2019})\BibitemShut {NoStop}%
\bibitem [{\citenamefont {Jordan}\ \emph {et~al.}(2006)\citenamefont {Jordan},
  \citenamefont {Farhi},\ and\ \citenamefont {Shor}}]{PhysRevA.74.052322}%
  \BibitemOpen
  \bibfield  {author} {\bibinfo {author} {\bibfnamefont {Stephen~P.}\
  \bibnamefont {Jordan}}, \bibinfo {author} {\bibfnamefont {Edward}\
  \bibnamefont {Farhi}}, \ and\ \bibinfo {author} {\bibfnamefont {Peter~W.}\
  \bibnamefont {Shor}},\ }\bibfield  {title} {\enquote {\bibinfo {title}
  {Error-correcting codes for adiabatic quantum computation},}\ }\href
  {\doibase 10.1103/PhysRevA.74.052322} {\bibfield  {journal} {\bibinfo
  {journal} {Phys. Rev. A}\ }\textbf {\bibinfo {volume} {74}},\ \bibinfo
  {pages} {052322} (\bibinfo {year} {2006})}\BibitemShut {NoStop}%
\end{thebibliography}%
\end{document}